\documentclass{article}

\usepackage{microtype}
\usepackage{graphicx}
\usepackage{subcaption}
\usepackage{booktabs} 
\usepackage{adjustbox}
\usepackage{rotating}
\usepackage{multirow}

\usepackage{hyperref}

\usepackage[preprint]{icml2026}

\usepackage{amsmath}
\usepackage{amssymb}
\usepackage{mathtools}
\usepackage{amsthm}

\usepackage[capitalize,noabbrev]{cleveref}

\theoremstyle{plain}

\theoremstyle{definition}

\theoremstyle{remark}

\usepackage[textsize=tiny]{todonotes}

\icmltitlerunning{Batch Effects In Brain Foundation Model Embeddings}

\begin{document}

\twocolumn[
  \icmltitle{Batch Effects In Brain Foundation Model Embeddings}



  \icmlsetsymbol{equal}{*}

  \begin{icmlauthorlist}
    \icmlauthor{Ye Tao}{equal,rutgers}
    \icmlauthor{Bradley T.~Baker}{equal,trends}
    \icmlauthor{Yu Wu}{rutgers}
    \icmlauthor{Anand D.~Sarwate}{rutgers}
    \icmlauthor{Sandeep Panta}{trends}
    \icmlauthor{Sergey Plis}{trends}
    \icmlauthor{Vince D.~Calhoun}{trends}
  \end{icmlauthorlist}

  \icmlaffiliation{rutgers}{Department of Electrical and Computer Engineering, Rutgers University, NJ, USA}
  \icmlaffiliation{trends}{TReNDS Center, Georgia State, Georgia Tech, and Emory, GA, USA}

  \icmlcorrespondingauthor{Ye Tao}{yt371@rutgers.edu}
  \icmlcorrespondingauthor{Anand D.~Sarwate}{anand.sarwate@rutgers.edu}

  \icmlkeywords{Machine Learning, ICML}

  \vskip 0.3in
]



\printAffiliationsAndNotice{\icmlEqualContribution}

\begin{abstract}
Foundation models show strong potential for large-scale, high-dimensional biomedical applications, yet their ability to capture relevant neurobiological characteristics remains underexplored. We systematically evaluate embeddings from two neuroimaging foundation models, BrainLM and SwiFT, across multi-site fMRI datasets using a comprehensive evaluation framework. Our results show that foundation model embeddings encode substantial batch-related variability, often dominating diagnosis-related information across heterogeneous datasets. We further investigate how harmonization, applied to reduce batch effects, influences these embeddings. In addition, we find that BrainLM prefers to capture fine-grained regional activity, whereas SwiFT tends to represent interactions between regions, consistent with their respective model architectures. Our study highlights the importance of accounting for batch effects in foundation models and motivates future work on disentangling biologically meaningful signals from acquisition-related variability.
\end{abstract}

\section{Introduction}
Neuroimaging data, particularly functional magnetic resonance imaging (fMRI), can present unique challenges to deep learning models due to their complex spatiotemporal structure. Transformer-based foundation models have recently emerged as a promising approach for modeling neuroimaging data~\cite{caro2023brainlm, kim2023swift, zheng20254dfcf, wang2026graph}. Complementing traditional approaches using decompositions (e.g., independent component analysis; ICA~\cite{beckmann2004probabilistic, calhoun2001method}) and summary representations (e.g., functional network connectivity; FNC~\cite{jafri2008method, smith2013functional}), these models operate directly on high-dimensional spatiotemporal fMRI signals, enabling end-to-end learning of complex, context-dependent dependencies that are difficult to capture within fixed representational frameworks.

Foundation models are often adapted to downstream tasks using finetuning or by extracting learned representations to apply to other downstream models~\cite{bommasani2021opportunities, devlin2019bert, he2022masked, chen2020simple}. 
Some recent work 
using such embeddings in neuroimaging applications suggests they can be used for zero-shot generalization~\cite{caro2023brainlm, kim2023swift}, although the intrinsic properties of these embeddings are not yet fully understood. It remains unclear whether these representations primarily capture biologically meaningful signals or are influenced by confounding factors present in the data. Recent studies in other domains have shown that embeddings extracted from foundation models can reveal latent, source-specific signals~\cite{vargas2025understanding}, suggesting that a similar effect may occur when analyzing neuroimaging embeddings. This issue is particularly important in 
neuroimaging, where data collection is costly and large-scale studies require 
collaborations across institutions using approaches such as federated learning~\cite{plis2016coinstac}. In multi-site settings, systematic differences across scanners, acquisition protocols, and preprocessing pipelines introduce substantial non-biological variability, commonly referred to as site effects or batch effects~\cite{johnson2007adjusting, fortin2017harmonization, han2006reliability, yu2018statistical}.

In this paper we conduct a systematic empirical study of two neuroimaging foundations models and multiple multi-site fMRI datasets to better understand what 
their embeddings capture.
We show embeddings exhibit strong batch-dependent structures that can dominate clinically relevant signals. This means models may achieve strong predictive performance by exploiting batch-related signals as shortcuts, so high accuracy on downstream tasks does not necessarily indicate that clinically relevant patterns are being captured. Moreover, the presence of batch information in the embedding space may reveal sensitive metadata, raising potential privacy and data sharing concerns~\cite{song2020information}. We further find that while traditional harmonization techniques~\cite{fortin2017harmonization, chen2022mitigating, moyer2020scanner, fortin2018harmonization} applied to the embeddings can mitigate batch-related effects, they do not necessarily enhance clinically relevant signals. Finally, we explore the interpretability of these embeddings to study which biological signals are emphasized by different foundation models, revealing systematic differences consistent with model architecture. These analyses highlight the importance of understanding foundation model embeddings, emphasizing the need for robust, batch- or data source-aware approaches in multi-site neuroimaging studies, as well as in other domains where models are trained on heterogeneous datasets.

\begin{figure*}[htb!]
\centering     
\includegraphics[width=17cm]{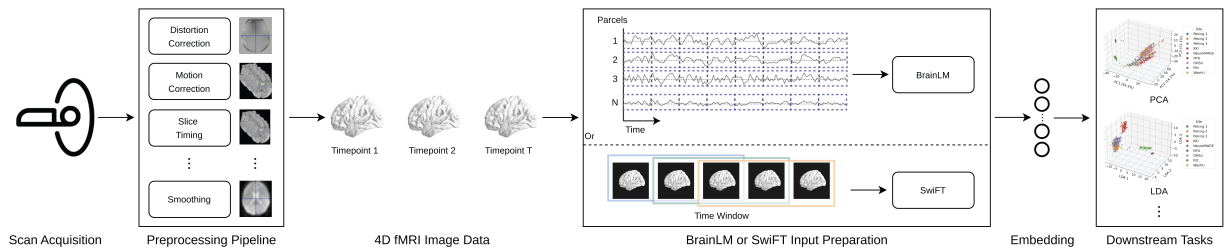}
\caption{Overview of the fMRI representation pipeline using foundation models. fMRI scans are encoded into low-dimensional embeddings by foundation models. These embeddings are used for dimensionality reduction (e.g., PCA, LDA) and predictive modeling.}
\label{fig:embedding_pipeline}
\vspace{-0.25cm}
\end{figure*}

\section{Background and Related Work}

\textbf{Fundamentals of fMRI Data.}
Resting-state fMRI captures the blood oxygenation level dependent (BOLD) signal while subjects remain in the scanner without performing specific tasks~\cite{poldrack2024handbook, buxton2009introduction}, producing a sequence of volumetric images over time in which each voxel encodes the temporal evolution of BOLD signal intensity, resulting in high-dimensional spatiotemporal data. Prior to analysis, standard preprocessing steps, including motion correction, spatial normalization, and temporal filtering, are typically applied to reduce noise and artifacts~\cite{esteban2019fmriprep, poldrack2024handbook}. To reduce dimensionality and improve interpretability, voxel-level signals are often aggregated into regions of interest (ROIs) based on predefined parcellation schemes~\cite{tzourio2002automated, craddock2012whole, gordon2016generation, du2020neuromark} or into overlapping brain networks via data-driven approaches such as independent component analysis (ICA)~\cite{mckeown1998analysis, beckmann2004probabilistic, calhoun2001method}.

Functional connectivity measures quantify statistical dependencies between brain regions, capturing how regions interact with one another over time. A widely used approach is FNC, which typically computes pairwise correlations between BOLD activations in different regions~\cite{mohanty2020rethinking}. Various statistical methods are employed to quantify these dependencies, including Pearson's correlation coefficient and mutual information~\cite{marrelec2006partial, hlinka2011functional}. In contrast, regional activity measures characterize the magnitude of spontaneous neural fluctuations within individual brain regions, reflecting the intrinsic activity of each region. A commonly used metric in this category is the amplitude of low-frequency fluctuations (ALFF)~\cite{yu2007altered, zou2008improved}, which quantifies the power of the BOLD signal within a low-frequency band by transforming voxel-wise timeseries into the frequency domain using a fast Fourier transform.

\textbf{Deep Learning for fMRI Analysis.}
Early deep learning approaches for neuroimaging analysis primarily rely on low-dimensional representations such as functional connectivity or FNC. These representations motivated the application of graph neural networks (GNNs) and convolutional neural networks (CNNs), which capture topological interactions between brain regions~\cite{ktena2018metric, li2021braingnn, kawahara2017brainnetcnn, li2022learning, thapaliya2025brain, hu2026brainib++}. For example, BrainNetCNN~\cite{kawahara2017brainnetcnn} introduced specialized edge-to-edge and edge-to-node filters to learn from connectivity matrices. While these methods are computationally efficient and relatively robust to local noise, they inherently abstract away fine-grained spatial heterogeneity and high-frequency temporal dynamics, potentially limiting their ability to capture richer patterns in fMRI timeseries. To better preserve the inherent spatiotemporal structure of neural activity, subsequent research shifted toward volumetric models that operate directly on 3D or 4D fMRI data. Advanced architectures, such as 3D convolutional neural networks (3D-CNNs) and hybrid CNN-LSTM frameworks, were designed to jointly extract hierarchical spatial representations and model long-term temporal dynamics~\cite{nie20163d, li2020detecting}. However, most existing volumetric models rely on supervised learning and large labeled datasets, which may limit their generalizability across clinical populations.

More recently, transformer-based architectures have enabled the development of brain foundation models, representing brain signals as sequences or graph-structured tokens to capture long-range spatiotemporal interactions and whole-brain dynamics. 
ROI-based transformer models, like BrainLM~\cite{caro2023brainlm}, employ masked autoencoding to reconstruct masked ROI timeseries, thereby learning intrinsic functional patterns. Graph transformer models~\cite{wang2026graph} extend this approach by modeling node features and edge interactions in a graph autoencoder framework, enabling connectivity-aware attention. In contrast, voxel-level transformer architectures, like SwiFT~\cite{kim2023swift} and 4DfCF~\cite{zheng20254dfcf}, preserve fine-grained spatiotemporal information through hierarchical and local/global attention mechanisms, maintaining computational scalability while capturing detailed voxel-level dynamics.

\textbf{Batch Effects and Harmonization Methods.} Presence of batch or site effects has long been recognized as a major confounder in multi-site neuroimaging studies. Prior work~\cite{han2006reliability, power2012spurious, bayer2022site} has documented non-biological variability arising from differences in scanner hardware, acquisition protocols, and acquisition environments, which can introduce shifts in signal intensity, contrast, and noise characteristics across sites. Additional variability may stem from subject characteristics and preprocessing choices. Such variability can obscure subtle neural signals, potentially leading to inconsistent or inflated findings in clinical analyses.

To mitigate batch effects, a range of harmonization techniques have been proposed. Early approaches relied on classical statistical models~\cite{bernal2013statistical}, such as generalized linear models and linear mixed-effects models, which treat site differences as systematic additive or multiplicative offsets. While computationally straightforward and highly interpretable, these methods typically assume linear relationships and are univariate, limiting their ability to capture complex, non-linear, or multivariate site-specific variations. Empirical Bayes methods~\cite{johnson2007adjusting, fortin2017harmonization, yu2018statistical}, most notably ComBat~\cite{fortin2017harmonization, yu2018statistical} and its extensions such as CovBat~\cite{chen2022mitigating}, explicitly model both mean and variance shifts across sites and improve stability in small-sample settings. They have been successfully applied to various neuroimaging metrics, including functional connectivity matrices, and remain a practical standard for harmonizing summary-level neuroimaging features. Deep learning-based harmonization techniques~\cite{moyer2020scanner, hu2024deepcombat, yan2022harmless, yan2023maximum}, such as variational autoencoders and generative adversarial networks, have been developed to capture more complex, non-linear site effects. These models have shown significant promise in structural MRI and static connectome settings. However, most frameworks operate on manually aggregated or static representations, and the harmonization of embeddings learned directly from raw spatiotemporal fMRI signals, such as those produced by foundation models, remains largely unexplored.

\section{Methods}
Given preprocessed fMRI timeseries, a pre-trained foundation model produces subject-level embeddings. We aim to understand what information they capture, in particular whether they encode biologically meaningful signals, batch-specific variability, or other spurious factors.

We extract embeddings from two representative pre-trained foundation models, BrainLM and SwiFT (see Figure~\ref{fig:embedding_pipeline}). For BrainLM~\cite{caro2023brainlm}, which is trained with a masked autoencoding objective, voxel-wise fMRI data are first parcellated into ROI-level timeseries using the AAL-424 atlas~\cite{rolls2020automated}. The transformer encoder outputs token-wise hidden states for each ROI along with a dedicated CLS token, and the hidden state of the CLS token is used as the subject-level embedding. For SwiFT~\cite{kim2023swift}, whose pre-training is based on a contrastive learning objective, volumetric fMRI inputs are divided into multiple temporal windows. Each 4D window is processed by the \href{https://github.com/Transconnectome/SwiFT/blob/main/pretrained_models/contrastive_pretrained.ckpt}{pre-trained model} to produce an embedding, and subject-level representations are obtained by averaging embeddings across all windows of the same subject. In this work, foundation models are treated as fixed feature extractors, allowing us to isolate the intrinsic properties of the learned representations without task-specific finetuning. 

Our analysis integrates visualization, statistical analysis, predictive assessment, controlled experimental designs, harmonization, and interpretability analyses. Visualization and statistical analyses reveal intrinsic patterns in the embedding space, predictive assessment quantifies how embeddings reflect site or diagnosis differences, controlled experiments disentangle confounding factors, and harmonization reduce source-specific effects. Interpretability analyses further indicate which biological signals are preferentially represented by different foundation models. 
Details are in Section~\ref{sec:exp_res}.

\section{Experiments and Results}
\label{sec:exp_res}

We use 
three widely used multi-site resting-state fMRI benchmark datasets: FBIRN~\cite{keator2016function}, ADHD-200~\cite{bellec2017neuro}, and ABIDE~I~\cite{nielsen2013multisite}. These datasets were collected across multiple sites and geographic regions. 
Details of data acquisition, preprocessing, and cohort statistics are in Appendix~\ref{appendix:datasets}.

\textbf{Intrinsic Encoding of Batch Effects in Embeddings.} To determine whether embeddings from pre-trained foundation models capture batch-related signals, subject-level embeddings are evaluated using both qualitative and quantitative approaches. For qualitative inspection, PCA is applied to visualize potential clustering patterns. Quantitative assessment of batch effects is performed via permutational multivariate analysis of variance (PERMANOVA) on Euclidean distances between embeddings. In addition, site identity predictability is evaluated using classifiers including linear discriminant analysis (LDA), logistic regression, and nonlinear kernel SVMs. These analyses reveal the extent to which pre-trained embeddings reflect batch-specific information.

\begin{figure}[htbp]
    \centering
    \begin{subfigure}{0.23\textwidth}
        \centering
        \includegraphics[width=\linewidth]{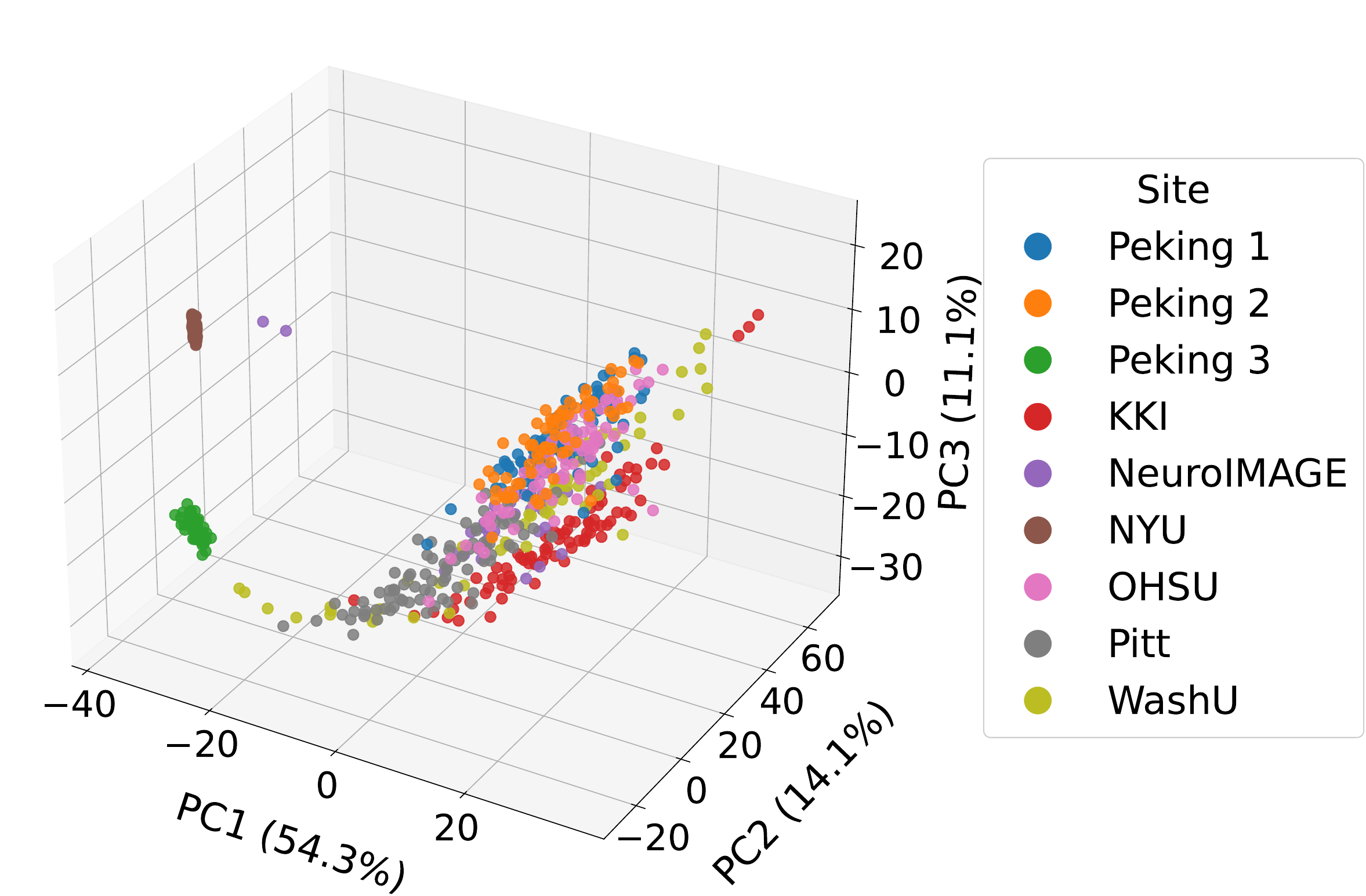}
        \caption{PCA on ADHD-200}
    \end{subfigure}
    \begin{subfigure}{0.23\textwidth}
        \centering
        \includegraphics[width=\linewidth]{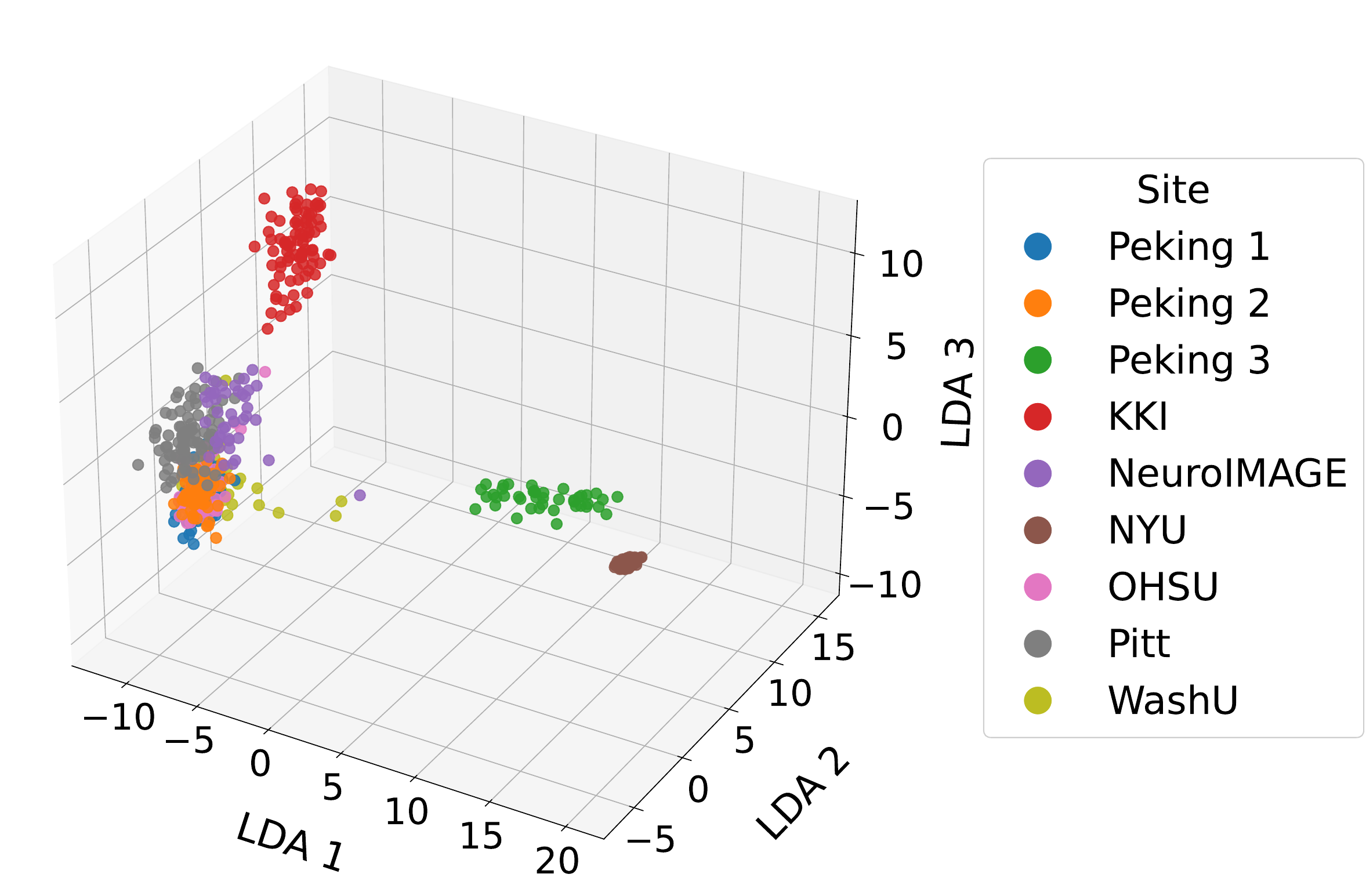}
        \caption{LDA on ADHD-200}
    \end{subfigure}
    \caption{Visualization of subject-level embeddings extracted from the pre-trained BrainLM model for the ADHD-200 dataset. 
    All points are colored according to site identity.}
    \label{fig:brainlm_adhd200_sites}
\end{figure}

Figure~\ref{fig:brainlm_adhd200_sites} (or Figure~\ref{fig:brainlm_sites} for full results) and Figure~\ref{fig:swift_sites} show embeddings from the pre-trained BrainLM and SwiFT models, projected into low-dimensional spaces using PCA and PCA followed by LDA for qualitative inspection. The projections reveal pronounced batch-dependent structures. In the ADHD-200 dataset, embeddings exhibit clustering aligned with site identity. KKI, the only site using a Philips 3T scanner, appears separated. In addition, although the Peking site used identical scanners and acquisition parameters across three batches, the third batch (Peking~3) forms the most distinct sub-cluster. These patterns indicate models capture not only coarse scanner differences but also subtle site- or batch-related variations. Similar clustering patterns are observed in the FBIRN and ABIDE~I datasets (see Appendix~\ref{appendix:supp_batch_effect}). Controlling for potential demographic confounds by regressing out age and gender has minimal effect on the batch-dependent structures, indicating that the observed clustering cannot be explained solely by demographic differences.

PERMANOVA results (see Table~\ref{tab:permanova} in Appendix~\ref{appendix:supp_batch_effect}) quantitatively support these observations. For BrainLM embeddings, batch effects consistently dominate over diagnosis across all datasets. In contrast, SwiFT embeddings show more comparable effects between site and diagnosis in FBIRN, while batch effects remain stronger in ADHD-200 and ABIDE~I. Consistent with the PERMANOVA results, classifiers trained to predict site identity from pre-trained embeddings achieved high accuracy in the ADHD-200 dataset, which includes nine sites (see Table~\ref{tab:site_adhd200_diag_prediction} and Table~\ref{tab:site_diag_prediction} for full results). Site prediction was very high, indicating that acquisition-specific characteristics are strongly and largely separable in the embedding space. In contrast, diagnosis prediction performance is consistently lower than site prediction, and while SwiFT embeddings show moderately improved diagnosis prediction compared with BrainLM, clinical signals remain weaker and less separable than batch-related variability in the learned embeddings. Additional results for FBIRN and ABIDE~I, along with full site and diagnosis prediction tables, are reported in Appendix~\ref{appendix:supp_batch_effect}.

\begin{table}[htbp]
\vspace{-0.1cm}
\centering
\tiny
\caption{ADHD-200 dataset site and diagnosis prediction accuracy using embeddings from pre-trained foundation models.}
\vspace{-0.1cm}
\begin{tabular}{lllcc}
\toprule
Feature & Classifier & Site Acc & Diagnosis Acc \\
\midrule
\multirow{3}{*}{BrainLM}
& LDA & 94.30 & 66.67 \\
& Logistic Regression & 93.86 & 65.35 \\
& RBF SVM & 89.91 & 67.54 \\
\midrule
\multirow{3}{*}{SwiFT}
& LDA & 89.47 & 71.05 \\
& Logistic Regression & 88.60 & 70.18 \\
& RBF SVM & 85.09 & 68.42 \\
\bottomrule
\end{tabular}
\label{tab:site_adhd200_diag_prediction}
\vspace{-0.1cm}
\end{table}

\textbf{Batch Effects Versus Biological Signals.} To isolate effects of biologically meaningful signals and confounding batch-related variability, we design four controlled evaluation settings: within-site diagnosis, multi-site diagnosis, site prediction, and prediction under site-diagnosis confounding (see Table~\ref{tab:ablation_adhd200_results} and Table~\ref{tab:ablation_results} for full results). When training and testing within a single site, diagnostic classification accuracy remain relatively modest. It is natural to expect incorporating additional data from other sites would improve performance; however, this expectation is not always observed in practice, as pooling data across sites may degrade prediction performance. For example, on ADHD-200, training on Peking 1 and KKI using SwiFT embeddings yields lower accuracy than training on either site individually. Moreover, in an extreme yet realistic scenario where each site contains only a single diagnostic group, classification accuracy becomes near-perfect. Such results suggest model exploits batch/site-specific confounding signals rather than learning biologically meaningful disease patterns.

\begin{table}[htbp]
\vspace{-0.1cm}
\centering
\tiny
\caption{ADHD-200 classification accuracy using SwiFT embeddings. Setting shows experimental scenario, Task means prediction target, and Data shows site-specific subsets, e.g., KKI (Control) includes healthy subjects, and KKI (All) includes all subjects. 
}
\vspace{-0.1cm}
\begin{tabular}{l l l c}
\toprule
Setting & Task & Data & Acc \\
\midrule
Within-site & Diagnosis & Peking 1 (All) & 76.92 \\
Within-site & Diagnosis & KKI (All) & 72.00 \\
Multi-site & Diagnosis & Peking 1 (All) \& KKI (All) & 60.78 \\
Confounded & Confounded & Peking 1 (Control) \& KKI (Patient) & 100.00 \\
Confounded & Confounded & Peking 1 (Patient) \& KKI (Control) & 100.00 \\
Site-only & Site & Peking 1 (Control) \& KKI (Control) & 100.00 \\
Site-only & Site & Peking 1 (Patient) \& KKI (Patient) & 100.00 \\
\bottomrule
\end{tabular}
\label{tab:ablation_adhd200_results}
\vspace{-0.1cm}
\end{table}

\textbf{Harmonization on Pre-trained Embeddings.}
Traditional harmonization methods are primarily designed for summary features like FNC and are not readily applicable to high-dimensional fMRI timeseries. In contrast, pre-trained foundation models generate spatiotemporal embeddings that may confound batch-specific and biological variability. 
To understand if standard harmonization approaches can still be used, we applied the widely used ComBat harmonization method to both FNC features and pre-trained foundation model embeddings. 
We measured the effect on downstream site and diagnosis prediction  (see Table~\ref{tab:combat_site_diag_prediction_adhd200} and Table~\ref{tab:combat_site_diag_prediction} for full results). For FNC, ComBat reduces site prediction accuracy while leaving diagnosis prediction largely unchanged (see Table~\ref{tab:fnc_site_diag_prediction} in Appendix~\ref{appendix:fnc_baseline}), consistent with prior work suggesting that harmonization does not necessarily improve task performance; however, the removal of batch effects can result in increased predictive model robustness and generalizability~\cite{hu2023image}. For pre-trained embeddings, site prediction also decreases, though to a lesser extent than for FNC, while diagnosis prediction remains essentially unaffected. These results indicate that harmonization can partially mitigate batch effects, but batch-related variability persists in foundation model embeddings. Consequently, incorporating batch-effect control during pre-training or finetuning may be more effective than relying solely on downstream harmonization to disentangle biological signals from batch-specific artifacts in fMRI foundation models.

\begin{table}[htbp]
\vspace{-0.1cm}
\centering
\tiny
\caption{Site and diagnosis prediction accuracy for ADHD-200 after ComBat harmonization. 
}
\vspace{-0.1cm}
\begin{tabular}{lllcc}
\toprule
Feature & Classifier & Site Acc & Diagnosis Acc \\
\midrule
\multirow{3}{*}{FNC}
& LDA & 20.61 & 64.04 \\
& Logistic Regression & 19.74 & 63.60 \\
& RBF SVM & 34.65 & 62.28 \\
\midrule
\multirow{3}{*}{BrainLM}
& LDA & 28.51 & 65.79 \\
& Logistic Regression & 27.19 & 65.35 \\
& RBF SVM & 64.04 & 68.42 \\
\midrule
\multirow{3}{*}{SwiFT}
& LDA & 25.44 & 60.96 \\
& Logistic Regression & 31.14 & 60.09 \\
& RBF SVM & 52.19 & 65.79 \\
\bottomrule
\end{tabular}
\label{tab:combat_site_diag_prediction_adhd200}
\vspace{-0.1cm}
\end{table}

\textbf{Decoding Biological Signals from Latent Representations.} To assess the biological information encoded in the learned embeddings before and after ComBat harmonization, we performed linear decoding of regional and network-level functional measures using ridge regression. Both ALFF and FNC were predicted from the embeddings. ALFF quantifies the magnitude of spontaneous low-frequency activity within individual brain regions, reflecting intrinsic regional neural dynamics, whereas FNC captures inter-regional functional interactions at the network level. Decoding these measures provides a direct assessment of whether embeddings preserve biologically meaningful information related to both local activity and large-scale connectivity.

When batch effects were not controlled, the resulting $R^2$ maps exhibited widespread increases in the brain. As summarized in Table~\ref{tab:decode_r2_adhd200} and Table~\ref{tab:decode_r2} for full results, ALFF decoding produced abnormally high whole-brain $R^2$ values, with BrainLM consistently outperforming SwiFT across all datasets. A similar pattern was observed for FNC decoding, where network-level $R^2$ values remained uniformly high, but SwiFT consistently achieved higher performance than BrainLM. These uniformly elevated decoding accuracies indicate that the embeddings primarily capture scanner- or acquisition-related batch signatures rather than region-specific neural signals. Furthermore, the contrasting trends suggest distinct representational biases: BrainLM appears more sensitive to regional activity patterns, whereas SwiFT preferentially encodes large-scale functional connectivity. To disentangle representational characteristics from batch-related confounds, the analysis was repeated after applying ComBat harmonization. The observed trends persisted, with BrainLM outperforming SwiFT for ALFF and SwiFT outperforming BrainLM for FNC (see Table~\ref{tab:decode_r2_adhd200} and Table~\ref{tab:decode_r2}), indicating that these differences reflect intrinsic model representations rather than residual site effects. 

Beyond whole-brain average $R^2$ statistics, the spatial distributions of predictive performance further characterize how each model captures regional activity. Figure~\ref{fig:alff_3x2} visualizes the top 30\% $R^2$ regions for ALFF decoding after harmonization. BrainLM exhibits high-$R^2$ clusters in specific regions such as the thalamus, whereas SwiFT displays more spatially diffuse predictive patterns broadly distributed across cortical regions. To further quantify these differences at the systems level, we summarize decoding performance within functional networks by averaging $R^2$ values within each network. The resulting network-level comparisons (see Figure~\ref{fig:fnc_1x3}) provide a more interpretable view of which large-scale systems are preferentially captured by each model. Across all datasets, SwiFT consistently achieves higher mean $R^2$ across most functional networks, whereas BrainLM exhibits substantially weaker connectivity-related performance.

\begin{table}[htbp]
\vspace{-0.1cm}
\centering
\tiny
\caption{Whole-brain decoding performance ($R^2$) for ALFF and FNC in ADHD-200 before and after ComBat harmonization. Values are reported as mean across regions or connections.}
\vspace{-0.1cm}
\begin{tabular}{lcccc}
\toprule
\multirow{2}{*}{Feature} 
& \multicolumn{2}{c}{ALFF $R^2$} 
& \multicolumn{2}{c}{FNC $R^2$} \\
\cmidrule(lr){2-3} \cmidrule(lr){4-5}
& Raw & ComBat & Raw & ComBat \\
\midrule
BrainLM & 0.285 & 0.115 & 0.179 & 0.004 \\
SwiFT   & 0.275 & 0.061 & 0.197 & 0.009 \\
\bottomrule
\end{tabular}
\label{tab:decode_r2_adhd200}
\vspace{-0.1cm}
\end{table}

These complementary trends align with the distinct architectures and pre-training objectives of the two models. BrainLM employs a masked autoencoding objective that emphasizes reconstruction of local signal structure, promoting representations that preserve region-specific information. In contrast, SwiFT uses a hierarchical transformer with contrastive learning to aggregate information across regions and enhance global, subject-invariant features. Consequently, SwiFT representations may better capture large-scale functional interactions, whereas BrainLM representations remain more sensitive to regional activity patterns.

\section{Conclusion}
We investigate representational characteristics of large-scale brain foundation models and show pretrained BrainLM and SwiFT embeddings encode substantial batch-related information. Compared with handcrafted functional network connectivity features, these embeddings amplify batch-specific signatures while exhibiting distinct sensitivities to regional and network-level activity, with BrainLM favoring fine-grained regional signals and SwiFT capturing large-scale functional interactions. These findings highlight importance of addressing batch effects during model development, as post-hoc harmonization only partially removes batch-related variability. Future work may explore incorporating batch-invariant objectives during pretraining or finetuning, as well as potential privacy risks associated with embeddings.

\clearpage


\section*{Impact Statement}
This paper presents work whose goal is to advance the field of Machine Learning. There are many potential societal consequences of our work, none which we feel must be specifically highlighted here.

\nocite{langley00}

\bibliography{reference}
\bibliographystyle{icml2026}

\newpage
\appendix
\onecolumn

\section{Datasets}
\label{appendix:datasets}

All three datasets, FBIRN, ADHD-200, and ABIDE~I, comprise resting-state fMRI scans acquired across multiple sites using heterogeneous scanner hardware (e.g., 3T Siemens and Philips systems) and varying acquisition parameters, including repetition time (TR) and echo time (TE).

\subsection{Data Acquisition and Cohorts} 
The FBIRN dataset consists of rs-fMRI scans acquired from individuals with schizophrenia and matched healthy controls under a standardized acquisition protocol designed to minimize inter-site variance. The ADHD-200 dataset includes samples from children and adolescents, reflecting diverse developmental stages and clinical presentations. The ABIDE~I dataset comprises rs-fMRI data from individuals with autism spectrum disorder and neurotypical controls aggregated, characterized by substantial heterogeneity in recruitment strategies and imaging protocols. After quality control, the final number of participants retained for analysis and site-wise cohort statistics are summarized in Table~\ref{tab:fbirn_demo}, Table~\ref{tab:adhd200_demo}, and Table~\ref{tab:abide_demo}.

\begin{table}[ht]
\centering
\footnotesize
\caption{Demographic and clinical characteristics of the FBIRN dataset across imaging sites, including healthy controls and schizophrenia patients.}
\begin{tabular}{c c c c c c c}
\hline
Site ID & Count & Age (Mean $\pm$ SD) & Male & Female & Control & Patient \\
\hline
3  & 47 & 35.57 $\pm$ 9.57  & 35 & 12 & 23 & 24 \\
7  & 13 & 41.23 $\pm$ 9.11  & 12 & 1  & 8  & 5  \\
9  & 61 & 43.44 $\pm$ 12.51 & 47 & 14 & 30 & 31 \\
10 & 56 & 36.54 $\pm$ 11.87 & 45 & 11 & 28 & 28 \\
12 & 28 & 37.43 $\pm$ 10.03 & 19 & 9  & 15 & 13 \\
13 & 57 & 36.37 $\pm$ 10.69 & 31 & 26 & 29 & 28 \\
18 & 59 & 38.68 $\pm$ 11.24 & 45 & 14 & 31 & 28 \\
\hline
\end{tabular}
\label{tab:fbirn_demo}
\end{table}

\begin{table}[ht]
\centering
\footnotesize
\caption{Demographic and clinical characteristics of the ADHD-200 dataset across imaging sites. The Peking University subset consists of three sub-cohorts (Peking 1, 2, and 3) acquired at the same institution. For diagnostic grouping, participants labeled as 1 (ADHD-Combined), 2 (ADHD-Hyperactive/Impulsive), and 3 (ADHD-Inattentive) were merged into a single ADHD patient group, while those labeled as 0 were treated as typically developing controls.}
\begin{tabular}{lcccccc}
\hline
Site Name & Count & Age (Mean $\pm$ SD) & Male & Female & Control & Patient \\
\hline
Peking 1 & 85  & 11.24 $\pm$ 1.83 & 34 & 51 & 61 & 24 \\
Peking 2 & 67  & 12.12 $\pm$ 1.83 & 66  & 1 & 32 & 35 \\
Peking 3 & 42  & 13.24 $\pm$ 1.14 & 42  & 0 & 23 & 19 \\
KKI & 83  & 10.24 $\pm$ 1.35 & 46 & 37 & 61 & 22 \\
NeuroIMAGE & 42  & 16.82 $\pm$ 2.77 & 25 & 17 & 23 & 19 \\
NYU & 214 & 11.70 $\pm$ 2.92 & 136 & 78 & 98 & 116 \\
OHSU & 79  & 8.83 $\pm$ 1.13  & 44 & 35 & 42 & 37 \\
Pitt & 89  & 15.11 $\pm$ 2.90 & 46 & 43 & 89 & 0  \\
WashU & 59  & 11.59 $\pm$ 3.88 & 32 & 27 & 59 & 0  \\
\hline
\end{tabular}
\label{tab:adhd200_demo}
\end{table}

\begin{table}[ht]
\centering
\footnotesize
\caption{Demographic and clinical characteristics of the ABIDE~I dataset across imaging sites, including healthy controls and individuals with autism spectrum disorder.}
\begin{tabular}{l c c c c c c}
\hline
Site Name & Count & Age (Mean $\pm$ SD) & Male & Female & Control & Patient \\
\hline
CMU      & 27  & 26.59 $\pm$ 5.69  & 21 & 6  & 13 & 14 \\
Caltech  & 37  & 27.72 $\pm$ 10.45 & 29 & 8  & 18 & 19 \\
KKI      & 48  & 10.01 $\pm$ 1.27  & 36 & 12 & 28 & 20 \\
Leuven   & 63  & 18.00 $\pm$ 4.99  & 55 & 8  & 34 & 29 \\
MaxMun   & 47  & 23.45 $\pm$ 10.34 & 44 & 3  & 28 & 19 \\
NYU      & 175 & 15.26 $\pm$ 6.57  & 139 & 36 & 100 & 75 \\
OHSU     & 26  & 10.71 $\pm$ 1.79  & 26 & 0  & 14 & 12 \\
Olin     & 34  & 16.59 $\pm$ 3.47  & 29 & 5  & 15 & 19 \\
Pitt     & 56  & 18.94 $\pm$ 6.93  & 48 & 8  & 27 & 29 \\
SBL      & 30  & 34.37 $\pm$ 8.60  & 30 & 0  & 15 & 15 \\
SDSU     & 36  & 14.41 $\pm$ 1.84  & 29 & 7  & 22 & 14 \\
Stanford & 39  & 9.98 $\pm$ 1.59   & 31 & 8  & 20 & 19 \\
Trinity  & 47  & 16.96 $\pm$ 3.47  & 47 & 0  & 25 & 22 \\
UCLA     & 98  & 13.00 $\pm$ 2.22  & 86 & 12 & 44 & 54 \\
UM       & 140 & 14.04 $\pm$ 3.20  & 113 & 27 & 74 & 66 \\
USM      & 70  & 22.65 $\pm$ 8.40  & 70 & 0  & 25 & 45 \\
Yale     & 56  & 12.71 $\pm$ 2.88  & 40 & 16 & 28 & 28 \\
\hline
\end{tabular}
\label{tab:abide_demo}
\end{table}

\begin{table}[b!]
\centering

\footnotesize
\caption{Post-hoc pairwise comparisons between imaging sites for age and gender in the FBIRN dataset. $p$-values are Bonferroni-corrected. Bold values indicate $p<0.05$.}
\setlength{\tabcolsep}{5pt}
\renewcommand{\arraystretch}{1.05}

\begin{subtable}{0.8\textwidth}
\centering
\caption{Age differences ($p$-values)}
\begin{adjustbox}{max width=\textwidth}
\begin{tabular}{lccccccc}
\toprule
Site & 3 & 7 & 9 & 10 & 12 & 13 & 18 \\
\midrule
3  & -- & 1.0000 & \textbf{0.0072} & 1.0000 & 1.0000 & 1.0000 & 1.0000 \\
7  & 1.0000 & -- & 1.0000 & 1.0000 & 1.0000 & 1.0000 & 1.0000 \\
9  & \textbf{0.0072} & 1.0000 & -- & 0.0573 & 0.3820 & \textbf{0.0263} & 0.6314 \\
10 & 1.0000 & 1.0000 & 0.0573 & -- & 1.0000 & 1.0000 & 1.0000 \\
12 & 1.0000 & 1.0000 & 0.3820  & 1.0000 & -- & 1.0000 & 1.0000 \\
13 & 1.0000 & 1.0000 & \textbf{0.0263} & 1.0000 & 1.0000 & -- & 1.0000 \\
18 & 1.0000 & 1.0000 & 0.6314 & 1.0000 & 1.0000 & 1.0000 & -- \\
\bottomrule
\end{tabular}
\end{adjustbox}
\end{subtable}

\vspace{6pt}

\begin{subtable}{0.8\textwidth}
\centering
\caption{Gender ratio differences ($p$-values)}
\begin{adjustbox}{max width=\textwidth}
\begin{tabular}{lccccccc}
\toprule
Site & 3 & 7 & 9 & 10 & 12 & 13 & 18 \\
\midrule
3  & -- & 1.0000 & 1.0000 & 1.0000 & 1.0000 & 1.0000 & 1.0000 \\
7  & 1.0000 & -- & 1.0000 & 1.0000 & 1.0000 & 0.5562 & 1.0000 \\
9  & 1.0000 & 1.0000 & -- & 1.0000 & 1.0000 & 0.3403 & 1.0000 \\
10 & 1.0000 & 1.0000 & 1.0000 & -- & 1.0000 & 0.1287 & 1.0000 \\
12 & 1.0000 & 1.0000 & 1.0000 & 1.0000 & -- & 1.0000 & 1.0000 \\
13 & 1.0000 & 0.5562 & 0.3403 & 0.1287 & 1.0000 & -- & 0.4700 \\
18 & 1.0000 & 1.0000 & 1.0000 & 1.0000 & 1.0000 & 0.4700 & -- \\
\bottomrule
\end{tabular}
\end{adjustbox}
\end{subtable}
\label{tab:fbirn_posthoc}
\end{table}

\begin{table}[t]
\centering
\footnotesize
\caption{Post-hoc pairwise comparisons between imaging sites for age, gender, and diagnosis in the ADHD-200 dataset. $p$-values are Bonferroni-corrected. Bold values indicate $p<0.05$.}
\setlength{\tabcolsep}{3.5pt}
\renewcommand{\arraystretch}{1.05}

\begin{subtable}{0.98\textwidth}
\centering
\caption{Age differences ($p$-values)}
\begin{adjustbox}{max width=0.98\textwidth}
\begin{tabular}{lccccccccc}
\toprule
Site & Peking 1 & Peking 2 & Peking 3 & KKI & NeuroIMAGE & NYU & OHSU & Pitt & WashU \\
\midrule
Peking 1 & -- & 0.1360 & \textbf{$<$0.0001} & \textbf{0.0031} & \textbf{$<$0.0001} & 1.0000 & \textbf{$<$0.0001} & \textbf{$<$0.0001} & 1.0000 \\
Peking 2 & 0.1360 & -- & \textbf{0.0053} & \textbf{$<$0.0001} & \textbf{$<$0.0001} & 1.0000 & \textbf{$<$0.0001} & \textbf{$<$0.0001} & 1.0000 \\
Peking 3 & \textbf{$<$0.0001} & \textbf{0.0053} & -- & \textbf{$<$0.0001} & \textbf{$<$0.0001} & \textbf{$<$0.0001} & \textbf{$<$0.0001} & \textbf{$<$0.0001} & 0.1054 \\
KKI      & \textbf{0.0031} & \textbf{$<$0.0001} & \textbf{$<$0.0001} & -- & \textbf{$<$0.0001} & \textbf{$<$0.0001} & \textbf{$<$0.0001} & \textbf{$<$0.0001} & 0.4487 \\
NeuroIMAGE & \textbf{$<$0.0001} & \textbf{$<$0.0001} & \textbf{$<$0.0001} & \textbf{$<$0.0001} & -- & \textbf{$<$0.0001} & \textbf{$<$0.0001} & 0.0587 & \textbf{$<$0.0001} \\
NYU      & 1.0000 & 1.0000 & \textbf{$<$0.0001} & \textbf{$<$0.0001} & \textbf{$<$0.0001} & -- & \textbf{$<$0.0001} & \textbf{$<$0.0001} & 1.0000 \\
OHSU     & \textbf{$<$0.0001} & \textbf{$<$0.0001} & \textbf{$<$0.0001} & \textbf{$<$0.0001} & \textbf{$<$0.0001} & \textbf{$<$0.0001} & -- & \textbf{$<$0.0001} & \textbf{0.0001} \\
Pitt     & \textbf{$<$0.0001} & \textbf{$<$0.0001} & \textbf{$<$0.0001} & \textbf{$<$0.0001} & 0.0587 & \textbf{$<$0.0001} & \textbf{$<$0.0001} & -- & \textbf{$<$0.0001} \\
WashU    & 1.0000 & 1.0000 & 0.1054 & 0.4487 & \textbf{$<$0.0001} & 1.0000 & \textbf{0.0001} & \textbf{$<$0.0001} & -- \\
\bottomrule
\end{tabular}
\end{adjustbox}
\end{subtable}

\vspace{6pt}

\begin{subtable}{0.98\textwidth}
\centering
\caption{Gender ratio differences ($p$-values)}
\begin{adjustbox}{max width=0.98\textwidth}
\begin{tabular}{lccccccccc}
\toprule
Site & Peking 1 & Peking 2 & Peking 3 & KKI & NeuroIMAGE & NYU & OHSU & Pitt & WashU \\
\midrule
Peking 1 & -- & \textbf{$<$0.0001} & \textbf{$<$0.0001} & 1.0000 & 1.0000 & \textbf{0.0124} & 1.0000 & 1.0000 & 1.0000 \\
Peking 2 & \textbf{$<$0.0001} & -- & 1.0000 & \textbf{$<$0.0001} & \textbf{$<$0.0001} & \textbf{$<$0.0001} & \textbf{$<$0.0001} & \textbf{$<$0.0001} & \textbf{$<$0.0001} \\
Peking 3 & \textbf{$<$0.0001} & 1.0000 & -- & \textbf{$<$0.0001} & \textbf{0.0005} & \textbf{0.0002} & \textbf{$<$0.0001} & \textbf{$<$0.0001} & \textbf{$<$0.0001} \\
KKI      & 1.0000 & \textbf{$<$0.0001} & \textbf{$<$0.0001} & -- & 1.0000 & 1.0000 & 1.0000 & 1.0000 & 1.0000 \\
NeuroIMAGE & 1.0000 & \textbf{$<$0.0001} & \textbf{0.0005} & 1.0000 & -- & 1.0000 & 1.0000 & 1.0000 & 1.0000 \\
NYU      & \textbf{0.0124} & \textbf{$<$0.0001} & \textbf{0.0002} & 1.0000 & 1.0000 & -- & 1.0000 & 1.0000 & 1.0000 \\
OHSU     & 1.0000 & \textbf{$<$0.0001} & \textbf{$<$0.0001} & 1.0000 & 1.0000 & 1.0000 & -- & 1.0000 & 1.0000 \\
Pitt     & 1.0000 & \textbf{$<$0.0001} & \textbf{$<$0.0001} & 1.0000 & 1.0000 & 1.0000 & 1.0000 & -- & 1.0000 \\
WashU    & 1.0000 & \textbf{$<$0.0001} & \textbf{$<$0.0001} & 1.0000 & 1.0000 & 1.0000 & 1.0000 & 1.0000 & -- \\
\bottomrule
\end{tabular}
\end{adjustbox}
\end{subtable}

\vspace{6pt}

\begin{subtable}{0.98\textwidth}
\centering
\caption{Diagnosis ratio differences ($p$-values)}
\begin{adjustbox}{max width=0.98\textwidth}
\begin{tabular}{lccccccccc}
\toprule
Site & Peking 1 & Peking 2 & Peking 3 & KKI & NeuroIMAGE & NYU & OHSU & Pitt & WashU \\
\midrule
Peking 1 & -- & 0.1587 & 1.0000 & 1.0000 & 1.0000 & \textbf{0.0030} & 0.7704 & \textbf{$<$0.0001} & \textbf{0.0008} \\
Peking 2 & 0.1587 & -- & 1.0000 & 0.0800 & 1.0000 & 1.0000 & 1.0000 & \textbf{$<$0.0001} & \textbf{$<$0.0001} \\
Peking 3 & 1.0000 & 1.0000 & -- & 1.0000 & 1.0000 & 1.0000 & 1.0000 & \textbf{$<$0.0001} & \textbf{$<$0.0001} \\
KKI      & 1.0000 & 0.0800 & 1.0000 & -- & 1.0000 & \textbf{0.0011} & 0.4172 & \textbf{$<$0.0001} & \textbf{0.0017} \\
NeuroIMAGE & 1.0000 & 1.0000 & 1.0000 & 1.0000 & -- & 1.0000 & 1.0000 & \textbf{$<$0.0001} & \textbf{$<$0.0001} \\
NYU      & \textbf{0.0030} & 1.0000 & 1.0000 & \textbf{0.0011} & 1.0000 & -- & 1.0000 & \textbf{$<$0.0001} & \textbf{$<$0.0001} \\
OHSU     & 0.7704 & 1.0000 & 1.0000 & 0.4172 & 1.0000 & 1.0000 & -- & \textbf{$<$0.0001} & \textbf{$<$0.0001} \\
Pitt     & \textbf{$<$0.0001} & \textbf{$<$0.0001} & \textbf{$<$0.0001} & \textbf{$<$0.0001} & \textbf{$<$0.0001} & \textbf{$<$0.0001} & \textbf{$<$0.0001} & -- & 1.0000 \\
WashU    & \textbf{0.0008} & \textbf{$<$0.0001} & \textbf{$<$0.0001} & \textbf{0.0017} & \textbf{$<$0.0001} & \textbf{$<$0.0001} & \textbf{$<$0.0001} & 1.0000 & -- \\
\bottomrule
\end{tabular}
\end{adjustbox}
\end{subtable}
\label{tab:adhd200_posthoc}
\end{table}

\subsection{Demographic and Clinical Comparisons Across Sites}
To characterize demographic and clinical characteristics of the FBIRN, ADHD-200, and ABIDE~I datasets, we conducted a series of statistical analyses. To evaluate potential batch effects, one-way analysis of variance (ANOVA) was used to assess age variation across imaging sites, while chi-square tests were applied to examine differences in gender distribution and diagnostic composition. Between-site heterogeneity was further investigated through post-hoc pairwise comparisons, including two-sample Welch's t-tests for age and pairwise chi-square tests for gender and diagnosis, where assumptions of the chi-square test were met. In addition, group matching with respect to diagnosis was assessed by testing whether age and gender differed across diagnostic categories, using one-way ANOVA for age and chi-square tests for gender.

\begin{sidewaystable}[htb!]
\centering
\footnotesize
\caption{Post-hoc pairwise comparisons between imaging sites for age and gender in the ABIDE~I dataset. $p$-values are Bonferroni-corrected. Bold values indicate $p<0.05$.}
\setlength{\tabcolsep}{4pt}
\renewcommand{\arraystretch}{1.05}

\begin{subtable}{0.98\textheight}
\centering
\caption{Age differences ($p$-values)}
\begin{adjustbox}{max width=0.98\textheight}
\begin{tabular}{lccccccccccccccccc}
\toprule
Site & CMU & Caltech & KKI & Leuven & MaxMun & NYU & OHSU & Olin & Pitt & SBL & SDSU & Stanford & Trinity & UCLA & UM & USM & Yale \\
\midrule
CMU     & -- & 1.0000 & \textbf{$<$0.0001} & \textbf{$<$0.0001} & 1.0000 & \textbf{$<$0.0001} & \textbf{$<$0.0001} & \textbf{$<$0.0001} & \textbf{0.0002} & \textbf{0.0230} & \textbf{$<$0.0001} & \textbf{$<$0.0001} & \textbf{$<$0.0001} & \textbf{$<$0.0001} & \textbf{$<$0.0001} & 1.0000 & \textbf{$<$0.0001} \\
Caltech & 1.0000 & -- & \textbf{$<$0.0001} & \textbf{0.0004} & 1.0000 & \textbf{$<$0.0001} & \textbf{$<$0.0001} & \textbf{$<$0.0001} & \textbf{0.0046} & 0.7832 & \textbf{$<$0.0001} & \textbf{$<$0.0001} & \textbf{0.0001} & \textbf{$<$0.0001} & \textbf{$<$0.0001} & 1.0000 & \textbf{$<$0.0001} \\
KKI     & \textbf{$<$0.0001} & \textbf{$<$0.0001} & -- & \textbf{$<$0.0001} & \textbf{$<$0.0001} & \textbf{$<$0.0001} & 1.0000 & \textbf{$<$0.0001} & \textbf{$<$0.0001} & \textbf{$<$0.0001} & \textbf{$<$0.0001} & 1.0000 & \textbf{$<$0.0001} & \textbf{$<$0.0001} & \textbf{$<$0.0001} & \textbf{$<$0.0001} & \textbf{$<$0.0001} \\
Leuven  & \textbf{$<$0.0001} & \textbf{0.0004} & \textbf{$<$0.0001} & -- & 0.1967 & 0.1120 & \textbf{$<$0.0001} & 1.0000 & 1.0000 & \textbf{$<$0.0001} & \textbf{0.0002} & \textbf{$<$0.0001} & 1.0000 & \textbf{$<$0.0001} & \textbf{$<$0.0001} & \textbf{0.0201} & \textbf{$<$0.0001} \\
MaxMun  & 1.0000 & 1.0000 & \textbf{$<$0.0001} & 0.1967 & -- & \textbf{0.0005} & \textbf{$<$0.0001} & \textbf{0.0111} & 1.0000 & \textbf{0.0005} & \textbf{$<$0.0001} & \textbf{$<$0.0001} & \textbf{0.0194} & \textbf{$<$0.0001} & \textbf{$<$0.0001} & 1.0000 & \textbf{$<$0.0001} \\
NYU     & \textbf{$<$0.0001} & \textbf{$<$0.0001} & \textbf{$<$0.0001} & 0.1120 & \textbf{0.0005} & -- & \textbf{$<$0.0001} & 1.0000 & 0.0989 & \textbf{$<$0.0001} & 1.0000 & \textbf{$<$0.0001} & 1.0000 & \textbf{0.0065} & 1.0000 & \textbf{$<$0.0001} & \textbf{0.0096} \\
OHSU    & \textbf{$<$0.0001} & \textbf{$<$0.0001} & 1.0000 & \textbf{$<$0.0001} & \textbf{$<$0.0001} & \textbf{$<$0.0001} & -- & \textbf{$<$0.0001} & \textbf{$<$0.0001} & \textbf{$<$0.0001} & \textbf{$<$0.0001} & 1.0000 & \textbf{$<$0.0001} & \textbf{0.0002} & \textbf{$<$0.0001} & \textbf{$<$0.0001} & \textbf{0.0341} \\
Olin    & \textbf{$<$0.0001} & \textbf{$<$0.0001} & \textbf{$<$0.0001} & 1.0000 & \textbf{0.0111} & 1.0000 & \textbf{$<$0.0001} & -- & 1.0000 & \textbf{$<$0.0001} & 0.2793 & \textbf{$<$0.0001} & 1.0000 & \textbf{0.0002} & \textbf{0.0401} & \textbf{0.0001} & \textbf{0.0001} \\
Pitt    & \textbf{0.0002} & \textbf{0.0046} & \textbf{$<$0.0001} & 1.0000 & 1.0000 & 0.0989 & \textbf{$<$0.0001} & 1.0000 & -- & \textbf{$<$0.0001} & \textbf{0.0023} & \textbf{$<$0.0001} & 1.0000 & \textbf{$<$0.0001} & \textbf{0.0005} & 1.0000 & \textbf{$<$0.0001} \\
SBL     & \textbf{0.0230} & 0.7832 & \textbf{$<$0.0001} & \textbf{$<$0.0001} & \textbf{0.0005} & \textbf{$<$0.0001} & \textbf{$<$0.0001} & \textbf{$<$0.0001} & \textbf{$<$0.0001} & -- & \textbf{$<$0.0001} & \textbf{$<$0.0001} & \textbf{$<$0.0001} & \textbf{$<$0.0001} & \textbf{$<$0.0001} & \textbf{$<$0.0001} & \textbf{$<$0.0001} \\
SDSU    & \textbf{$<$0.0001} & \textbf{$<$0.0001} & \textbf{$<$0.0001} & \textbf{0.0002} & \textbf{$<$0.0001} & 1.0000 & \textbf{$<$0.0001} & 0.2793 & \textbf{0.0023} & \textbf{$<$0.0001} & -- & \textbf{$<$0.0001} & \textbf{0.0071} & 0.0541 & 1.0000 & \textbf{$<$0.0001} & 0.1140 \\
Stanford& \textbf{$<$0.0001} & \textbf{$<$0.0001} & 1.0000 & \textbf{$<$0.0001} & \textbf{$<$0.0001} & \textbf{$<$0.0001} & 1.0000 & \textbf{$<$0.0001} & \textbf{$<$0.0001} & \textbf{$<$0.0001} & \textbf{$<$0.0001} & -- & \textbf{$<$0.0001} & \textbf{$<$0.0001} & \textbf{$<$0.0001} & \textbf{$<$0.0001} & \textbf{$<$0.0001} \\
Trinity & \textbf{$<$0.0001} & \textbf{0.0001} & \textbf{$<$0.0001} & 1.0000 & \textbf{0.0194} & 1.0000 & \textbf{$<$0.0001} & 1.0000 & 1.0000 & \textbf{$<$0.0001} & \textbf{0.0071} & \textbf{$<$0.0001} & -- & \textbf{$<$0.0001} & \textbf{0.0004} & \textbf{0.0003} & \textbf{$<$0.0001} \\
UCLA    & \textbf{$<$0.0001} & \textbf{$<$0.0001} & \textbf{$<$0.0001} & \textbf{$<$0.0001} & \textbf{$<$0.0001} & \textbf{0.0065} & \textbf{0.0002} & \textbf{0.0002} & \textbf{$<$0.0001} & \textbf{$<$0.0001} & 0.0541 & \textbf{$<$0.0001} & \textbf{$<$0.0001} & -- & 0.4871 & \textbf{$<$0.0001} & 1.0000 \\
UM      & \textbf{$<$0.0001} & \textbf{$<$0.0001} & \textbf{$<$0.0001} & \textbf{$<$0.0001} & \textbf{$<$0.0001} & 1.0000 & \textbf{$<$0.0001} & \textbf{0.0401} & \textbf{0.0005} & \textbf{$<$0.0001} & 1.0000 & \textbf{$<$0.0001} & \textbf{0.0004} & 0.4871 & -- & \textbf{$<$0.0001} & 0.7775 \\
USM     & 1.0000 & 1.0000 & \textbf{$<$0.0001} & \textbf{0.0201} & 1.0000 & \textbf{$<$0.0001} & \textbf{$<$0.0001} & \textbf{0.0001} & 1.0000 & \textbf{$<$0.0001} & \textbf{$<$0.0001} & \textbf{$<$0.0001} & \textbf{0.0003} & \textbf{$<$0.0001} & \textbf{$<$0.0001} & -- & \textbf{$<$0.0001} \\
Yale    & \textbf{$<$0.0001} & \textbf{$<$0.0001} & \textbf{$<$0.0001} & \textbf{$<$0.0001} & \textbf{$<$0.0001} & \textbf{0.0096} & \textbf{0.0341} & \textbf{0.0001} & \textbf{$<$0.0001} & \textbf{$<$0.0001} & 0.1140 & \textbf{$<$0.0001} & \textbf{$<$0.0001} & 1.0000 & 0.7775 & \textbf{$<$0.0001} & -- \\
\bottomrule
\end{tabular}
\end{adjustbox}
\end{subtable}

\vspace{6pt}

\begin{subtable}{0.98\textheight}
\centering
\caption{Gender ratio differences ($p$-values)}
\begin{adjustbox}{max width=0.98\textheight}
\begin{tabular}{lccccccccccccccccc}
\toprule
Site & CMU & Caltech & KKI & Leuven & MaxMun & NYU & OHSU & Olin & Pitt & SBL & SDSU & Stanford & Trinity & UCLA & UM & USM & Yale \\
\midrule
CMU     & -- & 1.0000 & 1.0000 & 1.0000 & 1.0000 & 1.0000 & 1.0000 & 1.0000 & 1.0000 & 1.0000 & 1.0000 & 1.0000 & 0.4625 & 1.0000 & 1.0000 & \textbf{0.0430} & 1.0000 \\
Caltech & 1.0000 & -- & 1.0000 & 1.0000 & 1.0000 & 1.0000 & 1.0000 & 1.0000 & 1.0000 & 1.0000 & 1.0000 & 1.0000 & 0.3958 & 1.0000 & 1.0000 & \textbf{0.0346} & 1.0000 \\
KKI     & 1.0000 & 1.0000 & -- & 1.0000 & 1.0000 & 1.0000 & 1.0000 & 1.0000 & 1.0000 & 1.0000 & 1.0000 & 1.0000 & 0.1066 & 1.0000 & 1.0000 & \textbf{0.0055} & 1.0000 \\
Leuven  & 1.0000 & 1.0000 & 1.0000 & -- & 1.0000 & 1.0000 & 1.0000 & 1.0000 & 1.0000 & 1.0000 & 1.0000 & 1.0000 & 1.0000 & 1.0000 & 1.0000 & 0.9146 & 1.0000 \\
MaxMun  & 1.0000 & 1.0000 & 1.0000 & 1.0000 & -- & 1.0000 & 1.0000 & 1.0000 & 1.0000 & 1.0000 & 1.0000 & 1.0000 & 1.0000 & 1.0000 & 1.0000 & 1.0000 & 1.0000 \\
NYU     & 1.0000 & 1.0000 & 1.0000 & 1.0000 & 1.0000 & -- & 1.0000 & 1.0000 & 1.0000 & 1.0000 & 1.0000 & 1.0000 & 0.2043 & 1.0000 & 1.0000 & \textbf{0.0126} & 1.0000 \\
OHSU    & 1.0000 & 1.0000 & 1.0000 & 1.0000 & 1.0000 & 1.0000 & -- & 1.0000 & 1.0000 & 1.0000 & 1.0000 & 1.0000 & 1.0000 & 1.0000 & 1.0000 & 1.0000 & 0.8392 \\
Olin    & 1.0000 & 1.0000 & 1.0000 & 1.0000 & 1.0000 & 1.0000 & 1.0000 & -- & 1.0000 & 1.0000 & 1.0000 & 1.0000 & 1.0000 & 1.0000 & 1.0000 & 0.6952 & 1.0000 \\
Pitt    & 1.0000 & 1.0000 & 1.0000 & 1.0000 & 1.0000 & 1.0000 & 1.0000 & 1.0000 & -- & 1.0000 & 1.0000 & 1.0000 & 1.0000 & 1.0000 & 1.0000 & 0.5073 & 1.0000 \\
SBL     & 1.0000 & 1.0000 & 1.0000 & 1.0000 & 1.0000 & 1.0000 & 1.0000 & 1.0000 & 1.0000 & -- & 1.0000 & 1.0000 & 1.0000 & 1.0000 & 1.0000 & 1.0000 & 0.4261 \\
SDSU    & 1.0000 & 1.0000 & 1.0000 & 1.0000 & 1.0000 & 1.0000 & 1.0000 & 1.0000 & 1.0000 & 1.0000 & -- & 1.0000 & 0.7844 & 1.0000 & 1.0000 & 0.0901 & 1.0000 \\
Stanford& 1.0000 & 1.0000 & 1.0000 & 1.0000 & 1.0000 & 1.0000 & 1.0000 & 1.0000 & 1.0000 & 1.0000 & 1.0000 & -- & 0.5281 & 1.0000 & 1.0000 & 0.0517 & 1.0000 \\
Trinity & 0.4625 & 0.3958 & 0.1066 & 1.0000 & 1.0000 & 0.2043 & 1.0000 & 1.0000 & 1.0000 & 1.0000 & 0.7844 & 0.5281 & -- & 1.0000 & 0.3495 & 1.0000 & \textbf{0.0277} \\
UCLA    & 1.0000 & 1.0000 & 1.0000 & 1.0000 & 1.0000 & 1.0000 & 1.0000 & 1.0000 & 1.0000 & 1.0000 & 1.0000 & 1.0000 & 1.0000 & -- & 1.0000 & 0.8499 & 1.0000 \\
UM      & 1.0000 & 1.0000 & 1.0000 & 1.0000 & 1.0000 & 1.0000 & 1.0000 & 1.0000 & 1.0000 & 1.0000 & 1.0000 & 1.0000 & 0.3495 & 1.0000 & -- & \textbf{0.0274} & 1.0000 \\
USM     & \textbf{0.0430} & \textbf{0.0346} & \textbf{0.0055} & 0.9146 & 1.0000 & \textbf{0.0126} & 1.0000 & 0.6952 & 0.5073 & 1.0000 & 0.0901 & 0.0517 & 1.0000 & 0.8499 & \textbf{0.0274} & -- & \textbf{0.0009} \\
Yale    & 1.0000 & 1.0000 & 1.0000 & 1.0000 & 1.0000 & 1.0000 & 0.8392 & 1.0000 & 1.0000 & 0.4261 & 1.0000 & 1.0000 & \textbf{0.0277} & 1.0000 & 1.0000 & \textbf{0.0009} & -- \\
\bottomrule
\end{tabular}
\end{adjustbox}
\end{subtable}
\label{tab:abide_posthoc}
\end{sidewaystable}

\paragraph{FBIRN.} Significant site effects were observed for age ($p=0.0032$) and gender distribution ($p=0.0185$), whereas diagnostic composition did not differ across sites ($p=0.9894$). Post-hoc pairwise comparisons revealed significant differences in age and gender for several site pairs, whereas diagnostic ratios remained largely consistent across sites (see Table~\ref{tab:fbirn_posthoc}). Group matching analyses revealed no significant differences in age ($p=0.2041$) or gender ($p=0.6245$) between diagnostic groups, suggesting the dataset is reasonably well matched with respect to these potential confounders.

\paragraph{ADHD-200.} Significant site effects were observed for age ($p<0.001$), gender distribution ($p<0.001$), and diagnostic composition ($p<0.001$). Post-hoc pairwise comparisons revealed pronounced differences in age and gender for most site pairs, and diagnostic ratios also varied substantially across sites (see Table~\ref{tab:adhd200_posthoc}). Group matching analyses indicated significant differences between diagnostic groups in both age ($p=0.0024$) and gender ($p<0.001$), suggesting that the dataset exhibits notable confounding with respect to these variables across diagnostic categories.

\paragraph{ABIDE~I.} Significant site effects were observed for age ($p<0.001$) and gender distribution ($p<0.001$), whereas diagnostic composition did not differ significantly across sites ($p=0.4493$). Post-hoc pairwise comparisons revealed widespread age differences between most site pairs and heterogeneity in gender ratios (see Table~\ref{tab:abide_posthoc}). Group matching analyses indicated no significant age differences between diagnostic groups ($p=0.9691$), whereas a significant difference in gender distribution remained ($p=0.0139$).

\subsection{fMRI Preprocessing Pipeline}
Having described the demographic and clinical characteristics of the cohorts across sites, we next detail the fMRI preprocessing pipeline. A standardized procedure was implemented using SPM12, following the NeuroMark framework, to transform heterogeneous scans into a spatially and temporally aligned 4D format. Functional volumes underwent rigid body motion correction and slice-timing correction, followed by normalization to MNI space using an echo planar imaging (EPI) template and resampling to $3 \times 3 \times 3\,\mathrm{mm}^3$ isotropic voxels. Spatial smoothing was then applied using a Gaussian kernel with a full width at half maximum (FWHM) of 6\,mm to reduce high-frequency noise. Detailed preprocessing procedures are described in~\cite{du2020neuromark, feng2024functional}.

\subsection{Foundation Model Input Preparation}
After standard preprocessing, additional spatial transformations were applied to format the fMRI data for foundation model inputs. These steps were specific to model requirements and are not considered part of the general preprocessing pipeline. Two representative brain foundation models, BrainLM and SwiFT, were selected to evaluate the generalizability of the learned representations. For pretrained BrainLM, which operates on ROI-level timeseries, voxel-wise volumes were first resampled to the AAL-424 atlas space and parcellated into 424 predefined regions. The mean timeseries within each ROI were extracted to construct region-level temporal features as model input. For pretrained SwiFT, which accepts voxel-wise volumetric inputs, the preprocessed fMRI volumes were directly resampled to the MNI152 template resolution without additional parcellation. The resulting ROI-level or voxel-level timeseries were then fed into the corresponding foundation models to generate subject-level embeddings for downstream analyses.

\section{Additional Details on Experimental Results}

\subsection{Additional Analyses of Intrinsic Encoding of Batch Effects in Embeddings}
\label{appendix:supp_batch_effect}

\begin{figure}[htbp]
    \centering
    \begin{subfigure}{0.30\textwidth}
        \centering
        \includegraphics[width=\linewidth]{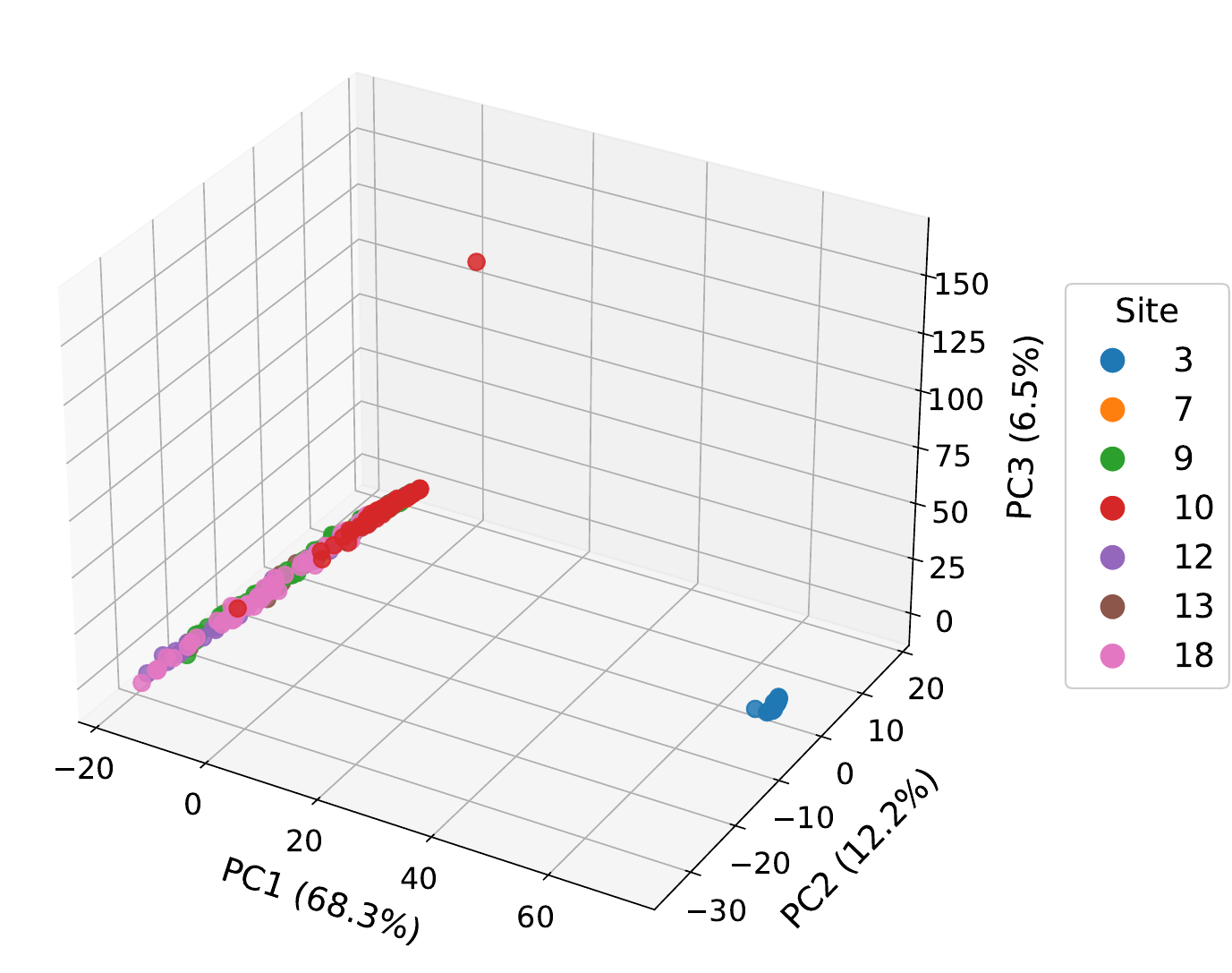}
        \caption{PCA on FBIRN}
    \end{subfigure}%
    \hspace{0.01\textwidth}
    \begin{subfigure}{0.34\textwidth}
        \centering
        \includegraphics[width=\linewidth]{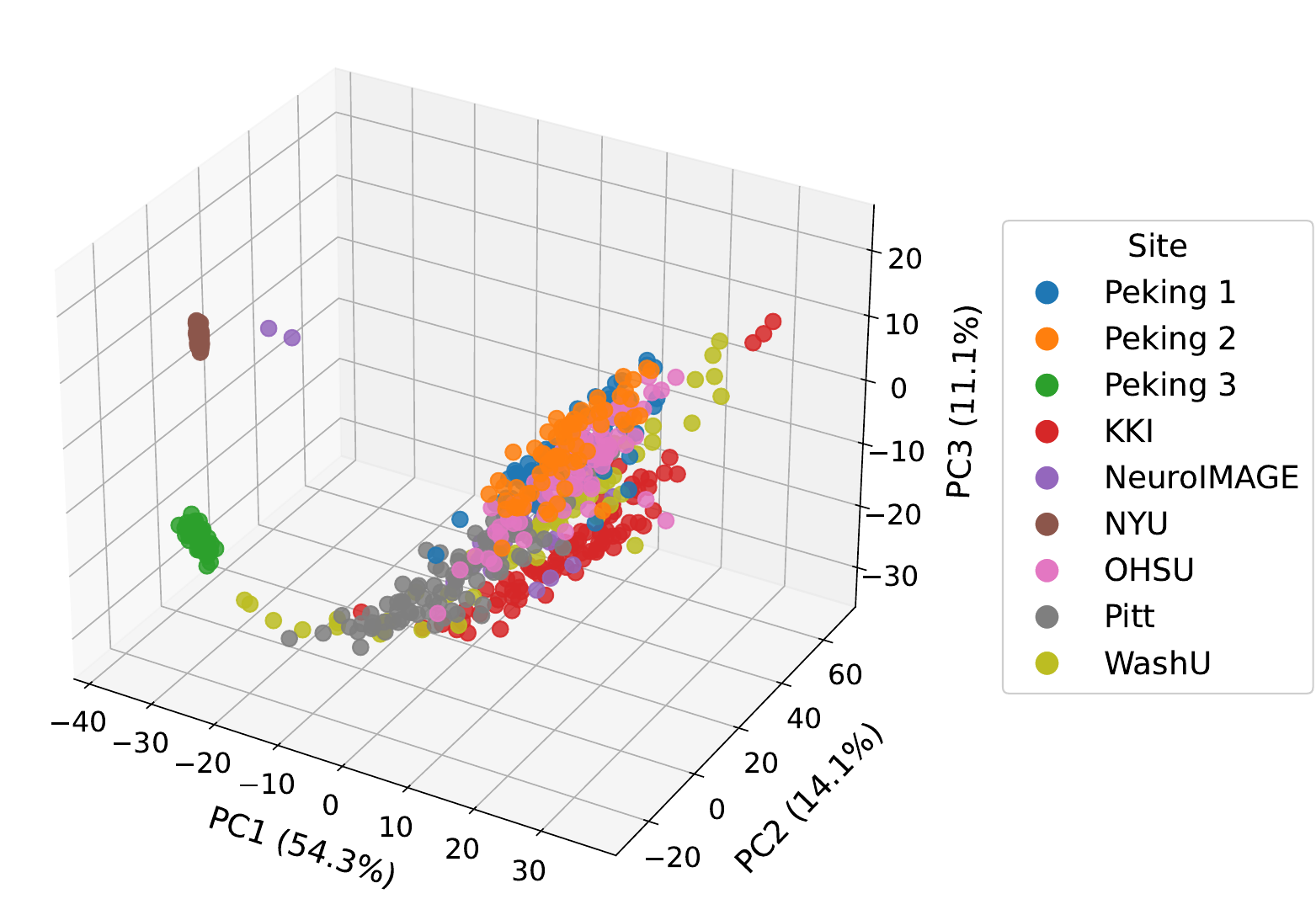}
        \caption{PCA on ADHD-200}
    \end{subfigure}%
    \hspace{0.01\textwidth}
    \begin{subfigure}{0.32\textwidth}
        \centering
        \includegraphics[width=\linewidth]{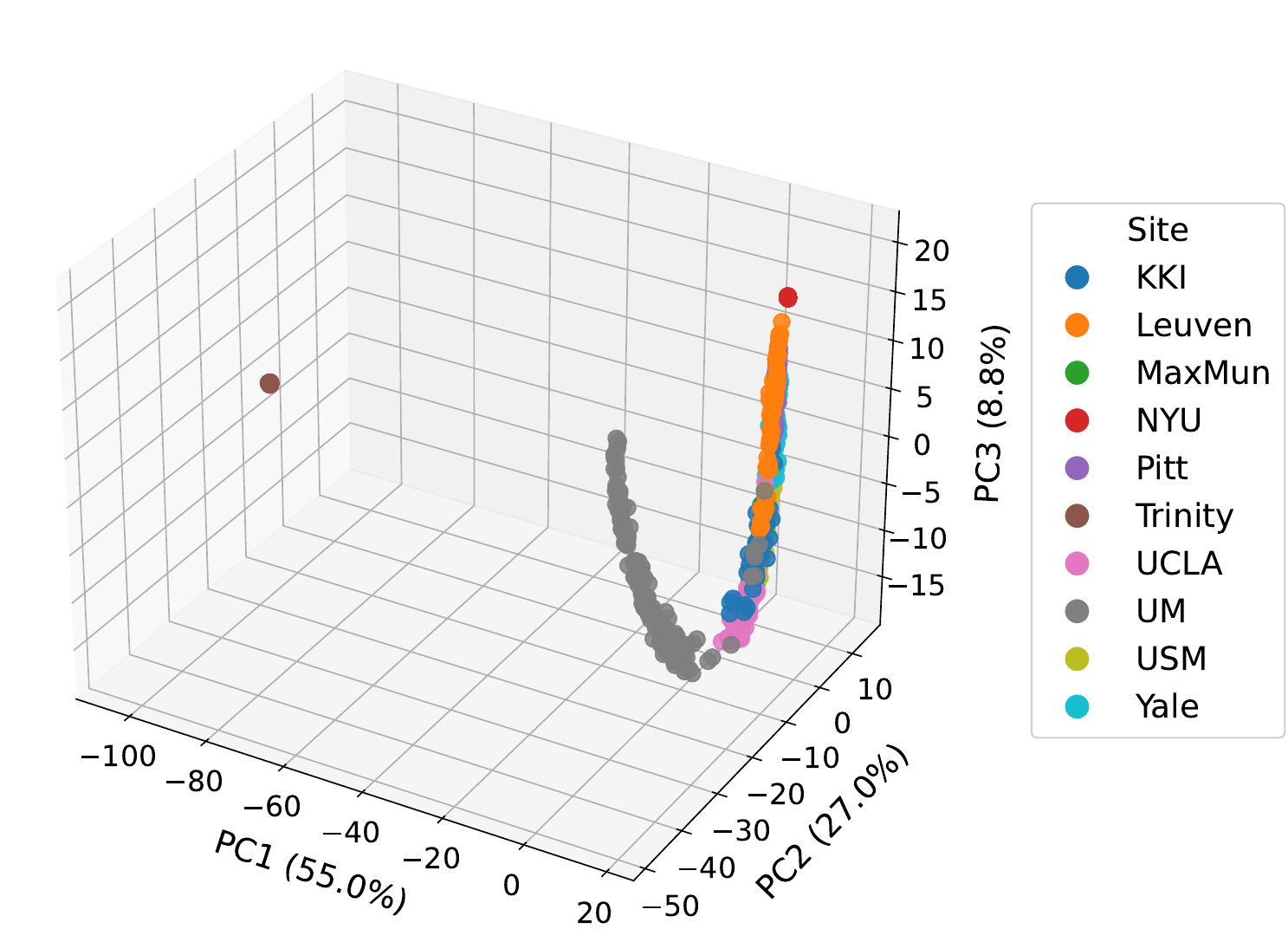}
        \caption{PCA on ABIDE I}
    \end{subfigure}%

    \vspace{0.3cm}
    \begin{subfigure}{0.30\textwidth}
        \centering
        \includegraphics[width=\linewidth]{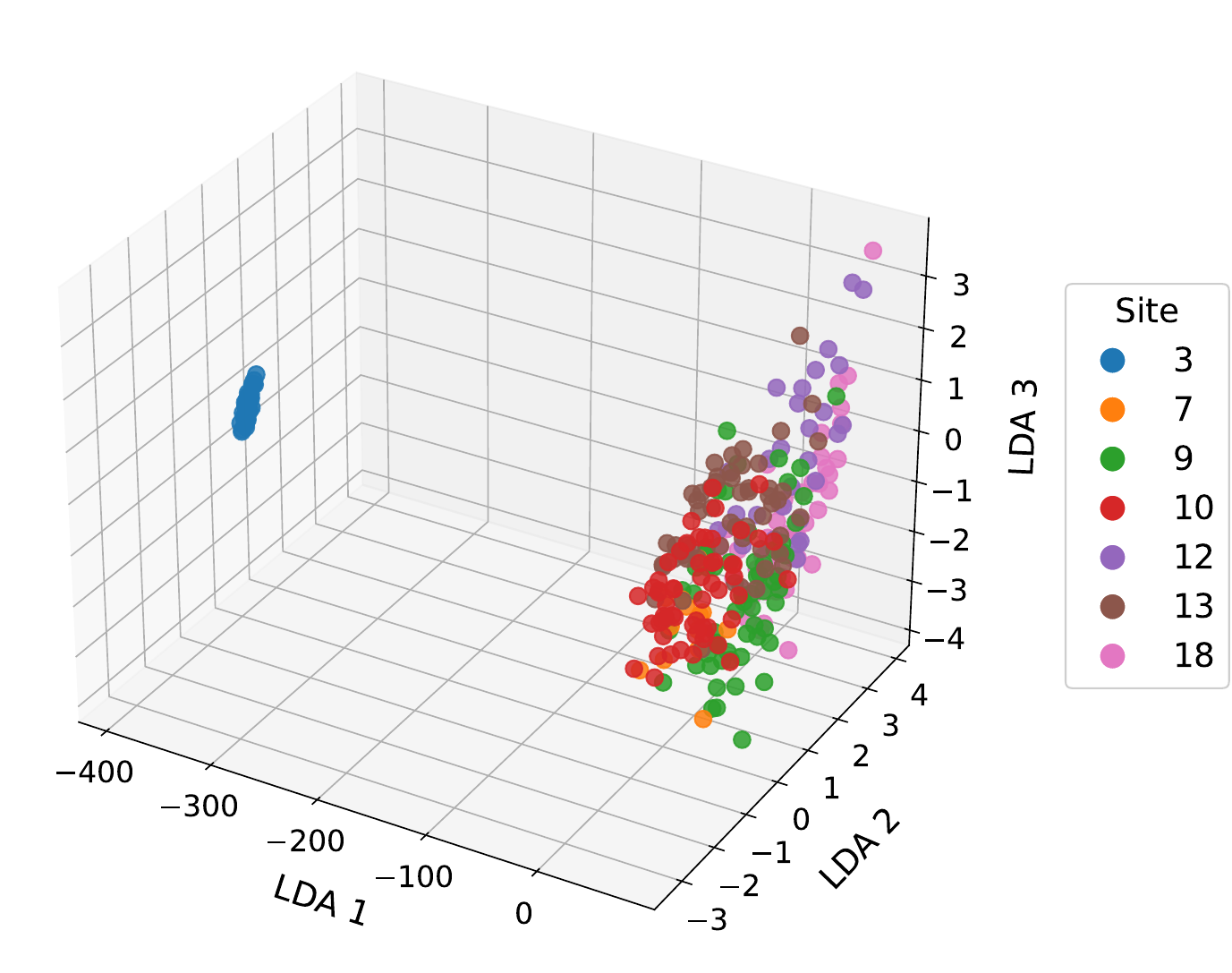}
        \caption{LDA on FBIRN}
    \end{subfigure}%
    \hspace{0.01\textwidth}
    \begin{subfigure}{0.34\textwidth}
        \centering
        \includegraphics[width=\linewidth]{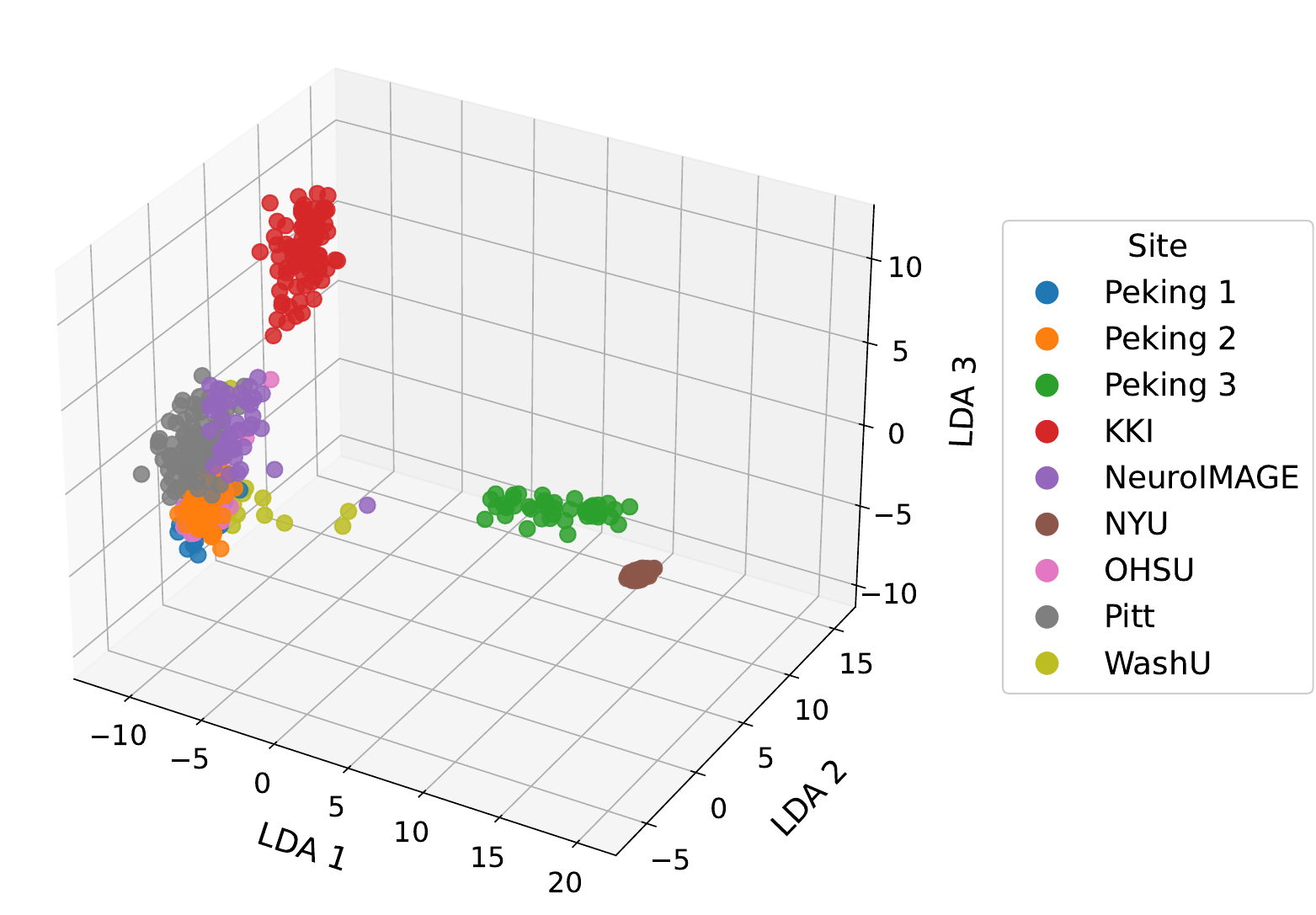}
        \caption{LDA on ADHD-200}
    \end{subfigure}%
    \hspace{0.01\textwidth}
    \begin{subfigure}{0.32\textwidth}
        \centering
        \includegraphics[width=\linewidth]{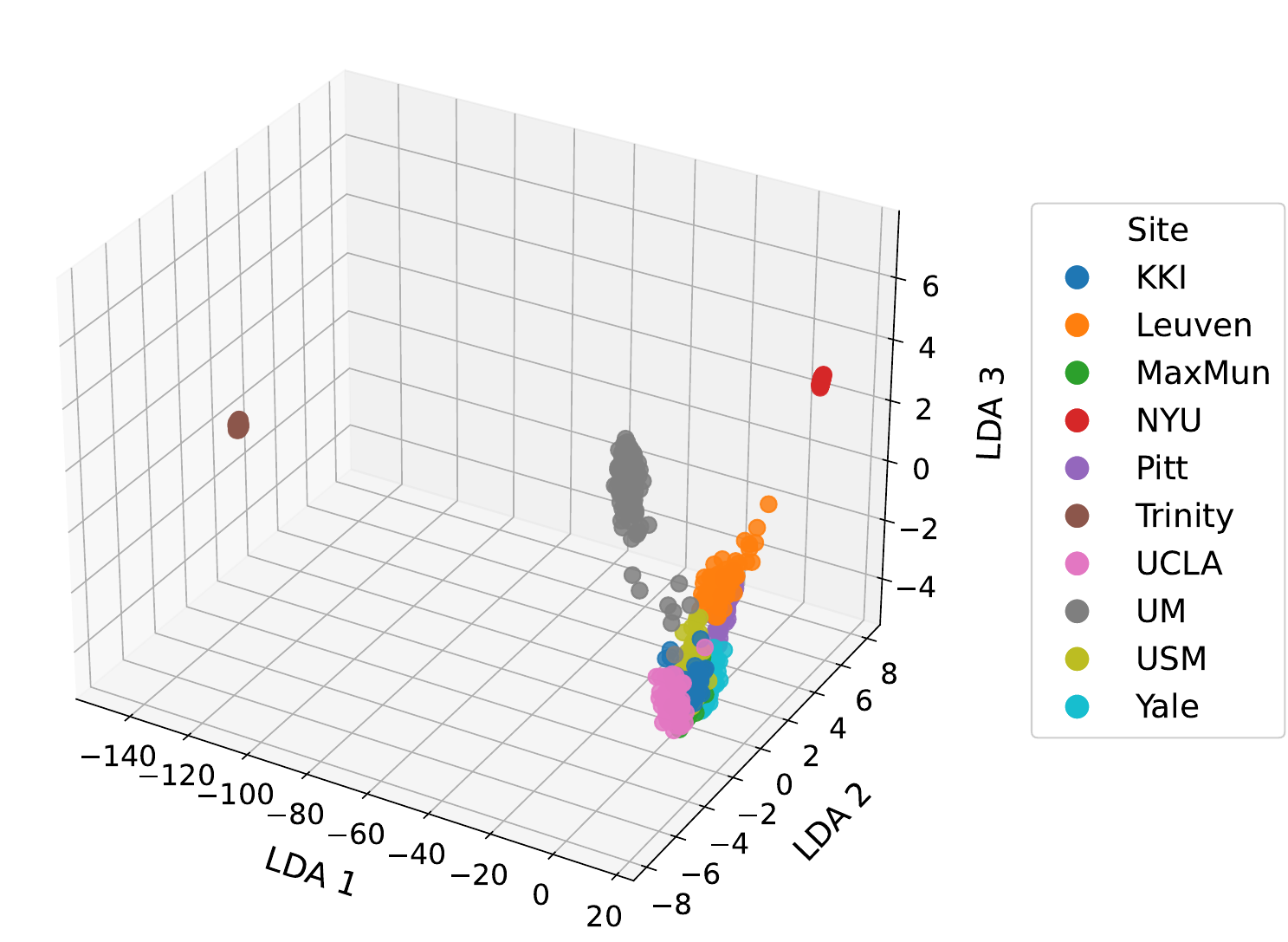}
        \caption{LDA on ABIDE I}
    \end{subfigure}

    \caption{Visualization of subject-level embeddings extracted from the pre-trained BrainLM model. In the last row, embeddings are first projected using PCA with 20 components, followed by further dimensionality reduction using LDA. All points are colored according to site identity. For the ABIDE I dataset, which includes 17 imaging sites, only the 10 sites with the largest sample sizes are shown for clarity.}
    \label{fig:brainlm_sites}
\end{figure}

\begin{figure}[htbp]
    \centering
    \begin{subfigure}{0.30\textwidth}
        \centering
        \includegraphics[width=\linewidth]{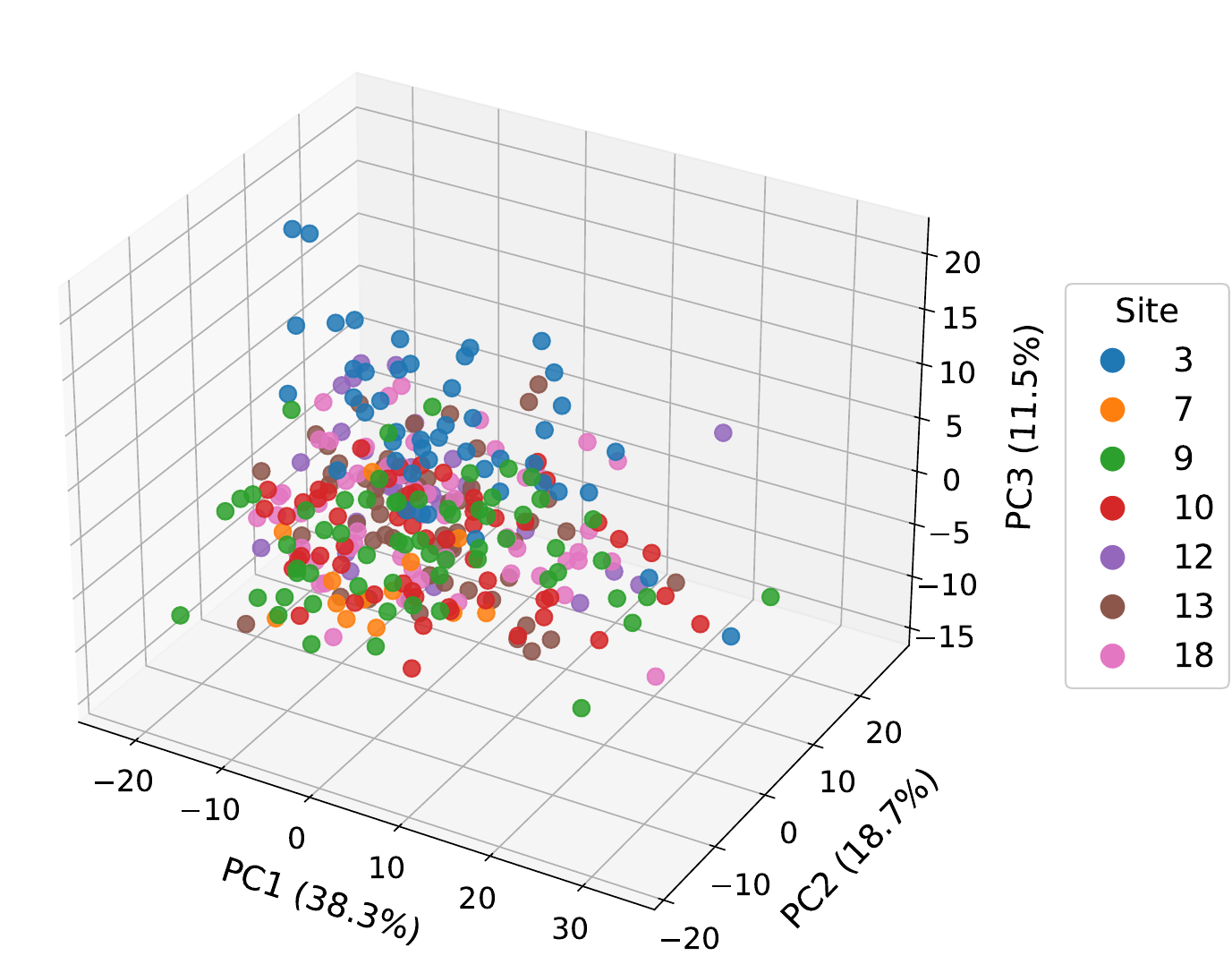}
        \caption{PCA on FBIRN}
    \end{subfigure}%
    \hspace{0.01\textwidth}
    \begin{subfigure}{0.34\textwidth}
        \centering
        \includegraphics[width=\linewidth]{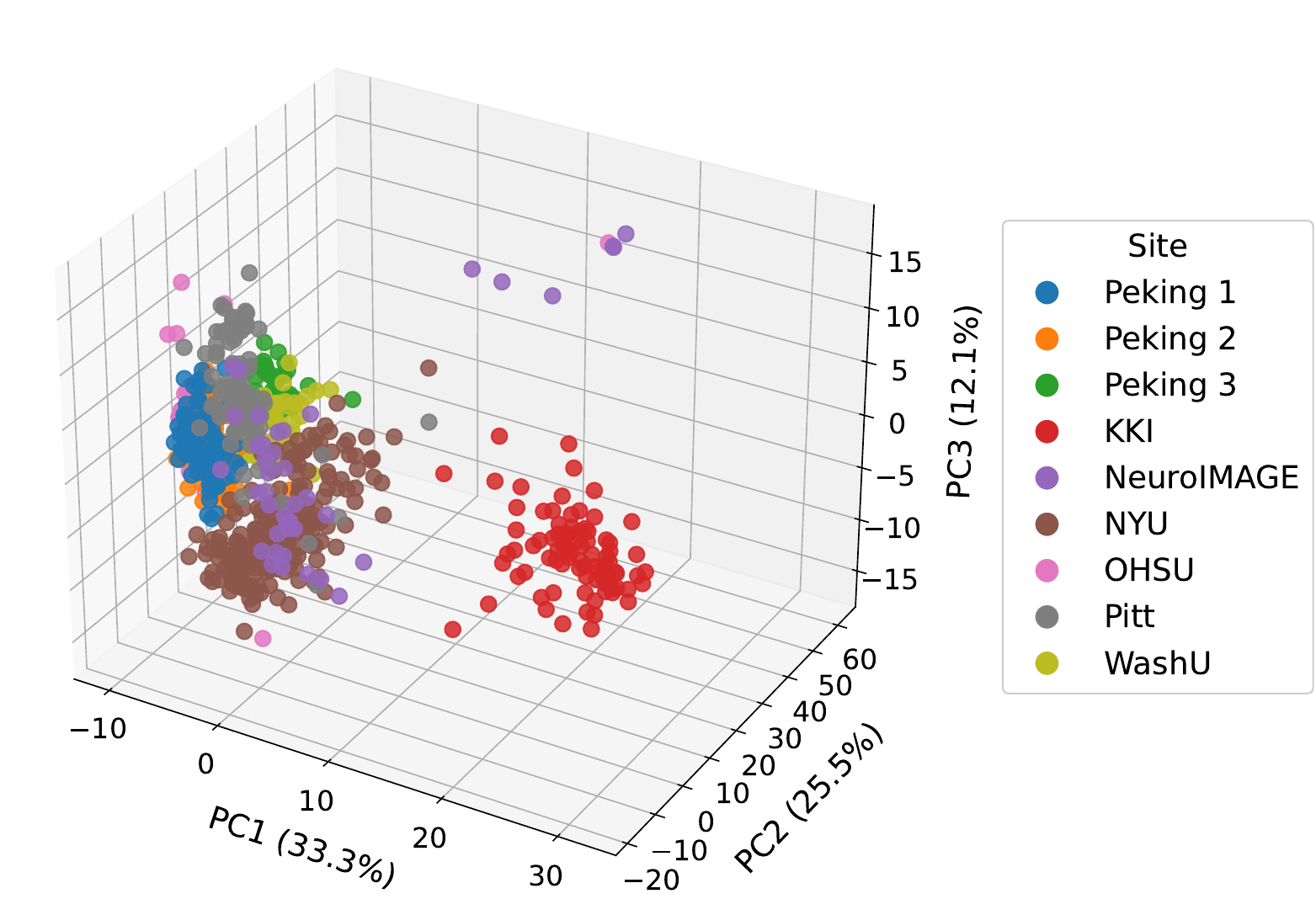}
        \caption{PCA on ADHD-200}
    \end{subfigure}%
    \hspace{0.01\textwidth}
    \begin{subfigure}{0.32\textwidth}
        \centering
        \includegraphics[width=\linewidth]{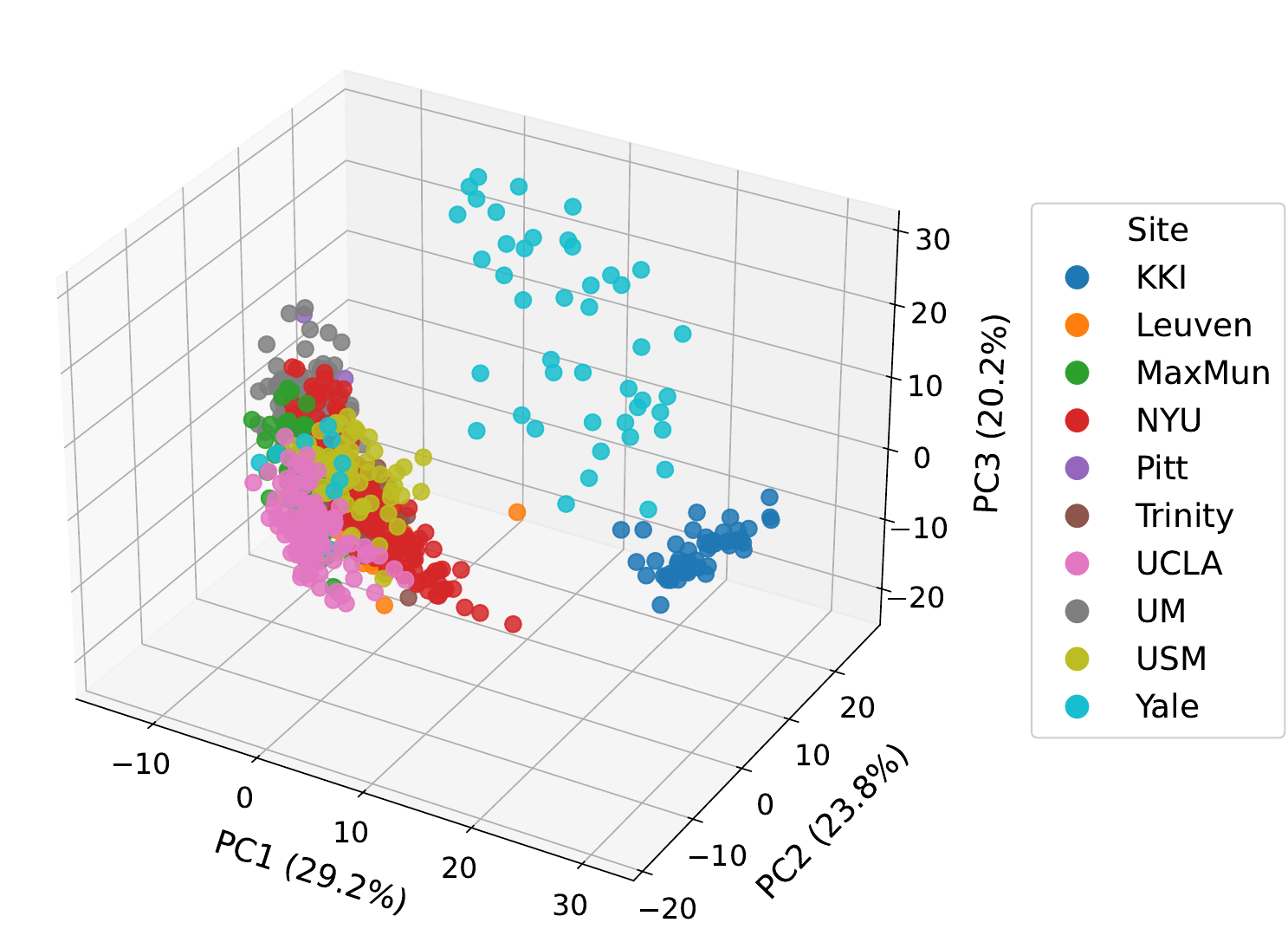}
        \caption{PCA on ABIDE I}
    \end{subfigure}%

    \vspace{0.3cm}
    \begin{subfigure}{0.30\textwidth}
        \centering
        \includegraphics[width=\linewidth]{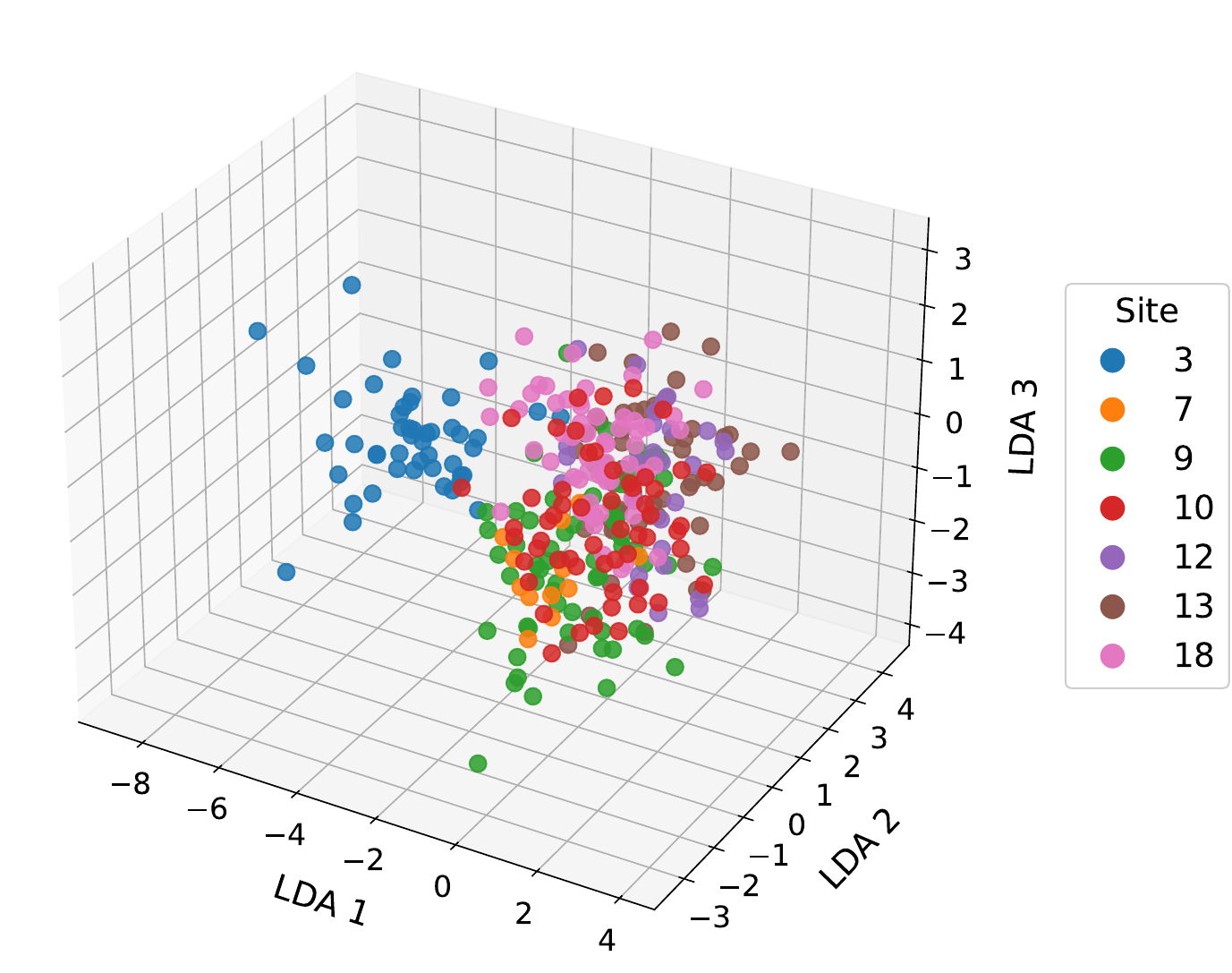}
        \caption{LDA on FBIRN}
    \end{subfigure}%
    \hspace{0.01\textwidth}
    \begin{subfigure}{0.34\textwidth}
        \centering
        \includegraphics[width=\linewidth]{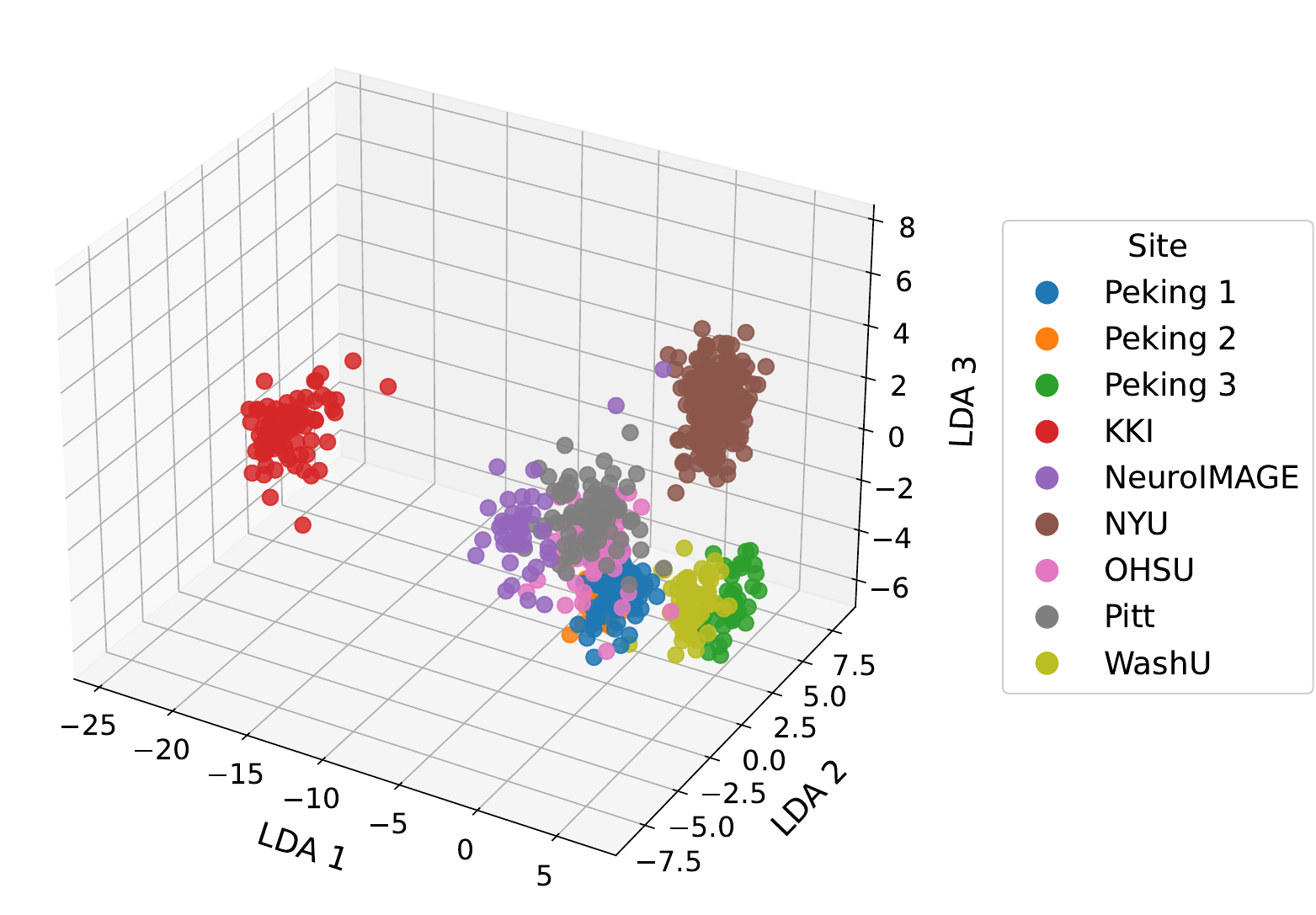}
        \caption{LDA on ADHD-200}
    \end{subfigure}%
    \hspace{0.01\textwidth}
    \begin{subfigure}{0.32\textwidth}
        \centering
        \includegraphics[width=\linewidth]{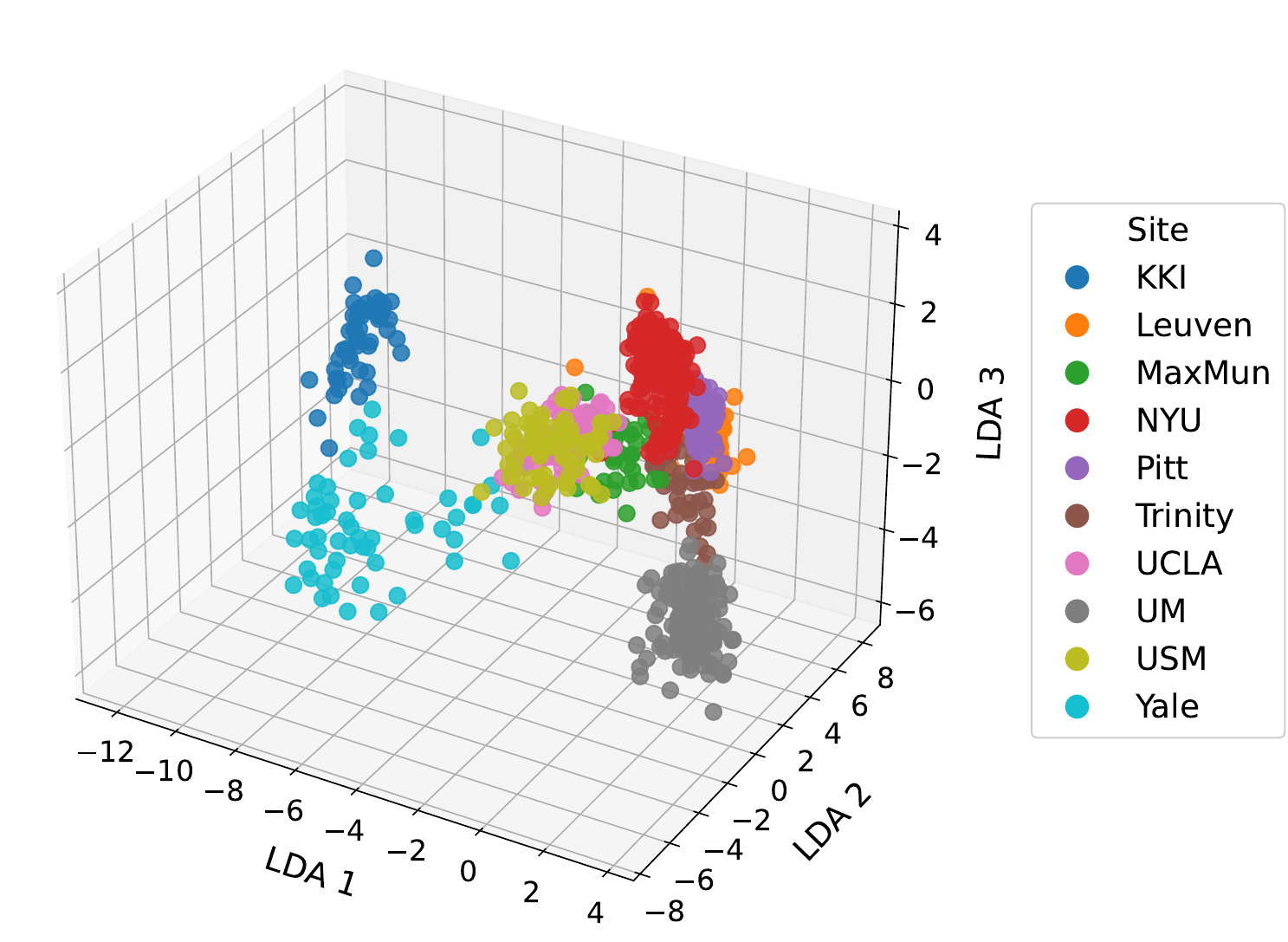}
        \caption{LDA on ABIDE I}
    \end{subfigure}

    \caption{Visualization of subject-level embeddings extracted from the pre-trained SwiFT model. In the last row, embeddings are first projected using PCA with 20 components, followed by further dimensionality reduction using LDA. All points are colored according to site identity. For the ABIDE I dataset, which includes 17 imaging sites, only the 10 sites with the largest sample sizes are shown for clarity.}
    \label{fig:swift_sites}
\end{figure}

Figure~\ref{fig:brainlm_sites} and Figure~\ref{fig:swift_sites} show embeddings from the pre-trained BrainLM and SwiFT models, projected into low-dimensional spaces using PCA and PCA followed by LDA for qualitative inspection. The projections reveal pronounced batch-dependent structures. In the FBIRN dataset, embeddings from both models separate clearly across sites, with Site~3 forming a distinct outlier. This site uniquely employed a GE LX scanner with spiral k-space trajectories, whereas most other sites used Siemens scanners with linear acquisitions, suggesting that scanner differences contribute to the divergence. A comparable trend is observed in ABIDE~I, where embeddings form clusters corresponding to site identity. To enhance clarity, only the 10 sites with the largest subject numbers are displayed. 

Table~\ref{tab:permanova} reports PERMANOVA results for all datasets. Effect sizes for batch and diagnosis are shown. Classifiers trained to predict site identity from pre-trained embeddings achieved high accuracy across all datasets (see Table~\ref{tab:site_diag_prediction}). In FBIRN, although site prediction accuracy is lower than in other datasets, it remains well above chance ($1/7 \approx 14.29\%$), indicating substantial batch-related information in the embeddings. ABIDE~I shows a similar trend, with embeddings enabling highly predictable site identity, although visualization is restricted to the 10 largest sites for clarity.  

\begin{table}[htbp]
\centering
\footnotesize
\caption{PERMANOVA $F$-statistics quantifying site and diagnosis effects in FNC and embedding spaces. Boldface values indicate statistically significant effects ($p < 0.05$).}
\begin{tabular}{llcc}
\toprule
Dataset & Feature & Site Pseudo-$F$ & Diagnosis Pseudo-$F$ \\
\midrule

\multirow{2}{*}{FBIRN}
& FNC & \textbf{1.74} & \textbf{12.52} \\
\cmidrule(lr){2-4}
& BrainLM & \textbf{717.35} & 0.15 \\
\cmidrule(lr){2-4}
& SwiFT   & \textbf{6.78}  & \textbf{8.24} \\
\midrule

\multirow{2}{*}{ADHD-200}
& FNC & \textbf{29.65} & \textbf{11.80} \\
\cmidrule(lr){2-4}
& BrainLM & \textbf{523.81} & \textbf{26.41} \\
\cmidrule(lr){2-4}
& SwiFT   & \textbf{131.93} & \textbf{6.48} \\
\midrule

\multirow{2}{*}{ABIDE I}
& FNC & \textbf{5.11} & \textbf{3.92} \\
\cmidrule(lr){2-4}
& BrainLM & \textbf{441.64} & \textbf{0.38} \\
\cmidrule(lr){2-4}
& SwiFT   & \textbf{109.30} & 0.41 \\

\bottomrule
\end{tabular}
\label{tab:permanova}
\end{table}

\begin{table}[htbp]
\centering
\footnotesize
\caption{Site and diagnosis prediction performance using embeddings from pre-trained foundation models across datasets. Embeddings were first reduced via PCA to 20 components before classifier training.}
\begin{tabular}{lllcccccccc}
\toprule
Dataset & Feature & Classifier
& \multicolumn{4}{c}{Site Prediction}
& \multicolumn{4}{c}{Diagnosis Prediction} \\
\cmidrule(lr){4-7} \cmidrule(lr){8-11}
& & & Acc & Prec & Rec & F1
& Acc & Prec & Rec & F1 \\
\midrule

\multirow{6}{*}{FBIRN}
& \multirow{3}{*}{BrainLM}
& LDA
& 56.70 & 54.64 & 56.70 & 55.32
& 48.45 & 48.14 & 48.45 &  47.97 \\
& & Logistic Regression
& 54.64 & 52.38 & 54.64 & 52.97
& 48.45 & 48.14 & 48.45 & 47.97 \\
& & RBF SVM
& 50.52 & 45.76 & 50.52 & 46.25
& 56.70 & 56.66 & 56.70 & 56.65 \\
\cmidrule(lr){2-11}
& \multirow{3}{*}{SwiFT}
& LDA
& 49.48 & 50.07 & 49.48 & 48.90
& 64.95 & 64.93 & 64.95 & 64.91 \\
& & Logistic Regression
& 44.33 & 40.79 & 44.33 & 42.09
& 64.95 & 64.96 & 64.95 & 64.84 \\
& & RBF SVM
& 44.33 & 38.40 & 44.33 & 40.27
& 70.10 & 70.40 & 70.10 & 69.88 \\

\midrule

\multirow{6}{*}{ADHD-200}
& \multirow{3}{*}{BrainLM}
& LDA
& 94.30 & 95.13 & 94.30 & 94.19
& 66.67 & 66.16 & 66.67 & 66.37 \\
& & Logistic Regression
& 93.86 & 94.17 & 93.86 & 93.97
& 65.35 & 65.26 & 65.35 & 65.30 \\
& & RBF SVM
& 89.91 & 91.04 & 89.91 & 89.59
& 67.54 & 66.04 & 67.54 & 65.99 \\
\cmidrule(lr){2-11}
& \multirow{3}{*}{SwiFT}
& LDA
& 89.47 & 89.60 & 89.47 & 89.30
& 71.05 & 70.22 & 71.05 & 68.78 \\
& & Logistic Regression
& 88.60 & 88.89 & 88.60 & 88.61
& 70.18 & 69.19 & 70.18 & 67.62 \\
& & RBF SVM
& 85.09 & 84.65 & 85.09 & 83.64
& 68.42 & 67.02 & 68.42 & 65.23 \\

\midrule

\multirow{6}{*}{ABIDE I}
& \multirow{3}{*}{BrainLM}
& LDA
& 70.87 & 72.28 & 70.87 & 68.05
& 53.40 & 53.26 & 53.40 & 52.95 \\
& & Logistic Regression
& 84.47 & 85.10 & 84.47 & 84.32
& 52.10 & 51.93 & 52.10 & 51.70 \\
& & RBF SVM
& 60.84 & 53.54 & 60.84 & 53.81
& 51.13 & 51.30 & 51.13 & 51.06 \\
\cmidrule(lr){2-11}
& \multirow{3}{*}{SwiFT}
& LDA
& 87.70 & 88.98 & 87.70 & 87.91
& 48.22 & 47.96 & 48.22 & 47.85 \\
& & Logistic Regression
& 88.67 & 89.08 & 88.67 & 88.74
& 47.90 & 47.58 & 47.90 & 47.43 \\
& & RBF SVM
& 83.82 & 83.40 & 83.82 & 82.77
& 48.87 & 48.52 & 48.87 & 48.23 \\

\bottomrule
\end{tabular}
\label{tab:site_diag_prediction}
\end{table}

\subsection{Additional Analyses of Batch Effects Versus Biological Signals}
Table~\ref{tab:ablation_results} provides the full results corresponding to the controlled evaluation settings discussed in the main text. It complements Table~\ref{tab:ablation_adhd200_results} by reporting all metrics across datasets, models, and experimental configurations.

\begin{table}[htbp]
\centering
\scriptsize
\caption{Classification performance under four controlled settings. Setting denotes the experimental scenario, and Task specifies the prediction target. Data is represented using the format site (label subset). For example, KKI (Control) includes only healthy subjects from KKI, KKI (Patient) includes only patients, and KKI (All) includes all subjects. Frozen foundation model embeddings are standardized and reduced via PCA prior to LDA.}
\begin{tabular}{lllccccccc}
\toprule
Dataset & Feature & Setting 
& Task & Data
& Acc & Prec & Rec & F1 \\
\midrule

\multirow{12}{*}{FBIRN}
& \multirow{6}{*}{BrainLM}
& Within-site
& Diagnosis & Site 3 (All)
& 73.33 & 75.56 & 73.33 & 73.10 \\
& & Within-site
& Diagnosis & Site 18 (All)
& 61.11 & 78.12 & 61.11 & 54.18 \\
& & Multi-site
& Diagnosis & Site 3 (All) \& Site 18 (All)
& 62.50 & 62.70 & 62.50 & 62.35 \\
& & Confounded
& Confounded & Site 3 (Control) \& Site 18 (Patient)
& 100.00 & 100.00 & 100.00 & 100.00 \\
& & Confounded
& Confounded & Site 3 (Patient) \& Site 18 (Control)
& 100.00 & 100.00 & 100.00 & 100.00 \\
& & Site-only
& Site & Site 3 (Control) \& Site 18 (Control)
& 100.00 & 100.00 & 100.00 & 100.00 \\
& & Site-only
& Site & Site 3 (Patient) \& Site 18 (Patient)
& 100.00 & 100.00 & 100.00 & 100.00 \\
\cmidrule(lr){2-9}

& \multirow{6}{*}{SwiFT}
& Within-site
& Diagnosis & Site 3 (All)
& 80.00 & 80.37 & 80.00 & 79.82 \\
& & Within-site
& Diagnosis & Site 18 (All)
& 66.67 & 67.53 & 66.67 & 66.25 \\
& & Multi-site
& Diagnosis & Site 3 (All) \& Site 18 (All)
& 62.50 & 62.50 & 62.50 & 62.50 \\
& & Confounded
& Confounded & Site 3 (Control) \& Site 18 (Patient)
& 100.00 & 100.00 & 100.00 & 100.00 \\
& & Confounded
& Confounded & Site 3 (Patient) \& Site 18 (Control)
& 100.00 & 100.00 & 100.00 & 100.00 \\
& & Site-only
& Site & Site 3 (Control) \& Site 18 (Control)
& 94.12 & 94.65 & 94.12 & 94.03 \\
& & Site-only
& Site & Site 3 (Patient) \& Site 18 (Patient)
& 93.75 & 94.38 & 93.75 & 93.67 \\

\midrule

\multirow{12}{*}{ADHD-200}
& \multirow{6}{*}{BrainLM}
& Within-site
& Diagnosis & Peking 1 (All)
& 76.92 & 75.14 & 76.92 & 73.41 \\
& & Within-site
& Diagnosis & KKI (All)
& 60.00 & 49.09 & 60.00 & 54.00 \\
& & Multi-site
& Diagnosis & Peking 1 (All) \& KKI (All)
& 70.59 & 52.24 & 70.59 & 60.04 \\
& & Confounded
& Confounded & Peking 1 (Control) \& KKI (Patient)
& 100.00 & 100.00 & 100.00 & 100.00 \\
& & Confounded
& Confounded & Peking 1 (Patient) \& KKI (Control)
& 100.00 & 100.00 & 100.00 & 100.00 \\
& & Site-only
& Site & Peking 1 (Control) \& KKI (Control)
& 100.00 & 100.00 & 100.00 & 100.00 \\
& & Site-only
& Site & Peking 1 (Patient) \& KKI (Patient)
& 100.00 & 100.00 & 100.00 & 100.00 \\

\cmidrule(lr){2-9}

& \multirow{6}{*}{SwiFT}
& Within-site
& Diagnosis & Peking 1 (All)
& 76.92 & 82.46 & 76.92 & 69.84 \\
& & Within-site
& Diagnosis & KKI (All)
& 72.00 & 67.22 & 72.00 & 65.93 \\
& & Multi-site
& Diagnosis & Peking 1 (All) \& KKI (All)
& 60.78 & 54.05 & 60.78 & 56.91 \\
& & Confounded
& Confounded & Peking 1 (Control) \& KKI (Patient)
& 100.00 & 100.00 & 100.00 & 100.00 \\
& & Confounded
& Confounded & Peking 1 (Patient) \& KKI (Control)
& 100.00 & 100.00 & 100.00 & 100.00 \\
& & Site-only
& Site & Peking 1 (Control) \& KKI (Control)
& 100.00 & 100.00 & 100.00 & 100.00 \\
& & Site-only
& Site & Peking 1 (Patient) \& KKI (Patient)
& 100.00 & 100.00 & 100.00 & 100.00 \\

\midrule

\multirow{12}{*}{ABIDE~I}
& \multirow{6}{*}{BrainLM}
& Within-site
& Diagnosis & Caltech (All)
& 66.67 & 68.75 & 66.67 & 65.71 \\
& & Within-site
& Diagnosis & MaxMun (All)
& 66.67 & 67.86 & 66.67 & 66.97 \\
& & Multi-site
& Diagnosis & Caltech (All) \& MaxMun (All)
& 57.69 & 57.48 & 57.69 & 57.50 \\
& & Confounded
& Confounded & Caltech (Control) \& MaxMun (Patient)
& 100.00 & 100.00 & 100.00 & 100.00 \\
& & Confounded
& Confounded & Caltech (Patient) \& MaxMun (Control)
& 100.00 & 100.00 & 100.00 & 100.00 \\
& & Site-only
& Site & Caltech (Control) \& MaxMun (Control)
& 100.00 & 100.00 & 100.00 & 100.00 \\
& & Site-only
& Site & Caltech (Patient) \& MaxMun (Patient) & 100.00 & 100.00 & 100.00 & 100.00 \\

\cmidrule(lr){2-9}

& \multirow{6}{*}{SwiFT}
& Within-site
& Diagnosis & Caltech (All)
& 73.58 & 74.94 & 73.58 & 72.36 \\
& & Within-site
& Diagnosis & MaxMun (All)
& 63.64 & 60.61 & 63.64 & 59.85 \\
& & Multi-site
& Diagnosis & Caltech (All) \& MaxMun (All)
& 70.31 & 70.06 & 70.31 & 69.74 \\
& & Confounded
& Confounded & Caltech (Control) \& MaxMun (Patient)
& 97.14 & 97.71 & 97.14 & 97.28 \\
& & Confounded
& Confounded & Caltech (Patient) \& MaxMun (Control)
& 100.00 & 100.00 & 100.00 & 100.00 \\
& & Site-only
& Site & Caltech (Control) \& MaxMun (Control)
& 97.30 & 97.38 & 97.30 & 97.22 \\
& & Site-only
& Site & Caltech (Patient) \& MaxMun (Patient)
& 100.00 & 100.00 & 100.00 & 100.00 \\

\bottomrule
\end{tabular}
\label{tab:ablation_results}
\end{table}

\subsection{Additional Analyses of Comparison with Handcrafted FNC Features}
\label{appendix:fnc_baseline}
Batch effects have long posed a major challenge for traditional handcrafted neuroimaging representations such as FNC. This has motivated extensive research on harmonization and batch correction methods. To ensure a fair comparison with foundation model embeddings, we apply the same qualitative and quantitative evaluation pipeline to FNC features. In this study, FNC representations are constructed using the AAL-424 atlas for brain parcellation, with functional connectivity defined as Pearson correlations between ROI timeseries. We further analyze low-dimensional projections, quantify batch effects using PERMANOVA, and assess predictive performance on site classification tasks.

\begin{figure}[htbp]
    \centering
    \begin{subfigure}{0.30\textwidth}
        \centering
        \includegraphics[width=\linewidth]{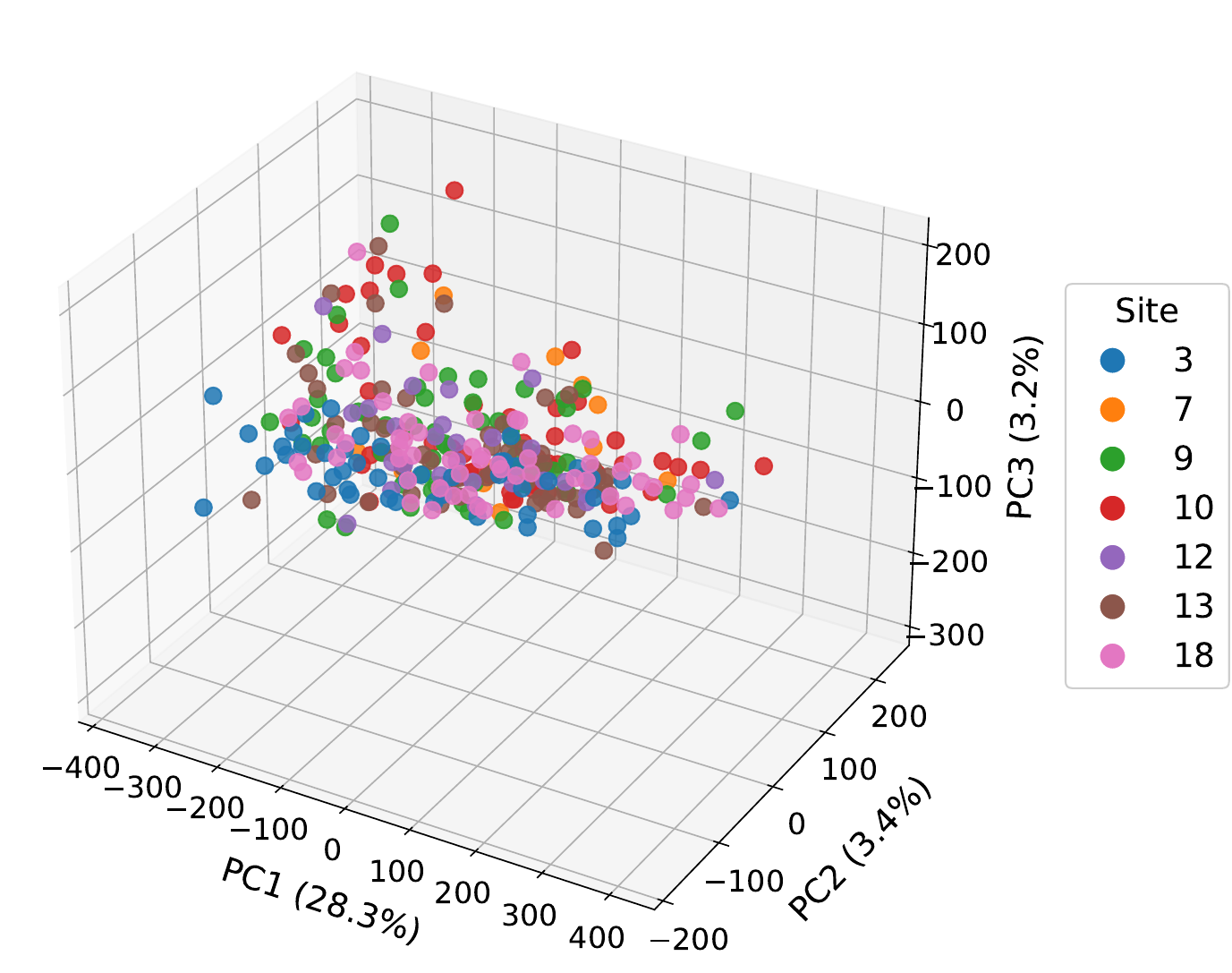}
        \caption{PCA on FBIRN}
    \end{subfigure}%
    \hspace{0.01\textwidth}
    \begin{subfigure}{0.34\textwidth}
        \centering
        \includegraphics[width=\linewidth]{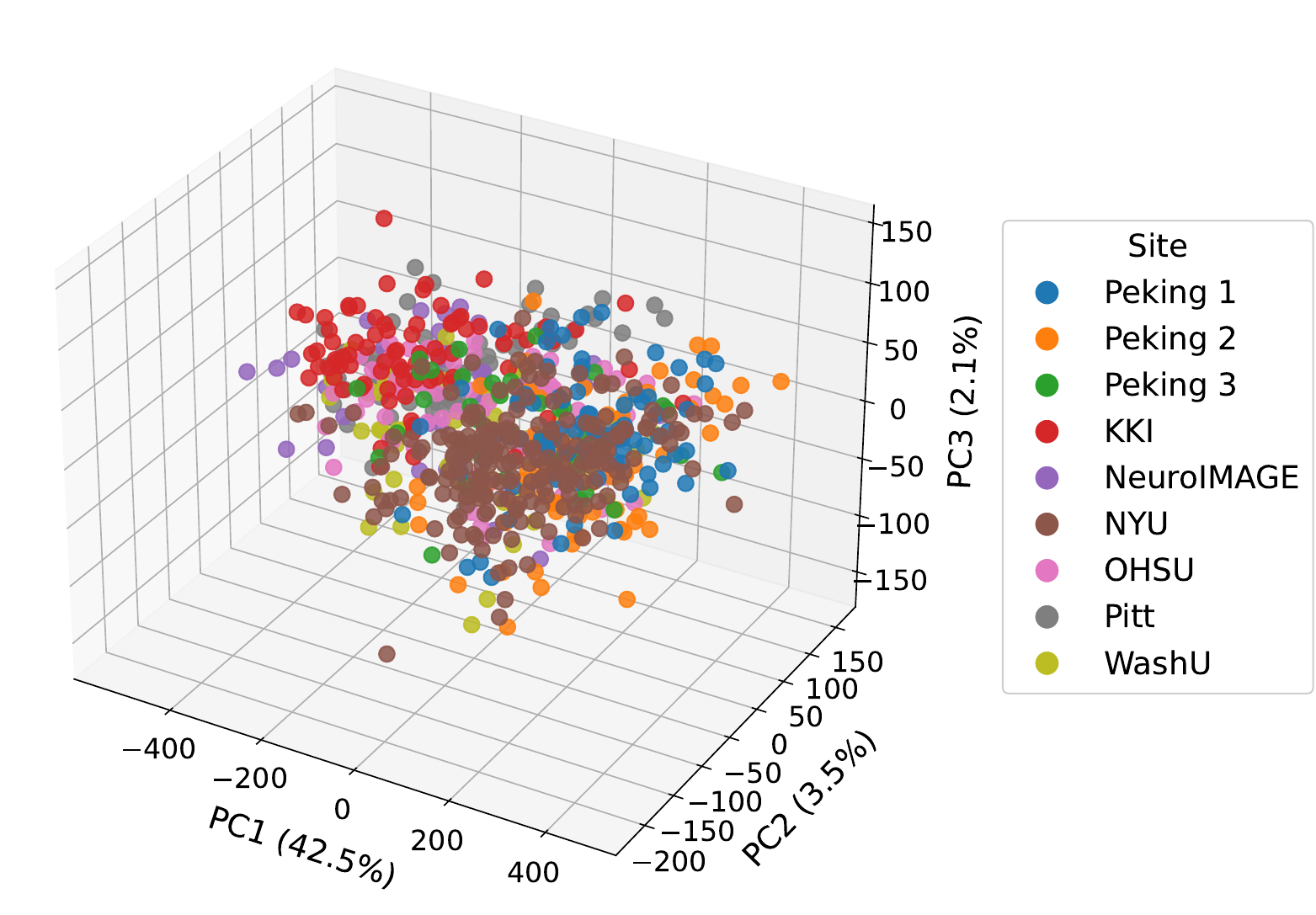}
        \caption{PCA on ADHD-200}
    \end{subfigure}%
    \hspace{0.01\textwidth}
    \begin{subfigure}{0.32\textwidth}
        \centering
        \includegraphics[width=\linewidth]{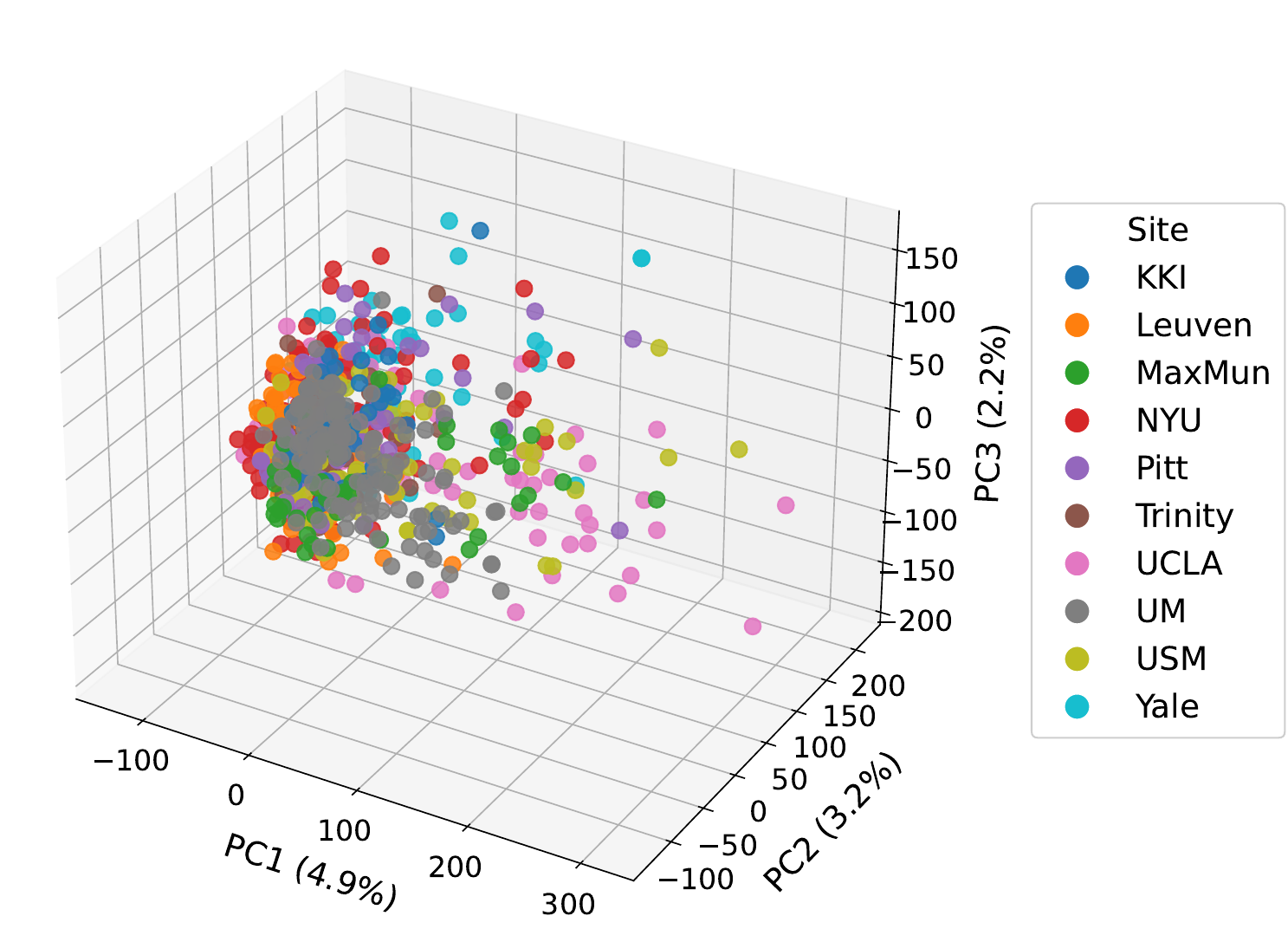}
        \caption{PCA on ABIDE I}
    \end{subfigure}%

    \vspace{0.3cm}
    \begin{subfigure}{0.30\textwidth}
        \centering
        \includegraphics[width=\linewidth]{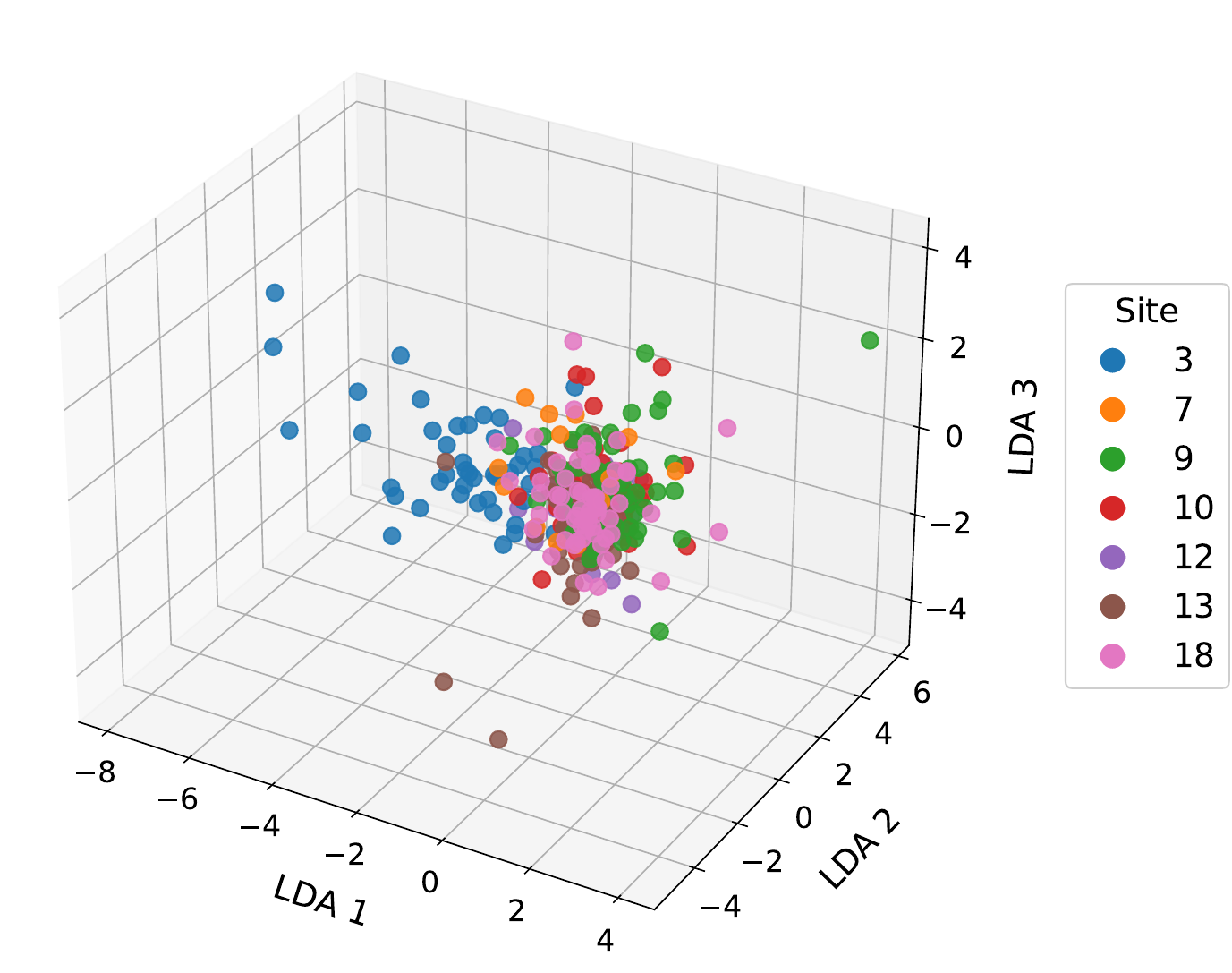}
        \caption{LDA on FBIRN}
    \end{subfigure}%
    \hspace{0.01\textwidth}
    \begin{subfigure}{0.34\textwidth}
        \centering
        \includegraphics[width=\linewidth]{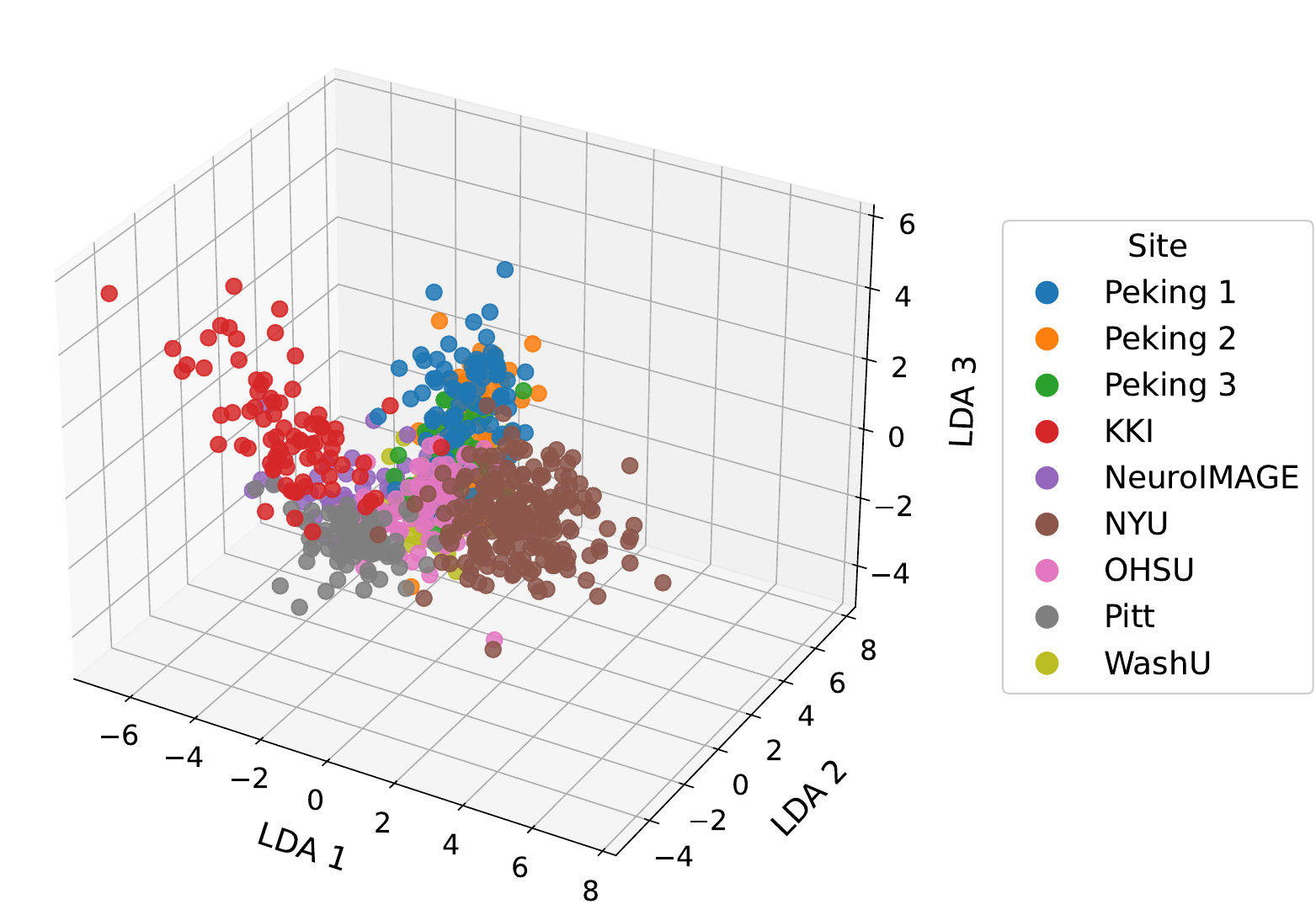}
        \caption{LDA on ADHD-200}
    \end{subfigure}%
    \hspace{0.01\textwidth}
    \begin{subfigure}{0.32\textwidth}
        \centering
        \includegraphics[width=\linewidth]{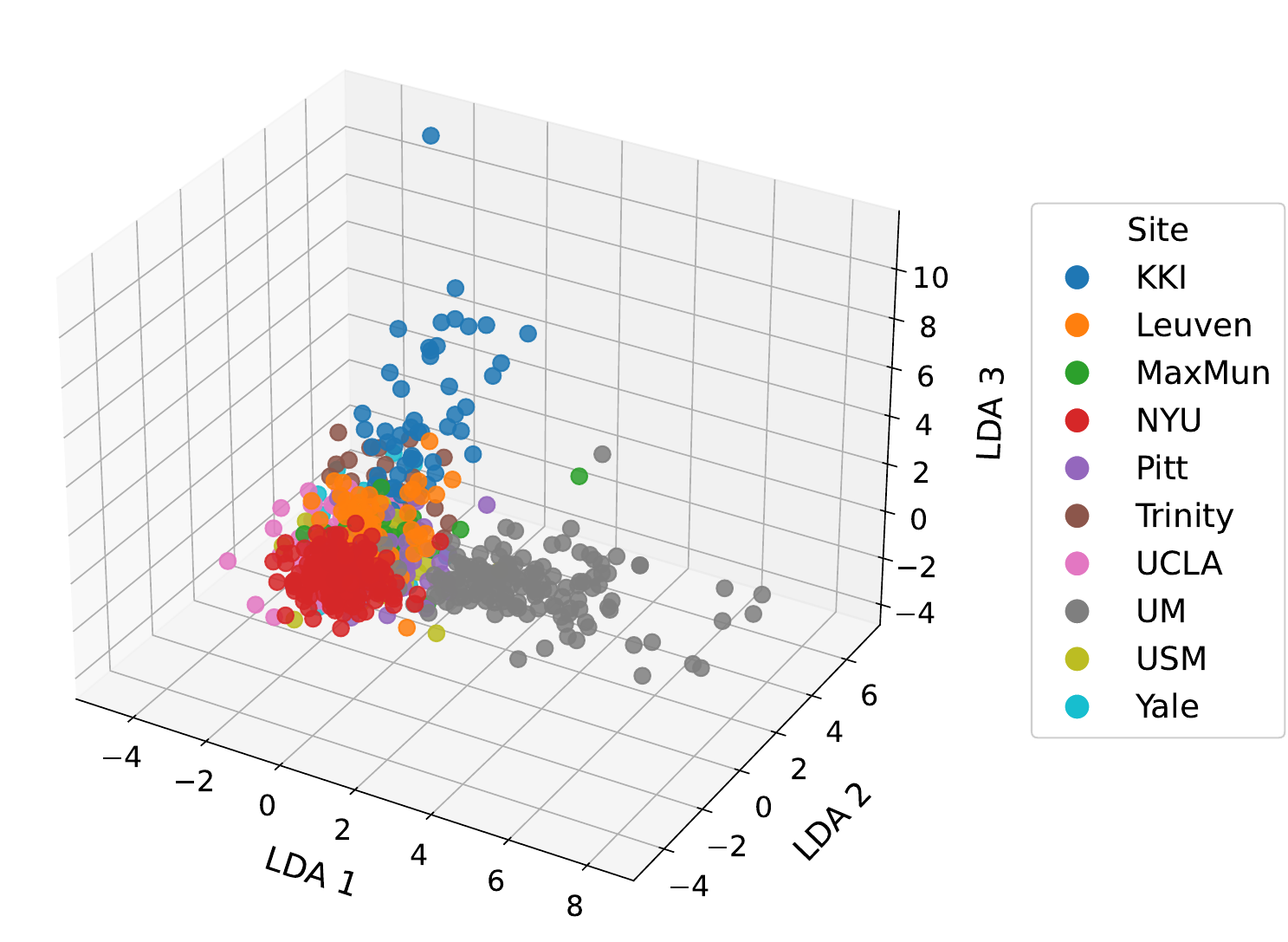}
        \caption{LDA on ABIDE I}
    \end{subfigure}

    \caption{Visualization of subject-level FNC features. In the last row, features are first projected using PCA with 20 components, followed by further dimensionality reduction using LDA. All points are colored according to site identity. For the ABIDE~I dataset, which includes 17 imaging sites, only the 10 sites with the largest sample sizes are shown for clarity.}
    \label{fig:fnc_sites}
\end{figure}

\begin{figure}[htbp]
    \centering
    \begin{subfigure}{0.32\textwidth}
        \centering
        \includegraphics[width=\linewidth]{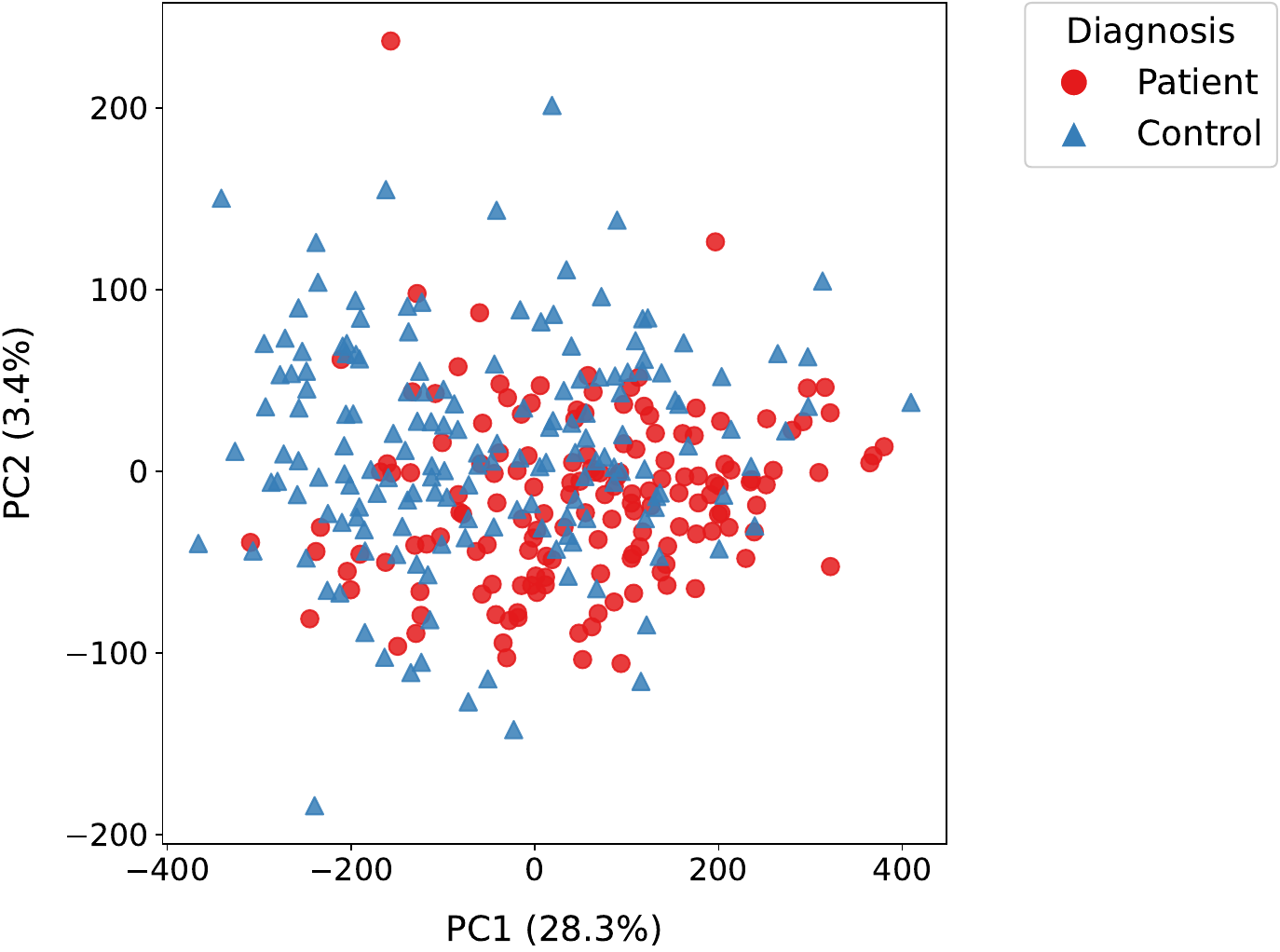}
        \caption{FBIRN FNC}
    \end{subfigure}%
    \hspace{0.01\textwidth}
    \begin{subfigure}{0.32\textwidth}
        \centering
        \includegraphics[width=\linewidth]{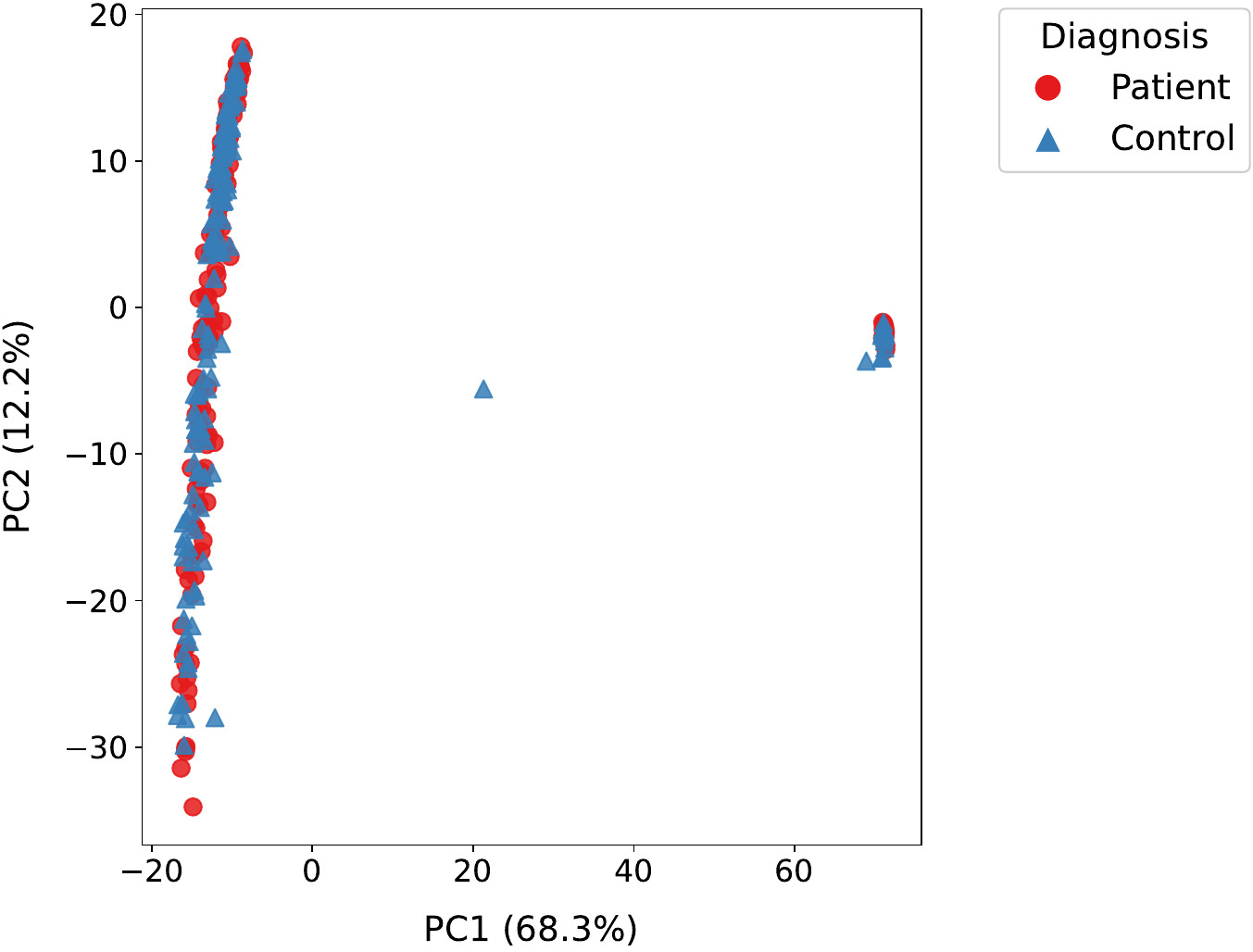}
        \caption{FBIRN BrainLM}
    \end{subfigure}%
    \hspace{0.01\textwidth}
    \begin{subfigure}{0.32\textwidth}
        \centering
        \includegraphics[width=\linewidth]{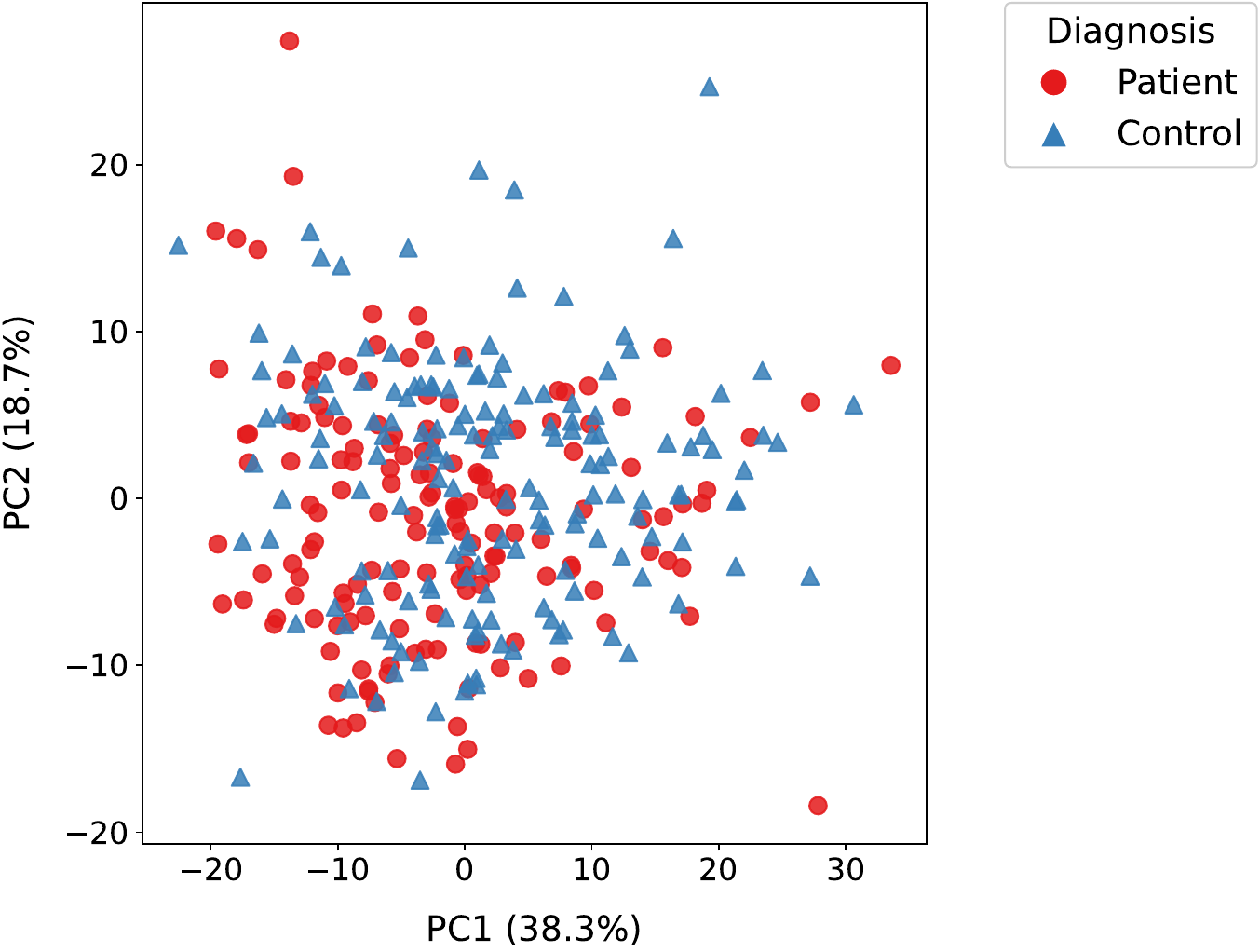}
        \caption{FBIRN SwiFT}
    \end{subfigure}%

    \vspace{0.3cm}
    \begin{subfigure}{0.32\textwidth}
        \centering
        \includegraphics[width=\linewidth]{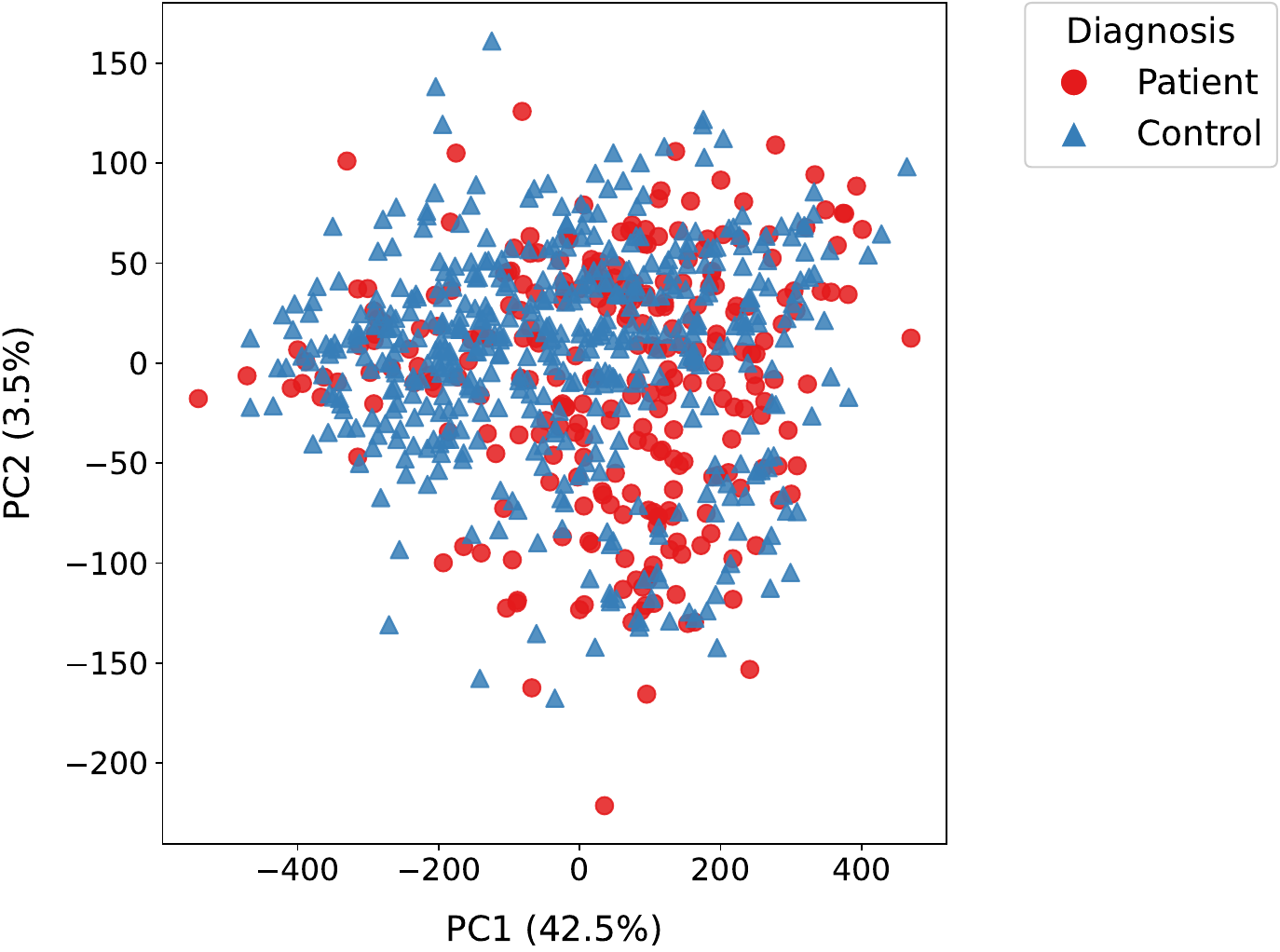}
        \caption{ADHD-200 FNC}
    \end{subfigure}%
    \hspace{0.01\textwidth}
    \begin{subfigure}{0.32\textwidth}
        \centering
        \includegraphics[width=\linewidth]{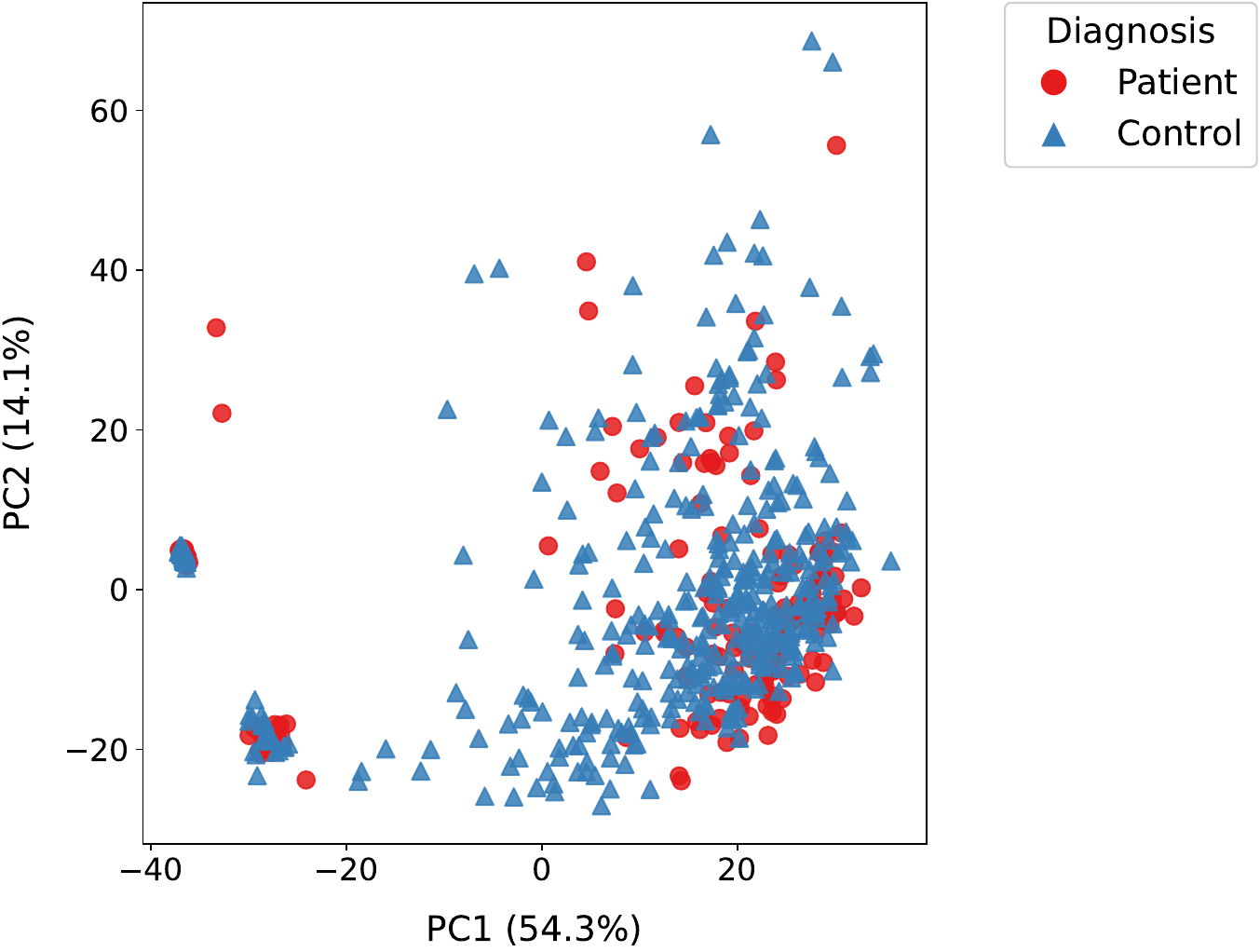}
        \caption{ADHD-200 BrainLM}
    \end{subfigure}%
    \hspace{0.01\textwidth}
    \begin{subfigure}{0.32\textwidth}
        \centering
        \includegraphics[width=\linewidth]{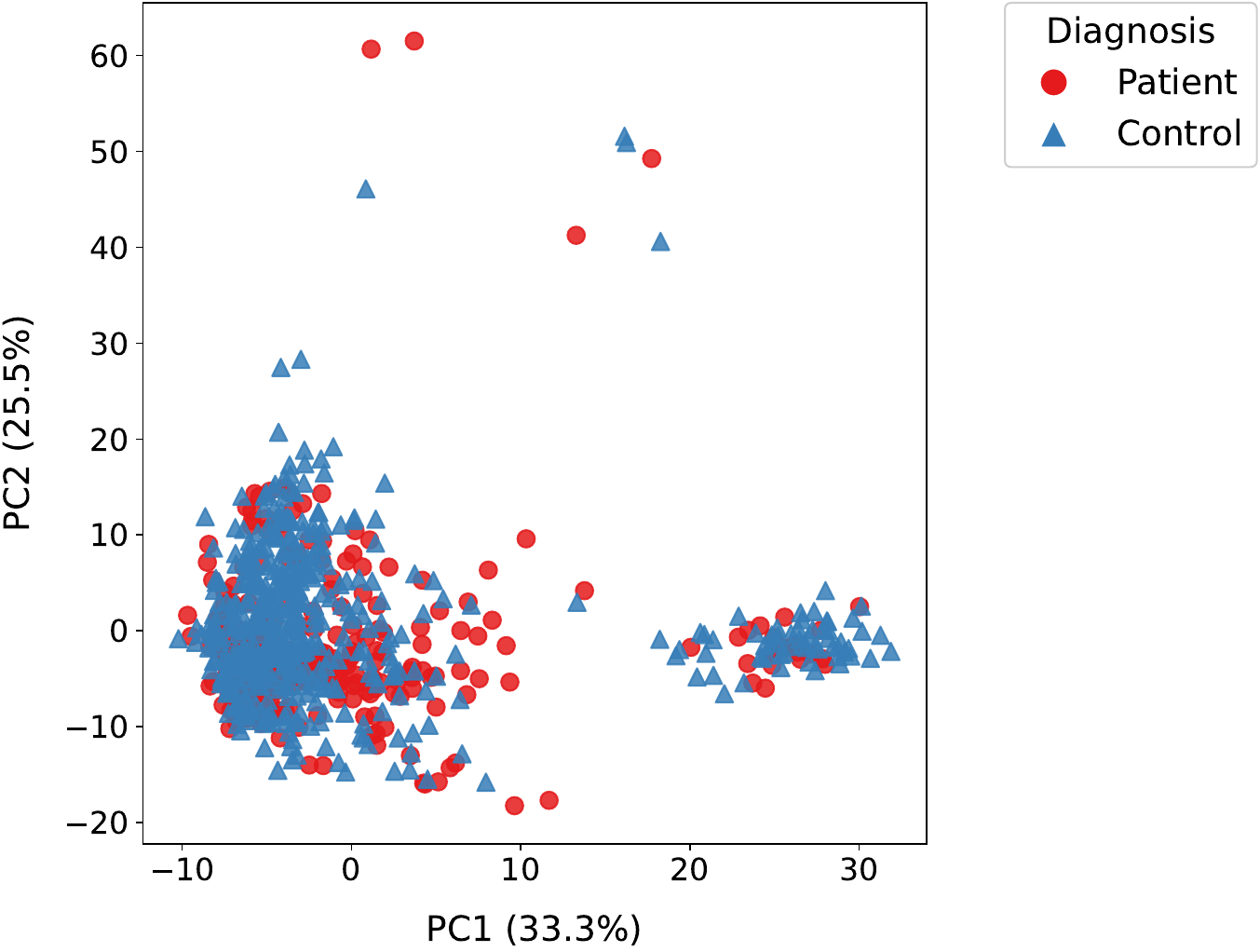}
        \caption{ADHD-200 SwiFT}
    \end{subfigure}

    \vspace{0.3cm}
    \begin{subfigure}{0.32\textwidth}
        \centering
        \includegraphics[width=\linewidth]{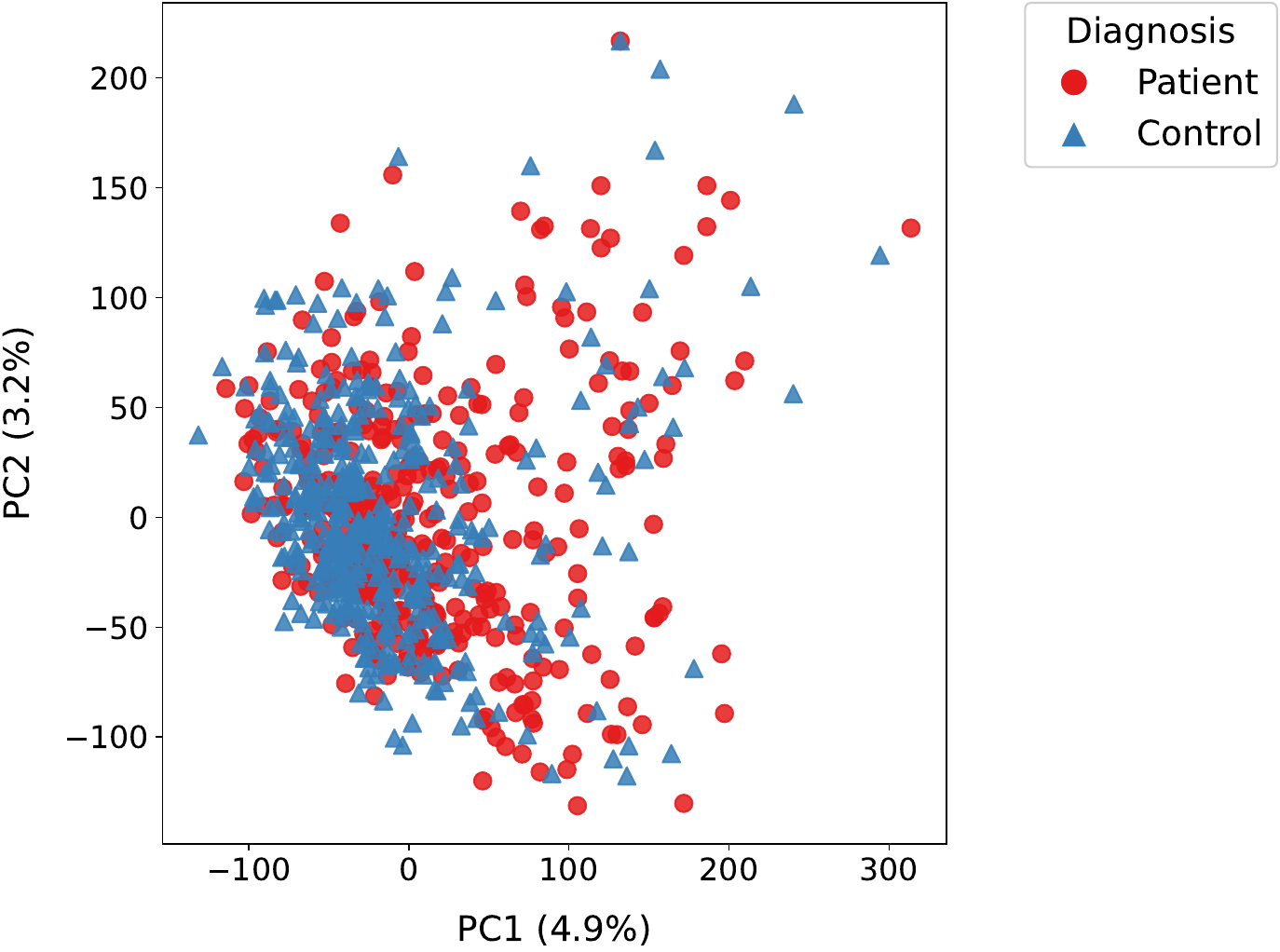}
        \caption{ABIDE I FNC}
    \end{subfigure}%
    \hspace{0.01\textwidth}
    \begin{subfigure}{0.32\textwidth}
        \centering
        \includegraphics[width=\linewidth]{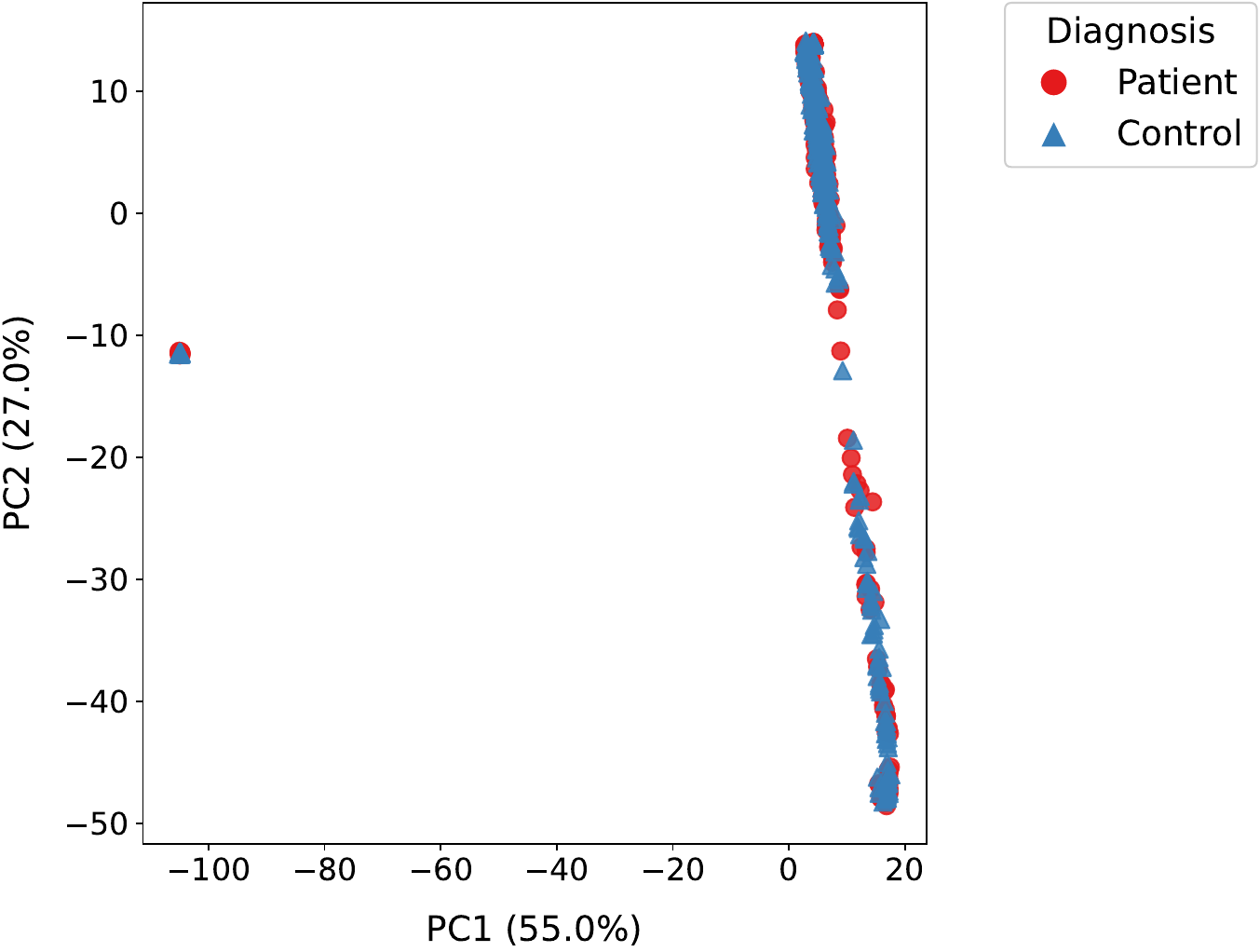}
        \caption{ABIDE I BrainLM}
    \end{subfigure}%
    \hspace{0.01\textwidth}
    \begin{subfigure}{0.32\textwidth}
        \centering
        \includegraphics[width=\linewidth]{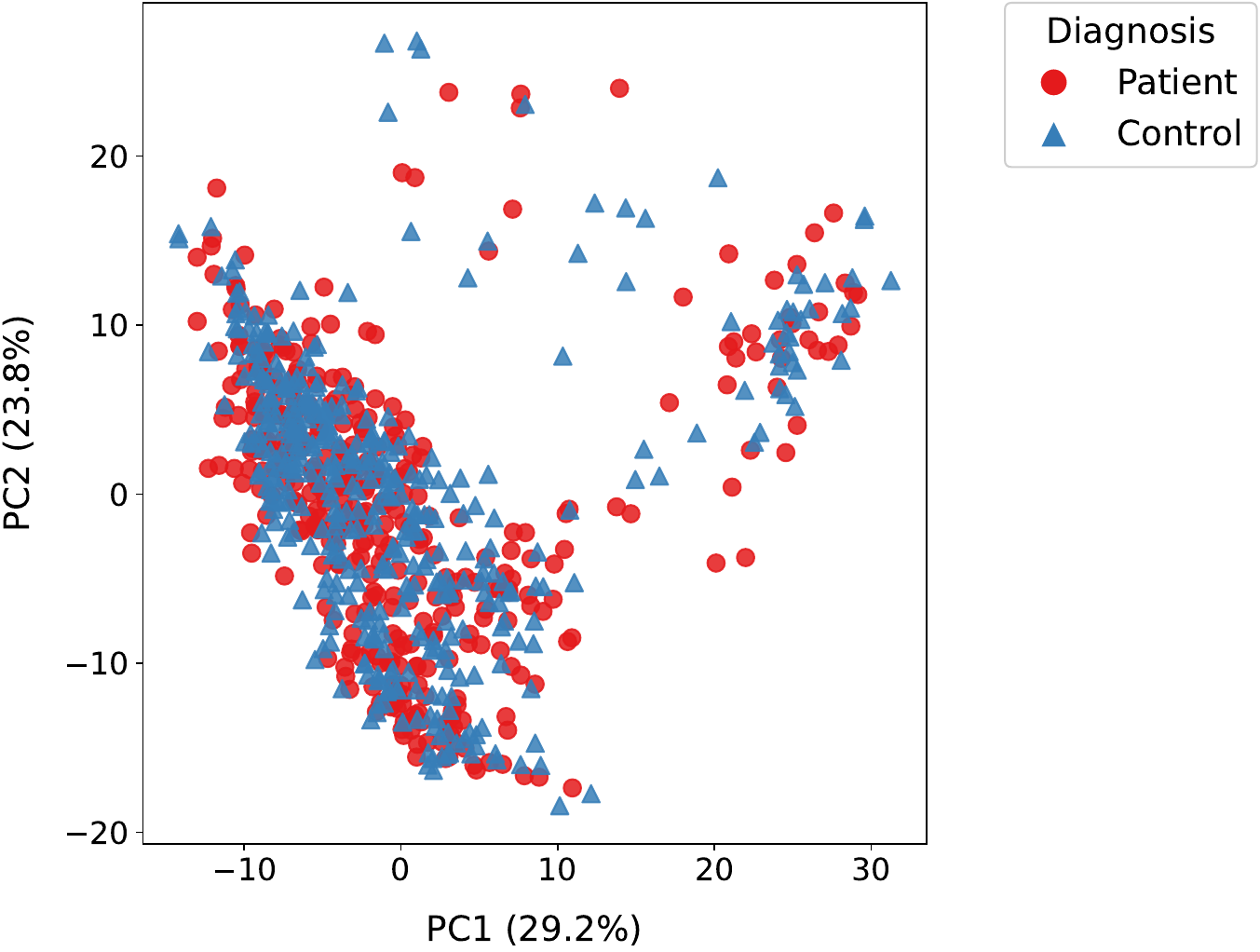}
        \caption{ABIDE I SwiFT}
    \end{subfigure}
    
    \caption{PCA visualization of subject-level representations colored by diagnostic labels. Top, middle, and bottom rows correspond to FBIRN, ADHD-200, and ABIDE~I, respectively. Columns represent different feature types: FNC, BrainLM embeddings, and SwiFT embeddings. This comparison illustrates how the various representations separate diagnostic groups across datasets.}
    \label{fig:diagnosis_pca_all_datasets}
\end{figure}

Qualitative analyses of FNC features are presented in Figure~\ref{fig:fnc_sites}, with corresponding results for BrainLM and SwiFT embeddings shown in Figure~\ref{fig:brainlm_sites} and Figure~\ref{fig:swift_sites}. On the ADHD-200 dataset, the leading low-dimensional components of FNC features do not fully separate samples from certain sites, such as Peking 3. In comparison, the first few principal components of pre-trained foundation model embeddings demonstrate clearer batch-specific clustering, even distinguishing between acquisition batches within the same site. When representations are colored according to diagnosis (see Figure~\ref{fig:diagnosis_pca_all_datasets}), FNC features exhibit a more coherent diagnostic structure than foundation model embeddings. These observations align with quantitative analyses, indicating that while FNC features capture stronger diagnosis-related variability, foundation model embeddings tend to encode batch-related information more prominently in their dominant subspaces.

For comparison, PERMANOVA was applied to FNC features (see Table~\ref{tab:permanova}). In FBIRN, FNC exhibits a modest site effect and a substantially stronger diagnostic effect, indicating that diagnosis-related variability predominates in the multivariate structure. In ADHD-200 and ABIDE~I, both site and diagnostic effects are observed, reflecting the greater heterogeneity and multi-site variability of these cohorts. These results contrast with those obtained from foundation model embeddings, emphasizing a key distinction: FNC features retain stronger diagnosis-related information, whereas pre-trained embeddings are more influenced by batch-specific variability.

To further assess how these differences impact downstream performance, the predictability of site and diagnosis was evaluated using both linear and nonlinear classifiers trained on the same features. Under identical evaluation protocols, FNC features exhibit moderate site effects. In contrast, pre-trained foundation model embeddings achieve substantially higher site classification accuracy but lower diagnosis prediction performance (see Table~\ref{tab:site_diag_prediction} and Table~\ref{tab:fnc_site_diag_prediction}). These results indicate that foundation model embeddings encode stronger batch-related variability, whereas FNC features preserve more diagnosis-relevant information.

\begin{table}[htbp]
\centering
\footnotesize
\caption{Site and diagnosis prediction performance using traditional FNC features across datasets. All features were reduced via PCA to 20 components before classifier training.}
\begin{tabular}{lllcccccccc}
\toprule
Dataset & Feature & Classifier
& \multicolumn{4}{c}{Site Prediction}
& \multicolumn{4}{c}{Diagnosis Prediction} \\
\cmidrule(lr){4-7} \cmidrule(lr){8-11}
& & & Acc & Prec & Rec & F1
& Acc & Prec & Rec & F1 \\
\midrule

\multirow{3}{*}{FBIRN}
& \multirow{3}{*}{FNC}
& LDA
& 31.96 & 35.00 & 31.96 & 32.44
& 72.16 & 72.49 & 72.16 & 72.14 \\
& & Logistic Regression
& 34.02 & 35.10 & 34.02 & 34.20
& 74.23 & 74.57 & 74.23 & 74.20 \\
& & RBF SVM
& 34.02 & 30.94 & 34.02 & 31.56
& 71.13 & 72.72 & 71.13 & 70.80 \\

\midrule

\multirow{3}{*}{ADHD-200}
& \multirow{3}{*}{FNC}
& LDA
& 70.18 & 73.43 & 70.18 & 70.64
& 66.23 & 64.18 & 66.23 & 63.69 \\
& & Logistic Regression
& 73.68 & 73.81 & 73.68 & 73.44
& 66.23 & 64.18 & 66.23 & 63.69 \\
& & RBF SVM
& 69.30 & 71.70 & 69.30 & 67.23
& 64.91 & 61.85 & 64.91 & 59.04 \\

\midrule

\multirow{3}{*}{ABIDE I}
& \multirow{3}{*}{FNC}
& LDA
& 61.49 & 62.87 & 61.49 & 60.86
& 65.37 & 65.36 & 65.37 & 65.35 \\
& & Logistic Regression
& 67.31 & 69.59 & 67.31 & 67.43
& 64.08 & 64.06 & 64.08 & 64.04 \\
& & RBF SVM
& 66.02 & 66.70 & 66.02 & 64.63
& 61.81 & 61.93 & 61.81 & 61.81 \\

\bottomrule
\end{tabular}
\label{tab:fnc_site_diag_prediction}
\end{table}

\subsection{Additional Analyses of Harmonization on Pre-trained Embeddings}
For reference, site and diagnosis prediction performance for FNC features without harmonization is reported in Appendix~\ref{appendix:fnc_baseline}. Table~\ref{tab:combat_site_diag_prediction} reports full site and diagnosis prediction performance after applying ComBat harmonization to both FNC features and embeddings from pre-trained foundation models. These results complement the summary observations presented in the main text.

\begin{table}[htbp]
\centering
\footnotesize
\caption{Site and diagnosis prediction performance for FNC features and pre-trained foundation model embeddings after ComBat harmonization. Features were reduced via PCA to 20 components before classifier training.}
\begin{tabular}{lllcccccccc}
\toprule
Dataset & Feature & Classifier
& \multicolumn{4}{c}{Site Prediction}
& \multicolumn{4}{c}{Diagnosis Prediction} \\
\cmidrule(lr){4-7} \cmidrule(lr){8-11}
& & & Acc & Prec & Rec & F1
& Acc & Prec & Rec & F1 \\
\midrule

\multirow{9}{*}{FBIRN}
& \multirow{3}{*}{FNC}
& LDA & 10.31 & 9.09 & 10.31 & 9.59 & 72.16 & 72.30 & 72.16 & 72.16 \\
& & Logistic Regression & 9.28 & 7.95 & 9.28 & 8.49 & 73.20 & 73.66 & 73.20 & 73.14 \\
& & RBF SVM & 20.62 & 14.50 & 20.62 & 16.56 & 70.10 & 71.40 & 70.10 & 69.82 \\
\cmidrule(lr){2-11}
& \multirow{3}{*}{BrainLM}
& LDA & 15.46 & 16.46 & 15.46 & 15.00 & 62.89 & 62.86 & 62.89 & 62.85 \\
& & Logistic Regression & 17.53 & 17.31 & 17.53 & 16.44 & 53.61 & 53.47 & 53.61 & 53.26 \\
& & RBF SVM & 53.61 & 54.12 & 53.61 & 52.28 & 56.70 & 56.71 & 56.70 & 56.09 \\
\cmidrule(lr){2-11}
& \multirow{3}{*}{SwiFT}
& LDA & 13.40 & 11.28 & 13.40 & 11.92 & 64.95 & 65.02 & 64.95 & 64.96 \\
& & Logistic Regression & 14.43 & 12.05 & 14.43 & 12.64 & 65.98 & 66.01 & 65.98 & 65.99 \\
& & RBF SVM & 21.65 & 14.97 & 21.65 & 17.17 & 68.04 & 68.13 & 68.04 & 67.90 \\

\midrule

\multirow{9}{*}{ADHD-200}
& \multirow{3}{*}{FNC}
& LDA & 20.61 & 8.89 & 20.61 & 10.94 & 64.04 & 60.45 & 64.04 & 58.78 \\
& & Logistic Regression & 19.74 & 10.58 & 19.74 & 10.77 & 63.60 & 59.80 & 63.60 & 58.46 \\
& & RBF SVM & 34.65 & 33.66 & 34.65 & 22.91 & 62.28 & 51.62 & 62.28 & 51.25 \\
\cmidrule(lr){2-11}
& \multirow{3}{*}{BrainLM}
& LDA & 28.51 & 19.82 & 28.51 & 17.68 & 65.79 & 63.54 & 65.79 & 59.67 \\
& & Logistic Regression & 27.19 & 15.19 & 27.19 & 16.63 & 65.35 & 62.64 & 65.35 & 59.74 \\
& & RBF SVM & 64.04 & 62.14 & 64.04 & 61.84 & 68.42 & 72.41 & 68.42 & 60.70 \\
\cmidrule(lr){2-11}
& \multirow{3}{*}{SwiFT}
& LDA & 25.44 & 21.17 & 25.44 & 14.22 & 60.96 & 56.69 & 60.96 & 56.86 \\
& & Logistic Regression & 31.14 & 38.88 & 31.14 & 21.64 & 60.09 & 56.03 & 60.09 & 56.52 \\
& & RBF SVM & 52.19 & 58.81 & 52.19 & 47.73 & 65.79 & 63.67 & 65.79 & 59.26 \\

\midrule

\multirow{9}{*}{ABIDE I}
& \multirow{3}{*}{FNC}
& LDA & 13.92 & 5.99 & 13.92 & 7.13 & 62.78 & 62.83 & 62.78 & 62.79 \\
& & Logistic Regression & 12.94 & 5.42 & 12.94 & 6.87 & 61.49 & 61.46 & 61.49 & 61.46 \\
& & RBF SVM & 33.66 & 22.94 & 33.66 & 21.78 & 61.49 & 61.54 & 61.49 & 61.50 \\
\cmidrule(lr){2-11}
& \multirow{3}{*}{BrainLM}
& LDA & 18.77 & 19.22 & 18.77 & 12.72 & 53.40 & 53.34 & 53.40 & 53.33 \\
& & Logistic Regression & 17.48 & 12.86 & 17.48 & 10.72 & 54.37 & 54.34 & 54.37 & 54.35 \\
& & RBF SVM & 66.99 & 63.76 & 66.99 & 62.75 & 52.75 & 52.56 & 52.75 & 51.66 \\
\cmidrule(lr){2-11}
& \multirow{3}{*}{SwiFT}
& LDA & 13.59 & 5.66 & 13.59 & 7.21 & 53.40 & 53.26 & 53.40 & 52.82 \\
& & Logistic Regression & 13.92 & 5.83 & 13.92 & 7.41 & 53.40 & 53.26 & 53.40 & 52.82 \\
& & RBF SVM & 36.57 & 28.33 & 36.57 & 28.55 & 53.07 & 52.92 & 53.07 & 51.94 \\

\bottomrule
\end{tabular}
\label{tab:combat_site_diag_prediction}
\end{table}

\subsection{Additional Analyses of Decoding Biological Signals from Latent Representations}
Full decoding results for ALFF and FNC, with and without ComBat harmonization, are provided in Table~\ref{tab:decode_r2}.

\begin{table}[htbp]
\centering
\footnotesize
\caption{Whole-brain decoding performance ($R^2$) for ALFF and FNC before and after ComBat harmonization. Values are reported as mean $\pm$ standard deviation across regions or connections.}
\begin{tabular}{llcccc}
\toprule
\multirow{2}{*}{Dataset} & \multirow{2}{*}{Feature} 
& \multicolumn{2}{c}{ALFF $R^2$} 
& \multicolumn{2}{c}{FNC $R^2$} \\
\cmidrule(lr){3-4} \cmidrule(lr){5-6}
& & Raw & ComBat & Raw & ComBat \\
\midrule

\multirow{2}{*}{FBIRN}
& BrainLM & 0.484 $\pm$ 0.136 & 0.049 $\pm$ 0.040 & 0.002 $\pm$ 0.026 & 0.001 $\pm$ 0.021 \\
\cmidrule(lr){2-6}
& SwiFT   & 0.268 $\pm$ 0.104 & 0.008 $\pm$ 0.039 & 0.003 $\pm$ 0.031 & 0.004 $\pm$ 0.030 \\
\midrule

\multirow{2}{*}{ADHD-200}
& BrainLM & 0.285 $\pm$ 0.110 & 0.115 $\pm$ 0.025 & 0.179 $\pm$ 0.084 & 0.004 $\pm$ 0.013 \\
\cmidrule(lr){2-6}
& SwiFT   & 0.275 $\pm$ 0.078 & 0.061 $\pm$ 0.070 & 0.197 $\pm$ 0.090 & 0.009 $\pm$ 0.018 \\
\midrule

\multirow{2}{*}{ABIDE I}
& BrainLM & 0.802 $\pm$ 0.110 & 0.505 $\pm$ 0.052 & 0.016 $\pm$ 0.023 & 0.002 $\pm$ 0.009 \\
\cmidrule(lr){2-6}
& SwiFT   & 0.453 $\pm$ 0.067 & 0.103 $\pm$ 0.051 & 0.029 $\pm$ 0.032 & 0.003 $\pm$ 0.012 \\

\bottomrule
\end{tabular}
\label{tab:decode_r2}
\end{table}


\begin{figure}[htbp]
    \centering
    \begin{subfigure}{0.3\textwidth}
        \centering
        \includegraphics[width=\linewidth]{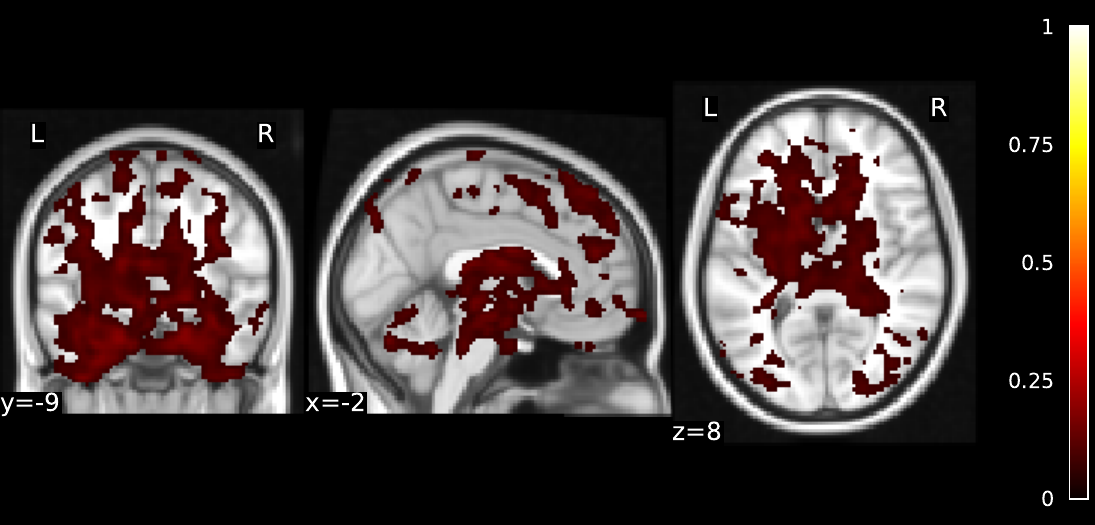}
        \caption{FBIRN BrainLM}
    \end{subfigure}
    \quad
    \begin{subfigure}{0.3\textwidth}
        \centering
        \includegraphics[width=\linewidth]{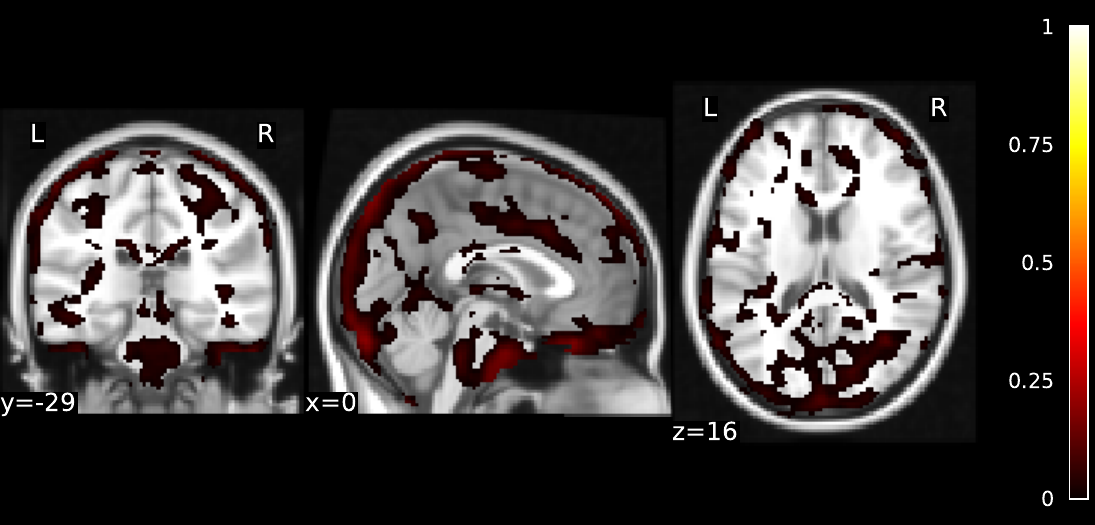}
        \caption{FBIRN SwiFT}
    \end{subfigure}

    \vspace{0.3cm}

    \begin{subfigure}{0.3\textwidth}
        \centering
        \includegraphics[width=\linewidth]{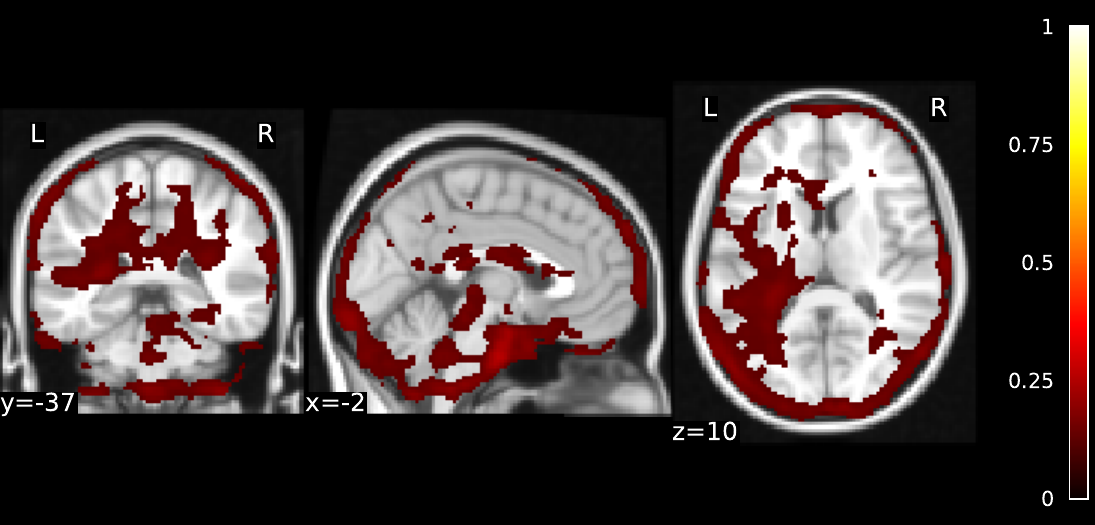}
        \caption{ADHD-200 BrainLM}
    \end{subfigure}
    \quad
    \begin{subfigure}{0.3\textwidth}
        \centering
        \includegraphics[width=\linewidth]{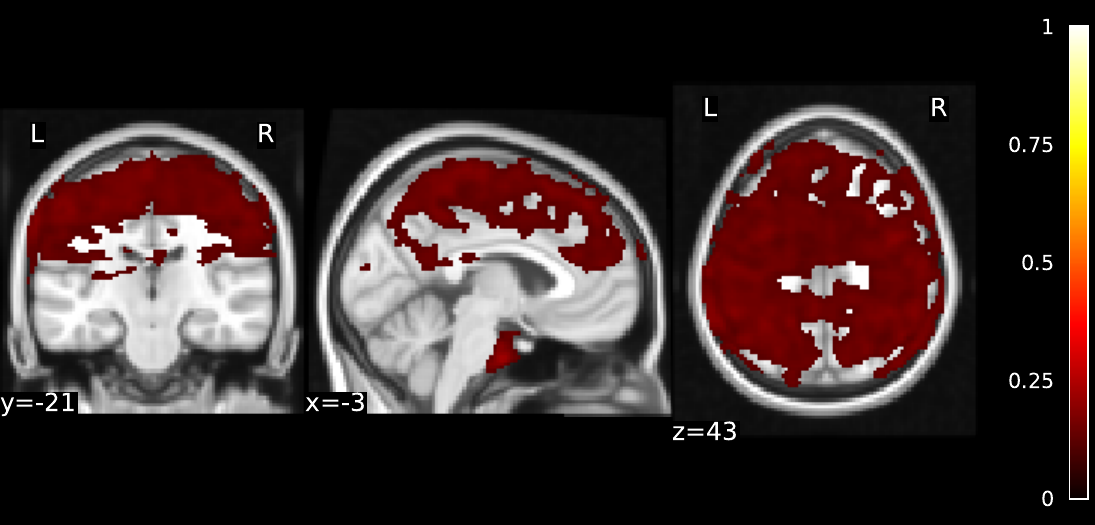}
        \caption{ADHD-200 SwiFT}
    \end{subfigure}

    \vspace{0.3cm}

    \begin{subfigure}{0.3\textwidth}
        \centering
        \includegraphics[width=\linewidth]{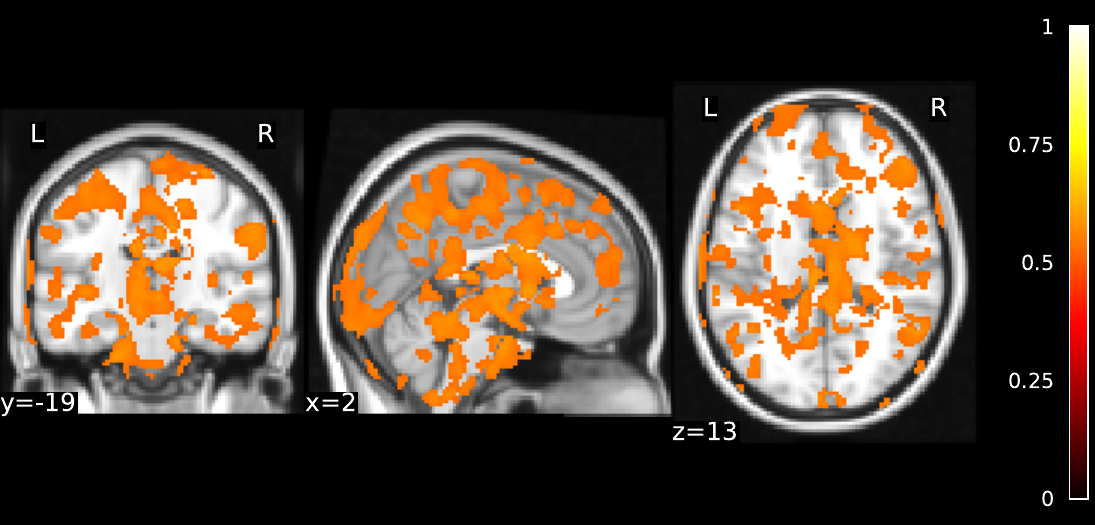}
        \caption{ABIDE~I BrainLM}
    \end{subfigure}
    \quad
    \begin{subfigure}{0.3\textwidth}
        \centering
        \includegraphics[width=\linewidth]{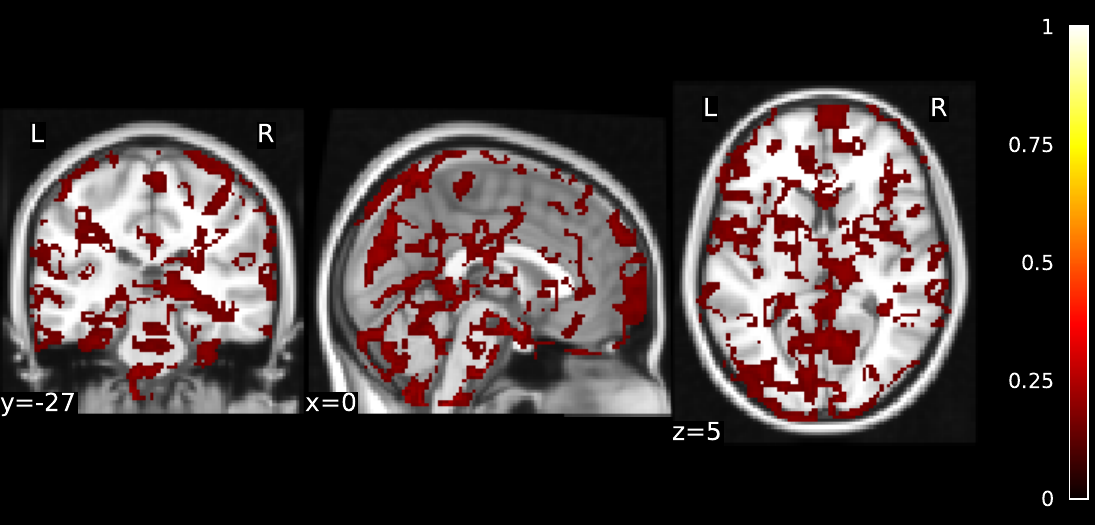}
        \caption{ABIDE~I SwiFT}
    \end{subfigure}
    \caption{Spatial maps of ALFF decoding performance ($R^2$) after ComBat harmonization. Only the top 30\% predictive regions are shown for visualization.}
    \label{fig:alff_3x2}
\end{figure}

\begin{figure}[htbp]
    \centering

    \begin{subfigure}{0.35\textwidth}
        \centering
        \includegraphics[width=\linewidth]{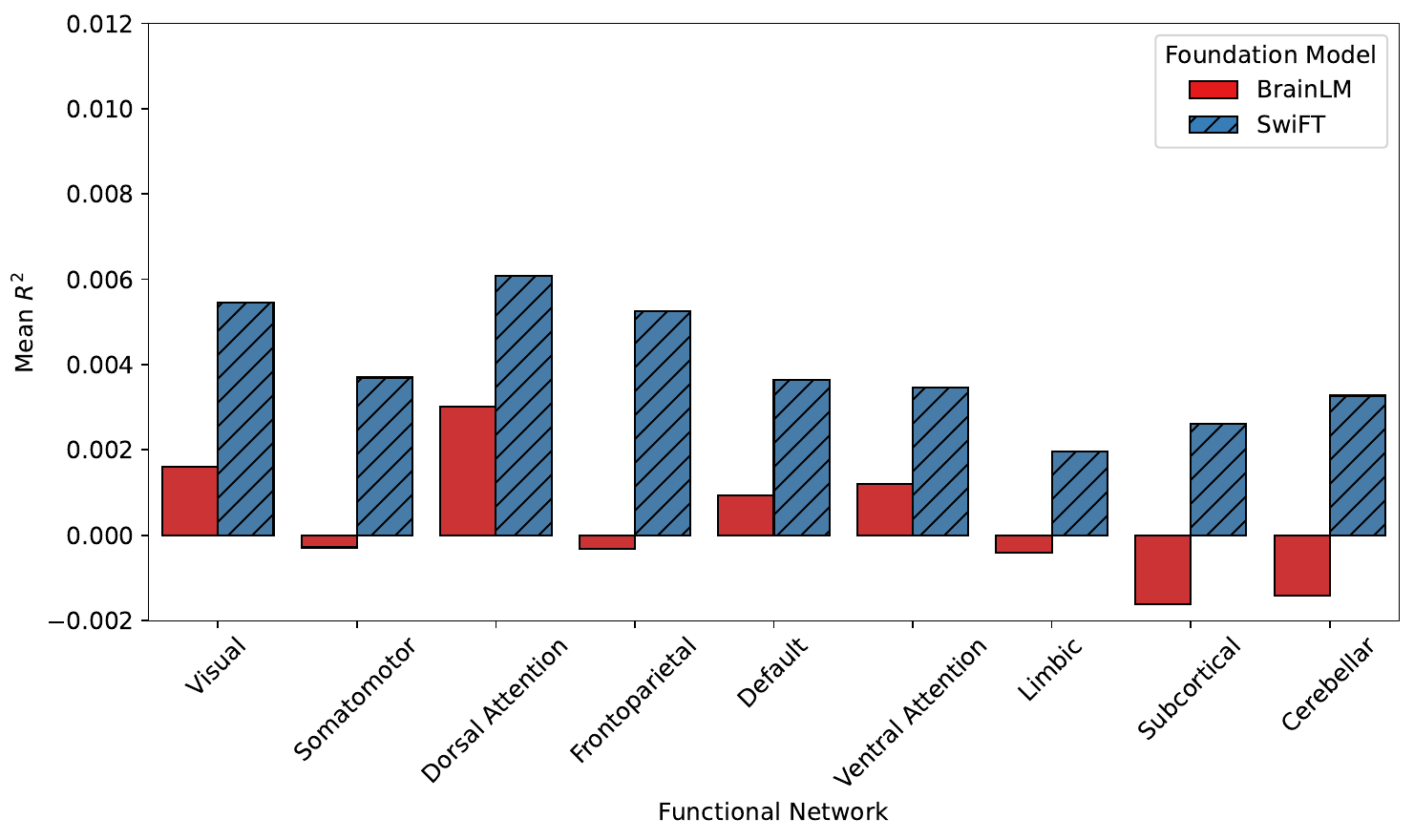}
        \caption{FBIRN}
        \label{fig:fbirn_fnc_sub}
    \end{subfigure}%
    \begin{subfigure}{0.35\textwidth}
        \centering
        \includegraphics[width=\linewidth]{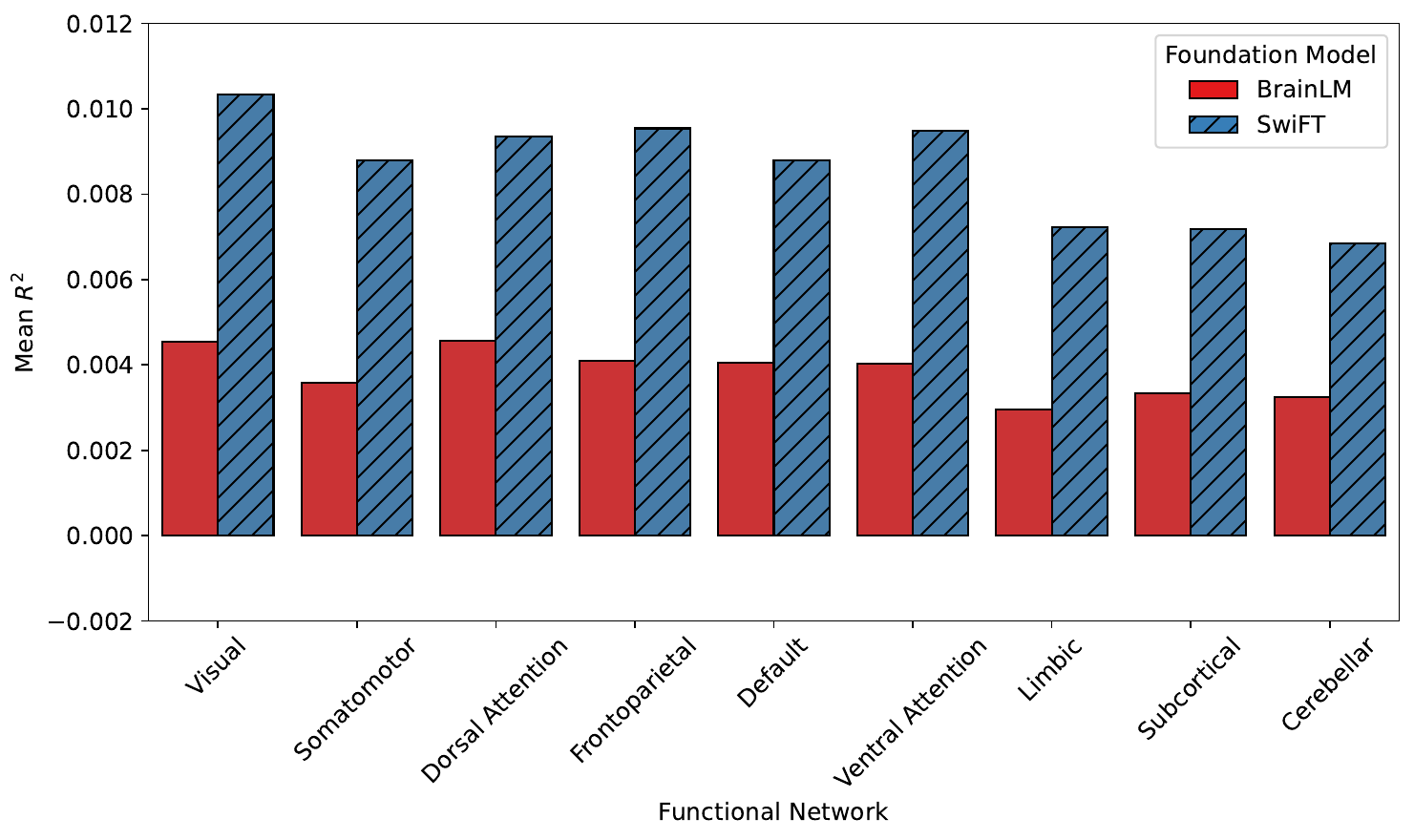}
        \caption{ADHD-200}
        \label{fig:adhd200_fnc_sub}
    \end{subfigure}%
    \begin{subfigure}{0.35\textwidth}
        \centering
        \includegraphics[width=\linewidth]{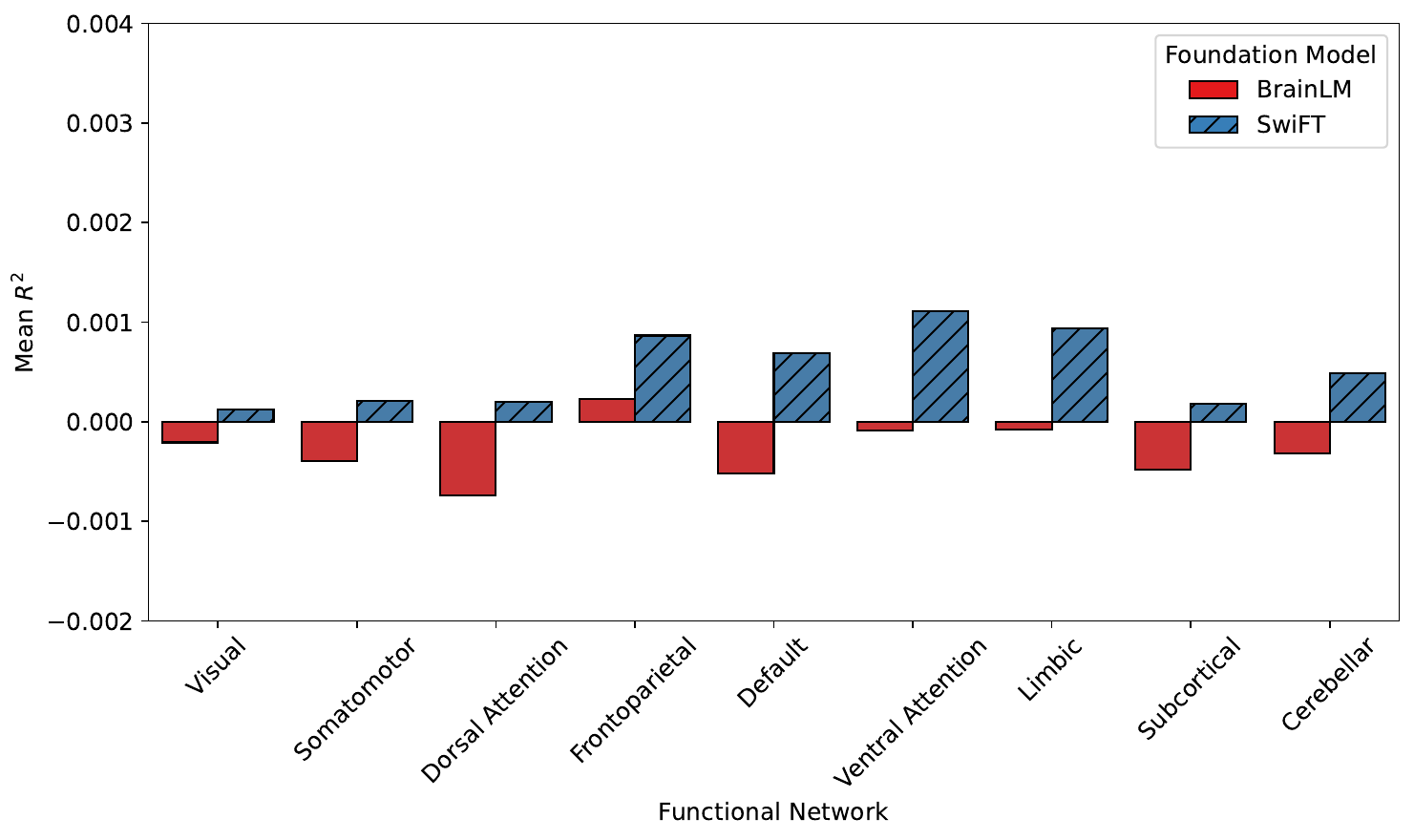}
        \caption{ABIDE-I}
        \label{fig:abide1_fnc_sub}
    \end{subfigure}

    \caption{Network-level mean $R^2$ of FNC decoding after ComBat harmonization. Bars show the average predictive performance within each functional network for BrainLM and SwiFT embeddings.}
    \label{fig:fnc_1x3}
\end{figure}


\end{document}